\documentclass[12pt,a4paper,twoside,fleqn]{book}
\usepackage{amssymb,axodraw}
\usepackage{psfig}
\usepackage[english,german]{babel}

\def\be{\begin{equation}}
\def\ee{\end{equation}}
\def\ba{\begin{eqnarray}}
\def\ea{\end{eqnarray}}

\begin{document}
\selectlanguage{english}

\frontmatter

\begin{titlepage}
\setlength{\topmargin}{-1.5cm}

Diss. ETH No.\ 13068 \\
\vspace{1.0cm}
\begin{samepage}
\begin{center}
  {\LARGE \bf Transport and Magnetism\\ in Mesoscopic Superconductors\\}
  \vspace{2cm}
  A dissertation submitted to the\\ \bigskip
  SWISS FEDERAL INSTITUTE OF TECHNOLOGY ZURICH\\ \smallskip
  (ETH Z\"urich)\\ \bigskip
  for the degree of\\ \medskip
  Doctor of Natural Sciences\\ 
  \vspace{1.8cm}
  presented by\\ \bigskip
  ALBAN LUC ANDR\'E FAUCH\`ERE\\ \bigskip
  Dipl.\ Phys.\ ETH\\ \medskip
  born May 21, 1971\\ \medskip
  citizen of Urdorf ZH, Evol\`ene VS\\
  \vspace{1.8cm}
  accepted on the recommendation of\\ \medskip
  Prof.\ Dr.\ G.\ Blatter, examiner\\ \medskip
  Prof.\ Dr.\ G.\ Sch\"on, co-examiner\\ \medskip
  Dr.\ V.\ Geshkenbein, co-examiner\\
  \vspace{1.5cm} \nopagebreak
  1999
\end{center}
\end{samepage}

\end{titlepage}

\thispagestyle{empty}


\chapter{Acknowledgments}

I am indebted to Gianni Blatter, who has supervised 
my thesis with great attention and diligence, for his choice of a fascinating 
problem to work on, for teaching me a scientific research style, for his 
availability for long hours of debate, for his expertise in writing, and 
for the friendly and team-oriented atmosphere in his group.

I am grateful to Prof.\ Gerd Sch\"on, for being my co-examiner and providing 
several opportunities to pursue scientific interests, at the Nato Advanced 
Study Institute and at the University of Karlsruhe. The 
long-standing collaboration and friendship with Wolfgang Belzig, with whom 
I have shared the passions on paramagnetism and proximity effect, 
has been most valuable to me.

I'm grateful to Maurice Rice for his interest in my work and his 
perspectives on science and life. Vadim Geshkenbein as the advocatus diavoli 
has provided me with sharp-minded criticism and most stimulating discussions.
I thank my friends and colleagues at the institute, Daniel Agterberg, 
Walter Aschbacher, Malek Bou-Diab, Thomas Chen, Matthew Dodgson, Beat 
Frischmuth, Pierre-Alain Genilloud, Denis Gorokhov, Stefan Haas, Elmar Heeb, 
Rolf Heeb, Karyn Le Hur, Thierry M\"uller, 
Anne van Otterlo, Kai Schnee, Andreas Sch\"onenberger, Manfred Sigrist, 
Thomas Siller, Matthias Troyer, Orlando Wagner, and Peter Widerin for fruitful 
exchanges, excellent collaboration on student exercises, late-night 
philosophical discussions and shared bottles of wine. I'm grateful to Klaus 
Hepp for promoting me within the Swiss Study Foundation. 
Urs Ledermann contributed to this thesis through his diploma project. 

I have drawn much motivation from the enthusiasm of Prof.\ Ana-Celia Mota 
for her experiments and have profited from numerous discussions with 
Roberto Frassanito and Bernd M\"uller-Allinger. I acknowledge fruitful 
discussions and collaborations with Christoph Bruder, Rosario Fazio, 
Mikhail Feigel'man, Christian Glattli, Lev Ioffe, Gordey Lesovik, 
Thomas Meyer, Marc Sanquer, and Andrei Zaikin.

Finally, I am most grateful to Claudia, Alexandre, Ivan, Marie-Pierre, Kay, 
Gis\`ele and Jean-Luc, who have been a wonderful family to me all these years.

\newpage
\thispagestyle{empty}
\cleardoublepage

\tableofcontents

\newpage

\chapter{Abstract}

Superconductivity, discovered by Kamerlingh Onnes in 1911, continues to be 
a fascinating subject of condensed matter physics today. Much interest 
has been devoted to the study of the superconductivity induced in a metal 
which by itself is not superconducting but is in electrical contact with a 
superconductor. As the carriers of superconductivity, the Cooper pairs, 
diffuse across the contact into the metal they remain correlated, although 
the pairing mechanism is lifted; we call this the proximity effect. 
The observation of these superconducting correlations has come within the 
reach of experiments in the last decade. With state-of-the-art fabrication 
techniques mesoscopic 
samples have been produced which are small and clean enough for the quantum 
mechanical coherence of the electrons to be preserved over the sample size. 
This thesis focuses on the variety of signatures of single-particle  
physics that appear in the electrical transport and the magnetic screening 
properties of these systems. 

The main contribution of this thesis to the field of mesoscopic 
superconductivity consists in the study of the thermodynamic properties 
in the presence of a magnetic induction. We map out the phase diagram 
of a normal-metal--superconductor (NS) hybrid system in the space spanned by 
the temperature ($T$) and the magnetic field ($H$). 
The region of the $H-T$ diagram 
of interest lies well below the transition temperature and the critical 
field line, i.e.\ in the Meissner phase, of the superconductor. The involved 
energy scales are determined by the geometric system size $d$ and the mean 
free path $l=v_F\tau$ of the normal metal and are given by the single-particle 
level spacing $\sim \hbar v_F/d$ or the Thouless energy $\sim\hbar v_F l/d^2$, 
respectively, in ballistic or diffusive systems.

We distinguish two phases in the $H-T$ diagram: a diamagnetic 
state (at low temperature and field) where the applied field is 
effectively screened, and a normal state where the magnetic field penetrates 
into the normal metal. The magnetic breakdown of the Meissner field expulsion 
occurring at the frontier of the two regimes is interpreted as the 
critical field of the proximity layer. We give 
a complete characterization of this first order transition for a ballistic 
NS structure, determining the spinodals, the magnetization jump, 
as well as the latent heat, and find a critical point, beyond which the two 
phases are connected by a continuous cross-over.

We study the magnetic screening in the diamagnetic phase for a NS hybrid 
layer. We find that the current--field relation is typically non-local in 
the proximity effect producing a high sensitivity of the diamagnetic 
susceptibility to the mean free path. We classify the behavior into several 
regimes between the clean ($l\to \infty$) and the dirty ($l\to 0$) limit. 
The reduction of the screening density and the non-locality range 
with decreasing purity affect the screening in an opposite way,  
competing in the diamagnetic susceptibility observed in the experiments. 

The screening properties can be traced back to the quasi-particle excitation 
spectrum in the absence of the field. We thus establish a connection to the 
transport properties of NS junctions: We explain the resonance structure 
in the conductance and the shot noise at finite voltage, which appears as 
a signature of the quantum behavior of the discrete electrons.

Back to the $H-T$ phase diagram: in the field penetration phase, 
where the screening is negligible, we analyze the geometry of a ballistic 
SNS Josephson link. We show that the combination of the non-locality and 
the finite size of the junction produces an anomalous critical-current--flux 
dependence. As a function of increasing field $H$ we find a cross-over from 
the usual single to a double flux quantum periodicity.

In the low-temperature low-field corner of the $H-T$ phase diagram we discover 
a new paramagnetic phase by considering the effect of a repulsive 
electron--electron interaction in the proximity metal. Quasi-particles are 
found to be trapped at the NS interface, accumulating to a density of states 
peak at the Fermi energy. This peak induces a paramagnetic instability towards 
spontaneous magnetic moments. On top of the Meissner phase we find a first 
order transition as the moments polarize in the applied field. The hysteresis 
and the dissipation accompanying a paramagnetic reentrance in the 
susceptibility provide the characteristic features observed in fascinating 
experiments on NS cylinders. 

Finally, we consider the transport in Josephson junctions with unconventional 
superconductors, where the current--phase relation exhibits a double frequency.
We propose to exploit this $\pi$-periodicity for the implementation of a 
quantum computer.

\newpage

\chapter{Zusammenfassung}

Supraleitung, in 1911 durch Kamerlingh Onnes entdeckt, is heute
noch ein faszinierendes Gebiet der Festk\"orperphysik. Die Untersuchung 
der Supraleitung, welche in einem Metall induziert wird, das selbst nicht 
supraleitend ist, jedoch im elektrischen Kontakt mit einem Supraleiter steht, 
hat ein grosses Interesse erfahren. Wenn die Tr\"ager der Supraleitung, die 
Cooper-Paare, durch den Kontakt in das Metall diffundieren, behalten sie ihre 
Korrelationen, obwohl der Paarungsmechanismus verschwindet. In der letzten 
Dekade ist die Beobachtung der supraleitenden Korrelationen zwischen einzelnen 
Quasi-Teilchen in die Reichweite der Experimente ger\"uckt. Die modernsten 
Fabrikationstechnologien haben die Herstellung von mesoskopischen Systemen 
erm\"oglicht, welche klein und sauber genug sind, um die quantenmechani\-sche 
Phasenkoh\"arenz der Elektronen \"uber die Probengr\"osse zu gew\"ahrleisten. 
Diese Dissertation widmet sich der Vielfalt der Einteilchen-Physik, welche 
im elektrischen Transport und der magnetischen Abschirmung in diesen Systemen 
auftritt.

Der Hauptbeitrag dieser Dissertation zum Feld der mesoskopischen Supraleitung 
besteht im Studium der thermodynamischen Eigenschaften unter dem Einfluss 
einer magne\-tischen Induktion. Wir bestimmen f\"ur eine Metall-Supraleiter 
(NS) Hybridstruktur das Verhalten im Phasendiagramm, welches durch die 
Temperatur ($T$) und das Magnetfeld ($H$) aufge\-spannt wird. Der untersuchte 
Bereich im $H-T$ Diagramm liegt deutlich unter der \"Ubergangstemperatur des 
Supraleiters sowie der kritischen-Feld-Grenze. Die geometri\-schen Masse $d$ 
und die freie Wegl\"ange $l= v_F \tau$ bestimmen die typi\-sche Energieskala
im Problem: die Einteilchen-Energie $\sim \hbar v_F/d$ beziehungsweise 
die Thouless-Energie $\sim \hbar v_F l/d^2$ in ballistischen oder diffusiven 
Systemen.

Wir unterscheiden zwei Phasen im $H-T$ Diagramm: einen diamagnetischen Zustand 
(bei kleinen Temperaturen und Feldern), der sich durch die Abschirmung der 
angelegten Felder auszeichnet, und einen normalleitenden Zustand, wo das 
Magnetfeld ungehindert in das Metall eindringt. Wir bestimmen das kritische 
Feld, das die beiden Phasen trennt, und charakterisieren diesen \"Ubergang 
erster Ordnung in einem ballistischen NS Schichtsystem durch die 
Spinodalen, den Sprung in der Magnetisierung sowie die latente W\"arme. 
Wir finden, dass die Phasengrenze oberhalb eines kritischen Punktes 
verschwindet.

In der Beschreibung der diamagnetischen Phase finden wir eine f\"ur die 
induzierte Supraleitung typische nicht-lokale Strom--Feld Relation, welche 
die diamagnetische Suszeptibilit\"at zu einem feinen Indikator der freien 
Wegl\"ange macht. Wir unterscheiden mehrere Verhalten zwischen dem 
reinen ($l\to \infty$) und dem schmutzigen ($l\to 0$) Grenzfall. Mit 
abnehmender Reinheit haben die Reduktion der abschirmenden Dichte sowie der 
Reichweite der Strom--Feld Relation eine entgegengesetzte Wirkung 
auf die Abschirmung und stehen daher im Wettbewerb in der experimentell 
gemessenen diamagnetischen Suszeptibilit\"at. 

Das Abschirmverhalten kann auch auf das Quasiteilchen-Spektrum 
zur\"uckgef\"uhrt werden. Dies ergibt eine Verbindung zu den 
Transport-Eigenschaften in NS Strukturen: Wir untersuchen die Resonanzen im 
Leitwert und finden sie wieder in den quantenmechanischen Stromfluk\-tuationen 
aufgrund der Diskretheit der Elektronenladung.

In der normalleitenden Phase, wo die Abschirmung vernachl\"assigt werden kann, 
untersuchen wir die Anordnung eines ballistischen SNS Josephson Kontaktes. 
Wir zeigen, dass die Verbindung von Nichtlokalit\"at und der endlichen 
Ausdehnung des Metalls zu einer anomalen Abh\"angigkeit des kritischen Stromes 
vom magnetischen Fluss f\"uhrt. Mit zunehmendem Feld finden wir 
einen \"Ubergang von der einfachen Periode eines Fluss-Quantums zur doppelten. 

Wir entdecken die Existenz einer neuen paramagnetischen Phase, welche durch 
eine repulsive Elektron-Elektron Wechselwirkung verursacht wird, in der 
N\"ahe des Ursprungs des $H-T$ Phasendiagramms. Quasiteilchen 
werden an der NS Grenz\-fl\"ache festgehalten und tragen zu einer Spitze 
in der Zustandsdichte bei der Fermi-Energie bei. Diese verursacht eine 
paramagnetische Instabilit\"at, welche spontane magnetische Momente 
hervorbringt. Dem diamagnetischen Verhalten \"uberlagert fin\-den wir  
das die Polarisation der magnetischen Momente zu einem Phasen\"ubergang 
erster Ordnung bei Feld Null f\"uhrt. Die Hysterese und Dissipation, welche 
das paramagnetische Signal in der Suszeptibilit\"at begleiten, sind in 
qualitativer \"Ubereinstimmung mit bisher noch unerkl\"arten Experimenten auf 
mesoskopischen Zylindern.

Schliesslich betrachten wir den Transport in Josephson Kontakten mit 
unkonventionellen Supraleitern, wo eine doppelte Josephson-Frequenz 
auftritt. Wir schlagen vor, diese $\pi$-Periodizit\"at in der Phase zur 
Implementation eines Quanten-Computers zu benutzen.

\mainmatter

\chapter{Introduction}
\label{introduction}

In this thesis we study the behavior of electrons in a metal which 
is in electrical contact with a superconductor. The superconductor (S) can be 
viewed as a reservoir where due to an effective attraction the electrons 
condense into Cooper pairs, while the normal metal (N) is idealized as 
a gas of free electrons. Through the diffusion across the NS contact the 
metallic electrons acquire superconducting correlations; we call this the 
proximity effect.

The Cooper pairs are essentially described by one wavefunction (the order 
parameter of the superconductor) producing quantum behavior on a macroscopic 
scale. The quantum interference effects between the free electrons instead 
rely on the coherence of the single-particle wavefunctions and are much more 
fragile. The observable quantum phenomena are limited to the mesoscopic scale 
given by the phase coherence length here. The proximity effect offers an 
experimental access to the coherent single-particle physics by combining it 
with the macroscopic superconducting correlations.

Here we study the electrical transport and the orbital magnetism of electrons 
in normal-metal -- superconductor hybrid structures. Our theoretical work 
is motivated by an intense experimental activity in this area of mesoscopic 
physics triggered by recent advances in nanostructure fabrication technology. 
For an introduction to the subject we refer the reader to the review on 
``Mesoscopic Electron Transport'' \cite{curacao}.

The microscopic process underlying the physics of proximity induced 
superconductivity is the Andreev reflection (AR) at a NS interface: 
An electron incident from the normal metal into the superconductor is 
retro-reflected into a hole, transmitting the double electron charge into 
the superconductor \cite{andreev}. In the following, we give an 
introduction to the basic properties of the AR and relate the works of this 
thesis to this fundamental process.\\
\pagebreak

The quasi-particle transfer at a NS interface is best described by the 
Bogoliubov-de-Gennes (BdG) equations \cite{deGennes}, which 
determine the two-component wavefunctions $\Phi$ of electron and hole 
close to the Fermi level,
\be \left( \begin{array}{cc}
   h_0 -\mu & \Delta \\
   \Delta^* & -h_0^* +\mu \end{array} 
\right) 
\left( \begin{array}{c}u\\  v\end{array} \right)
= \epsilon  \left( \begin{array}{c}u\\  v\end{array} \right),
\quad\quad \Phi = \left( \begin{array}{c}u\\  v\end{array} \right) 
e^{i {\bf k}_F{\bf x}} .
\label{linbdg}
\ee
We use the linearized Hamiltonian $h_0-\mu = -i\hbar {\bf v}_F \nabla$, 
where $v_F=\hbar k_F/m$ is the Fermi velocity, $\Delta$ provides the 
off-diagonal pair-potential and $\epsilon$ is the quasi-particle energy 
with respect to the chemical potential. Let us consider a 
NS junction as described by the pair potential 
\be 
\Delta(x) = \left\{ \begin{array}{lll}0,\quad\quad\quad & x<0\quad & (N),\\
                                      \Delta_0 e^{i\varphi} & x>0 & (S).
\end{array} \right. 
\ee
While in the normal metal the electron and hole states decouple ,
\be \Phi_+ = \left( \begin{array}{c} 1\\ 0\end{array}\right) e^{ik_+ x}, 
\quad\quad\quad
\Phi_- = \left( \begin{array}{c} 0\\ 1\end{array}\right) e^{ik_- x} 
\ee
($k_{\pm} = k_x \pm \epsilon/\hbar v_x$, $\hbar k_x/m=v_x=v_F\cos\vartheta$), 
the quasi-particles in the superconductor are coupled evanescent modes,
\be
\Phi_e = \left( \begin{array}{c} 1\\ \gamma e^{-i\varphi} \end{array}\right) 
e^{ik_e x},\quad\quad\quad
\Phi_h = \left( \begin{array}{c} \gamma e^{i\varphi} \\ 1 \end{array}\right) 
e^{ik_h x}
\ee
($\gamma = \exp [-i \arccos (\epsilon/\Delta)]$, 
$k_{e,h}  = 
k_x \pm i\sqrt{\Delta_0^2-\epsilon^2}/\hbar v_x$, $\epsilon < \Delta$ ).
By using the ansatz $\Phi_+ + r_A \Phi_-$ for $x<0$ and $\Phi_e$ for $x>0$ 
we easily find the Andreev reflection amplitude from the continuity of
the wavefunction,
\be 
r_A = \gamma e^{-i\varphi} = \exp (-i\arccos \frac{\epsilon}{\Delta}) 
\, e^{-i\varphi}.
\ee
For $\epsilon<\Delta$, the incident quasi-particle excitation is reflected 
with probability $|r_A|=1$. 
Similarly the Andreev amplitude describing the conversion of an incident hole
to a reflected electron is given by $\gamma \exp(i\varphi)$. Andreev thus
found that the incident quasi-particle excitation is prevented from 
entering the superconductor, and consequently heat (or energy) transport 
across the NS interface is suppressed \cite{andreev2}. 
On the other hand, the double charge $2e$ (of incident electron and reflected 
hole) is transferred across the NS interface. In the superconductor the  
decaying quasi-particle current is converted into supercurrent \cite{btk}.

The Andreev reflection (AR) coherently couples electron and hole states 
and thus induces the superconducting correlations $F(x)$ in the normal metal, 
\be
F(x) = \langle \hat{\Psi}_{\downarrow}(x) \hat{\Psi}_{\uparrow}(x) \rangle =
\sum_{\alpha} u_{\alpha}(x) v_{\alpha}^*(x) f(\epsilon_{\alpha}),
\label{fandre}
\ee
where the index $\alpha$ denotes the eigenstates and $f(\epsilon)$ is the Fermi 
occupation number, see also appendix \ref{bdgapp}. 
Instead of a Cooper pair condensate, the proximity metal exhibits correlated 
electron-hole pairs, which are mutually independent. Here lies the fragility 
of the proximity effect: The thermally excited Andreev pairs at energy 
$\epsilon \sim T$ are dephased over 
the thermal length $\xi_N(T) = \hbar v_F/2\pi T$ and the correlation 
function (\ref{fandre}) is suppressed as $\sim \exp[-x/\xi_N(T)]$ 
(see Chp.\ \ref{quasiclassprox}). \\

\begin{figure}[tb]
\centerline{\psfig{figure=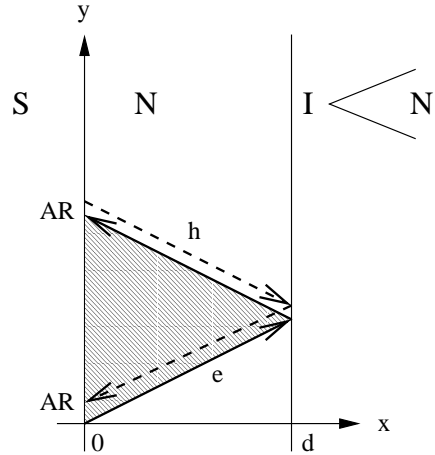,height=6cm}}
\caption{Trajectories of electron-hole Andreev levels in a proximity 
normal-metal 
layer: electron and hole travel back and forth and are converted into each 
other by the Andreev reflection (AR). Top: Coupling of the NS slab by a tunnel 
contact to a NINS junction.}
\label{introfigns}
\end{figure}

Let us consider in the following a normal metal slab of thickness $d$ 
on top of a bulk superconductor. Due to the AR process, the quasi-particles 
at subgap energies $\epsilon <\Delta$ are bound to the normal layer, 
see Fig.\ \ref{introfigns}. The
discrete energies of the Andreev bound states follow from a constructive 
interference condition for an electron and a hole traveling back and 
forth across the normal layer and being converted into each other at the 
NS interface ($n=0,1, ...$),
\ba
&& 2 k_+ d - 2 k_-d - 2 \arccos \frac{\epsilon}{\Delta} = 2 n\pi,
\label{intro:coninter} \\
&& \epsilon_n (v_x) = \frac{\hbar v_x}{2 d} 
(n\pi + \arccos \frac{\epsilon}{\Delta} ) \, 
\stackrel{\epsilon \ll \Delta}{\approx} \, 
\frac{\hbar v_x \pi}{2 d} ( n +\frac12 ).
\label{intro:spectrum}
\ea
The bound states are quantized at energies $\sim \hbar v_x/d$ depending only 
on the component of the Fermi velocity in forward direction 
($v_x=v_F\cos\vartheta$). The energy 
levels are repelled from the chemical potential, as a finite energy mismatch 
of the wave vector $k_{\pm}$ of electron and hole is needed to compensate 
for the phase $\arccos(\epsilon/\Delta) \approx \pi/2$ of the AR. 
The discrete Andreev levels contribute to the finite density of states at 
$\epsilon < \Delta$ filling the superconducting gap. 

The transport through the ballistic NS structure coupled by means of a 
tunneling barrier (I) to a NINS junction, see Fig.\ \ref{introfigns}, 
reveals a resonance structure at the energies of the Andreev (quasi)-bound 
states $\epsilon \sim \hbar v_F/d$ \cite{rowell}. 
A similar behavior is expected in a diffusive system, where the average 
path traveled $d$ is to be replaced by $d^2/l$ ($l$ mean free path). According 
to the constructive interference condition 
$ \epsilon d^2/\hbar v_F l \sim (2n+1)\pi$ the resonances are found at the 
Thouless energy $E_c \sim \hbar v_F l/ d^2$. 
In chapter \ref{nonlinearity} we analyze the spectral conductance $G(\epsilon)$ 
and the shot noise $S(\epsilon)$ in NS junctions with an arbitrary scattering 
region in the normal lead, 
which typically includes a barrier at the NS interface and additional 
disorder in the normal lead. We describe the single-particle resonance 
structure that follows from the interplay between the normal scattering and 
the AR processes and enhances the subgap transport. 

The linear transport coefficients $G(0)$ and $S(0)$ in a NS junction are 
suppressed by the transmission $T^2$ of the N conductor, reflecting the 
two-particle nature of the AR process \cite{btk,beenakker,Jong}. 
This explains why the first observation of an enhanced subgap 
conductance could be misinterpreted as being due to supercurrents 
\cite{kastalsky}.
Actually, the iterative scattering processes between the disorder and the 
{\it interface barrier} increase the possibility for an AR as compared to 
a simple tunneling junction \cite{melsen}. 
If, on the other hand, the interface barrier is weak, the resonance structure 
of the Andreev states, which are el-hole quasi-bound states between the 
disorder 
and the {\it superconductor}, determine the transport coefficients. 
The conductance $G(\epsilon)$ as well as the shot noise $S(\epsilon)$ exhibit 
peaks at the Andreev or the Thouless energy, respectively, for a ballistic 
or a diffusive conductor. Characteristically these resonances are 
pinned at a finite energy above the Fermi level, which we understand in 
terms of the $\pi/2$ phase shift in the AR process, and survive 
in the multichannel limit. The reentrance effect observed in NS junctions 
with a diffusive lead \cite{petrashov,charlat,sanquer} reflect the 
non-monotonicity of the spectral conductance due to the underlying
Andreev resonances. 
While theoretical works using the Green's function technique have explained 
these experiments accurately \cite{yip,nazarov,stoof,golubov,wilhelmns}, 
the Landauer-B\"uttiker approach offers an intuitive understanding here, 
based on a ballistic viewpoint. The diffusive system is described by 
averaging over an ensemble of impurity configurations, see 
\cite{marmokos,argaman:euro,raimondi}. In Chp.\ \ref{nonlinearity} we 
present new results for the shot noise, which are of interest with respect 
to ongoing noise experiments in NS junctions \cite{jehl,kozhevnikov}.\\

The quantization of the Andreev levels at discrete energies away from 
the Fermi level suppresses the density of states (DOS) of the proximity 
layer as compared to the bulk metal. From the low energy spectrum 
(\ref{intro:spectrum}) we obtain the Andreev DOS \cite{james}
\ba
N(E) &=& \frac{1}{d} \int \frac{dk_y dk_z}{2\pi^2} 
\delta[E-\epsilon_0(v_x)]
\\ 
&\sim&  N_0 \frac{\epsilon d}{\hbar v_F}
\label{intro:dos} 
\ea
($v_x=\hbar \sqrt{k_F^2-k_y^2-k_z^2}/m$, $N_0=m k_F/\hbar^2 \pi^2$). 
The excitation gap $\Delta$ is softened in the normal metal to a linearly 
vanishing DOS below $\hbar v_F/d$. 
We note that similar results for the DOS are observed in 
diffusive NS hybrid structures \cite{gueron,belzig:dos}.

The DOS suppression close to the Fermi level affects the response properties 
of the proximity layer. In chapter \ref{magresponse} we consider the magnetic 
response, i.e., the screening currents induced by an applied magnetic field. 
Generically, the linear current 
response $j=j_{\rm dia}+j_{\rm para}$ is given by the sum of the diamagnetic 
current describing the response of the rigid wavefunction and the paramagnetic 
current induced by the perturbation of the wavefunction in the applied field. 
While the diamagnetic current is trivially proportional to the bulk electron 
density, the paramagnetic current depends on the DOS at the Fermi surface 
\cite{schrieffer}. The solution of the screening problem requires the 
full knowledge of the dispersive relation $j(q)= -(c/\lambda^2) K(q)A(q)$ 
which has to be solved together with the Maxwell equation $j(q)=q^2A(q)$.
($\lambda = (4\pi n e^2/mc^2)^{-1/2}$ is the London penetration depth). 
The variation of the field on wavelength $1/q$ typically induces 
quasi-particle 
excitations at energy $E\sim \hbar q v_F$ giving a paramagnetic current.
Depending on the availability of the finite energy excitations, the 
paramagnetic current either vanishes or compensates for the diamagnetic 
current. 

In a BCS superconductor with excitation gap $\Delta$, the paramagnetic 
current is gapped (the DOS vanishes) for long wavelength and the response 
is purely diamagnetic, $K(q)\sim 1$ for $\hbar q v_F < \Delta$. Above the 
gap the paramagnetic current is found to cancel the diamagnetic current 
leaving a small response $K(q) \sim \Delta/\hbar q v_F$. In real space this 
implies a range of the current--field relations of the order of 
$\xi_0 \sim \hbar v_F/\Delta$. For a London length $\lambda>\xi_0$ we find a local 
response, while for $\lambda<\xi_0$ the current--field dependence is 
nonlocal according to a Pippard-type relation \cite{pippard}. In a bulk 
normal metal, the continuous density of states at the Fermi level implies 
a cancellation of dia- and paramagnetic currents, leaving no net screening 
response.

In a clean proximity layer, the suppression of the DOS produces a net 
diamagnetic current, but the lack of a quasi-particle gap translates 
to a diverging nonlocality range. It turns out that for finite $q$ the 
paramagnetic current is able to compensate the diamagnetic current, 
producing the kernel $K(q) \sim \delta(q)$, and thus the nonlocal current 
functional \cite{zaikin}
\be j(x) \sim  -\frac{c}{\lambda^2d} \int_0^d dx' A(x') .
\ee
We study this screening problem in more detail in Chp. \ref{magresponse}, 
where the magnetic response of a NS slab with arbitrary impurity 
concentration is discussed. 
We introduce the powerful quasi-classical Green's function 
technique to derive the general linear current functional. We find that
a finite mean free path provides a cutoff for the nonlocality range, but
that the current--field relations in the proximity effect nevertheless
remain typically nonlocal. We distinguish several regimes for the magnetic 
response depending on the geometric thickness $d$, the thermal length 
$\xi_N(T)$ and the mean free path $l$. The reduction of the screening density 
and the nonlocality range with decreasing purity affect the susceptibility 
in different ways, producing a non-monotonic dependence in the diamagnetic 
susceptibility with changing mean free path. This understanding provides 
the basis for a quantitative fit with the experiments \cite{mmb}.\\

The nonlocal field dependence is also found in the spectrum of the Andreev 
states. Replacing the gradient in (\ref{linbdg}) by 
$\nabla + ie {\bf A}(x)/\hbar c$ we can account for the vector potential by
adding an Aharonov-Bohm-type phase to the constructive interference 
condition (\ref{intro:coninter}), yielding the bound state energies
\be 
\epsilon_n = \frac{\hbar v_x}{4d} \left[ (2n+1)\pi  
- \frac{2\pi}{\Phi_0} \oint {\bf A} d{\bf x} \right]. 
\label{intro:efield}
\ee
The circular integral gives the flux $\Phi = \oint {\bf A} d{\bf x}$ 
enclosed between the quasi-classical trajectory and the NS interface, 
see Fig.\ \ref{introfigns} (hatched region). The flux enters 
modulo an integer number of flux quanta $\Phi_0=hc/2e$. This leads to a 
random distribution of the Andreev levels at large field $\Phi \gg \Phi_0$, 
suppressing the energy gap and inducing the magnetic breakdown in the 
proximity layer. The magnetic breakdown denotes the proximity layer's own 
critical field $H_b$, at which the Meissner field expulsion ceases to be
effective and the applied field penetrates into the normal layer
by a first order transition. While in a small applied magnetic field the Andreev 
pairs produce diamagnetic screening currents, at large fields the 
superconducting correlations are destroyed by the random Aharonov-Bohm 
phases in (\ref{intro:efield}) and the net screening current vanishes. 
In Chp.\ \ref{breakdown} we determine the $H-T$ phase diagram for this 
behavior, finding the first order transition line separating the 
diamagnetic phase of effective field expulsion and the normal state of 
field penetration. 

The magnetic breakdown has been observed recently in increasingly clean
samples \cite{oda:80,mota:82,pobell:87,visani} and we find good agreement 
of both temperature and thickness dependence of the breakdown field 
with experimental data on quasi-ballistic Ag-Nb samples \cite{mota:89,mmb}.\\

\begin{figure}[tb]
\centerline{\psfig{figure=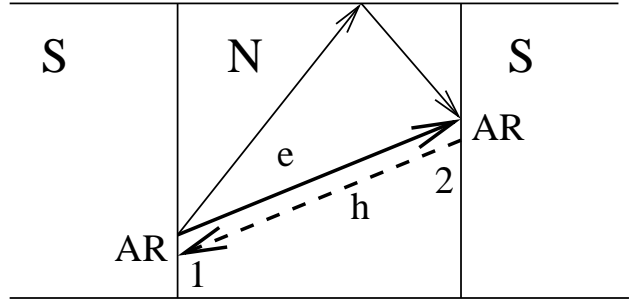,height=4cm}}
\caption{Trajectories of electron-hole Andreev levels in a normal-metal -- 
superconductor -- normal metal (SNS) junction. Note the existence of 
trajectories reflecting at the metal boundary.}
\label{introfigsns}
\end{figure}

We study a different aspect of nonlocality in Chp. \ref{nonlocalsns}, 
where we consider a ballistic normal metal layer connected by two 
superconducting reservoirs to a SNS junction. The two NS interfaces act 
as a confining potential for the electrons and holes in the normal 
interlayer. The junction exhibits current-carrying bound states consisting of 
an electron traveling forward and a hole traveling backwards which are 
converted into each other by the AR \cite{kulik}, see Fig.\ \ref{introfigsns}.  
The constructive interference condition, 
\be
(k_+-k_-)d \mp \Delta\varphi = (2n+1) \pi,
\label{intro:phasecondsns}
\ee
now features the phase difference $\Delta\varphi$ of the two superconducting 
reservoirs (the sign of $\Delta\varphi$ depending on the orientation of the 
quasi-particle current) and implies the spectrum of bound states
\be 
\epsilon_n (v_x) = \frac{\hbar v_x}{d} (n\pi+ \frac{\pi}{2} \pm \Delta\varphi).
\ee
These bound states are known to transmit the supercurrent between the 
two reservoirs: While at phase difference $\Delta\varphi=0$ the left- and 
right-going states are degenerate and their respective currents cancel, by 
changing $\Delta\varphi$ an asymmetry is introduced producing the supercurrents 
in the junction.

Including a magnetic induction as above, the bound state energies 
\be 
\epsilon_n = \frac{\hbar v_x}{d} \left[ n\pi + \frac{\pi}{2}
 \pm \underbrace{ \left( \Delta\varphi - \frac{2\pi}{\Phi_0} \int_1^2 {\bf A} 
\, d{\bf x} \right) }_{\gamma}  \right],
\ee
depend on the gauge invariant phase difference ($\gamma$) between the two 
endpoints
$1$ and $2$ of the quasiparticle trajectories connecting the superconducting 
reservoirs, see Fig.\ \ref{introfigsns}. 
The currents $j\propto \partial E/\partial\gamma$ depend 
non-locally on the field ${\bf A}({\bf x})$. The Josephson 
current through the junction is given by the sum of the currents carried 
along all the quasi-particle trajectories across the junction. 
In Chp.\ \ref{nonlocalsns} we find that in a junction of finite width  
the trajectories with multiple reflections at the metallic boundary alter the 
field dependence of the Josephson current, as they experience a different 
gauge invariant phase difference, see Fig.\ \ref{introfigsns}. 
The current--flux relation $I=I(\Phi)$ 
crosses over from the usual $\Phi_0$- to the anomalous $2\Phi_0$-periodicity 
as a function of increasing field. The cross-over takes place when the 
Josephson vortex distance 
$a_0=\Phi_0/Hd$ becomes smaller than the nonlocality range of the junction.
Our results explain the experimental data recently obtain on S-2DEG-S
junctions \cite{bib:Heida}, where the ballistic link was provided by 
a two-dimensional electron gas (2DEG).\\

\begin{figure}[tb]
\centerline{\psfig{figure=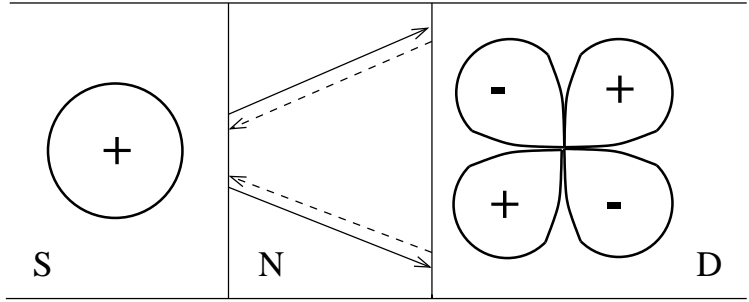,height=4cm}}
\caption{Trajectories of representatives from two Andreev level populations 
shifted by a phase difference $\pi$.}
\label{introfigsnd}
\end{figure}

Let us consider the Josephson relation 
$I=I(\Delta\varphi)$ in a junction between a conventional, S-wave 
superconductor (S) and a High-$T_c$ superconductor with D-wave symmetry (D).
The order parameter in the D-wave superconductor is direction-dependent 
as $\sin 2\vartheta$ exhibiting two lobes with positive sign and two lobes 
with negative sign, see Fig.\ \ref{introfigsnd}. The phase difference 
$\Delta\varphi$ across the SND 
junction thus depends on the orientation of the quasi-particle trajectory, 
varying by $\pi$ for trajectories connecting with the negative rather 
than the positive lobes of the D-wave material. For the alignment of the 
D-wave superconductor with a gap $\Delta \propto \sin 2\vartheta$ with 
respect to the $x$-axis, the two quasiparticle populations with $\vartheta>0$ 
and $\vartheta<0$, see Fig.\ \ref{introfigsnd}, 
correspond to two conventional Josephson junctions  shifted by $\pi$ with 
respect to each other. The spectrum of the bound states 
is given by Eq.\ (\ref{intro:phasecondsns}), where $\Delta\varphi$ differs 
by $\pi$ for the two quasi-particle populations. Therefore, the $2\pi$ 
periodicity of the critical current of each junction is reduced to a 
$\pi$ periodicity in the SND junction. 

In chapter \ref{quietsds} we propose to exploit the doubly periodic Josephson 
relation for the implementation of a quantum bit (qubit). The 
$\pi$-periodic junction can be viewed as a macroscopic two-level 
system exhibiting two degenerate ground states which carry no current
and are thus optimally isolated from the environment. The macroscopic 
quantum tunneling between these two states opens the possibility for a 
quantum time evolution of the junction, which we propose to use 
for quantum computation.\\

In chapter \ref{paramagnetic} we return to the orbital magnetism, synthesizing 
various elements of proximity induced superconductivity discussed above. 
The motivation for this work has been provided by the puzzling low-
temperature anomalies observed in normal-metal -- superconductor  
cylinders by Mota and co-workers \cite{mota:90,mota:94}. In the low-temperature 
low-field corner of the H-T phase diagram, the Ag-Nb and Cu-Nb samples 
develop a paramagnetic signal on top of the saturated Meissner response, 
which produces a reentrance in the magnetic susceptibility (for a detailed 
discussion of the experiments see also \cite{visanidiss,frassdiss}).
We note that 
the observed behavior is not believed to be related to the low-temperature 
reentrance in the conductance of a disordered NS junction discussed above, 
which is due to the finite energy properties entering the non-equilibrium 
transport. The magnetic response instead is a thermodynamic property of the 
system.

In chapter \ref{paramagnetic} we consider the consequence of a repulsive 
electron-electron interaction in the normal metal proximity layer. The 
repulsive coupling $V_N>0$ induces a finite gap $\Delta(x)=V_N F(x)$ 
in the normal layer which is proportional to the induced superconducting 
correlations (\ref{fandre}). The order parameter changes sign across the 
NS interface as the coupling constant changes from attractive to repulsive. 
The NS structure behaves like a Josephson junction with a phase difference 
of $\pi$ across the NS interface. We know from Eq.\ (\ref{intro:phasecondsns}) 
that a junction with $\Delta\varphi=\pi$ traps density of states at the 
Fermi energy $\epsilon=0$, according to 
\be
(k_+-k_-) L  = 2n \pi,
\ee
which remains true in the limit of a vanishing junction length 
$L \to 0$. The phase difference $\pi$ across the junction cancels the phase 
shifts $\pi/2$ of the two AR. As all transverse wavefunction $(k_y,k_z)$ 
exhibit a zero energy bound state, the DOS shows a peak a zero energy with 
macroscopic weight
\be
N(E) \sim N_0 k_F^2 \delta(E),
\ee
per unit surface. 

As we show in Chp.\ \ref{paramagnetic}, this peak implies 
a paramagnetic instability in the current response and leads to a 
spontaneous onset of currents along the interface. Spontaneous moments 
form at the NS interface producing a low-temperature reentrance in the 
magnetic susceptibility. The qualitative agreement of our results with 
the observed low-temperature anomaly is indicative of a repulsive 
electron-electron interaction in these systems.

\chapter{Nonlinearity in normal-metal -- superconductor transport: 
Conductance and shot noise}
\label{nonlinearity}

\section{Scattering matrix approach to normal-metal -- superconductor 
transport}
\label{scattapproach}

\markboth{CHAPTER \ref{nonlinearity}.  NONLINEARITY IN ...}{SCATTERING MATRIX 
APPROACH TO ... }

\subsection{Introduction}

The study of electronic transport in normal-metal--superconductor (NS)
or semicon\-ductor--superconductor (SmS) sandwiches has attracted a
considerable amount of interest in the past years. At sufficiently low
temperature and in high quality mesoscopic samples, the phase-breaking
length of the electrons is larger than the typical system size, 
resulting in directly observable quantum coherence effects. Of special
interest is the effect of the electronic phase-coherence in a normal 
metal--superconductor system. In the standard theory of the proximity 
effect, the influence of the superconductor on the normal metal can be 
understood in terms of the coherent coupling of electrons and holes in 
the metal as described by the Bogoliubov-de Gennes (BdG) equations
\cite{deGennes}. The correlation between
electrons and holes is produced by the process of Andreev
reflection \cite{andreev,btk} at the NS interface, by which an incident electron
is back-reflected into a hole, converting quasiparticle current to
supercurrent. This microscopic picture of electron-hole correlation is 
equivalent to that provided by a pair correlation function in the normal 
metal, 
which is induced by the superconductor through the contact at 
the NS interface. The scattering matrix approach makes use of the microscopic 
single particle picture of coupled electron and hole channels providing a
straightforward tool within a formalism of the 
Landauer--B\"uttiker type \cite{landauer,buttland,beenakker}. 

The quality of the interface as well as the phase breaking processes 
determine the strength of the proximity effect and naturally have their 
impact on the current--vol\-tage characteristics (CVC). A few fascinating 
transport experiments \cite{pothier,petrashov,pannetier} have been carried 
out recently, investigating temperature and voltage dependence, as well as
the flux modulation, of both NS and SmS junctions. Interestingly, the
relative strength of the interface barrier and the elastic scattering
in the normal region is crucial for the features of the CVC. The ratio
of the two determines whether subgap conductance peaks arise at zero or 
finite voltage \cite{volkov,marmokos,yip}.  The investigation of these 
so-called zero and finite bias anomalies in the subgap conductance have 
been the object of recent experiments
 \cite{kastalsky,nguyen,nitta,bakker,magnee,sanquer}. 
The present work draws much of its motivation from the 
ongoing discussions and experiments in this area, see also 
 \cite{glazmanSn}. In our case study of a
double barrier NINIS junction we observe zero and finite bias anomalies and 
shed light on the underlying mechanism.

Recently in \cite{nazarov,stoof} as well as \cite{golubov}, the carrier 
transport in disordered NS junctions has been described 
in terms of an energy-dependent diffusion 
constant, successfully explaining the recent experiments on reentrance in the
conductivity at low temperatures \cite{petrashov,charlat}. 
Their work uses the quasi-classical Green's 
functions technique \cite{rammersmith}, 
which allows to describe transport both close to 
equilibrium \cite{hekking} and far away from 
equilibrium \cite{volkov,volkov94,zhouspivak} 
and facilitates the averaging over disorder in diffusive conductors. 
This approach is quite indispensable if phase-breaking processes are to be 
included. An appealing alternative approach is the scattering matrix 
technique, which relies on the quasiparticle wavefunctions described by the 
BdG equations. 
While being valid in a general context, it describes the transport in
mesocopic systems from the intuitive ballistic point of view. The transport
through normal or superconducting leads is expressed through the properties of
a multichannel scattering matrix accounting for all elastic scattering 
processes, whether they be due to a geometric constriction, single impurities, 
or disorder (inelastic processes are excluded from such a description). 
By these means, the transport problem is reduced to solving a ballistic 
problem at the interfaces of the normal and superconducting leads. 
The current and the
conductance of the system are determined analytically in terms of the 
transmission and reflection amplitudes of the scattering matrix. 
Adhering to this formalism rather than the Green's function technique 
helps us to improve our understanding of zero and finite bias anomalies.

The study of normal--superconducting junctions goes back to the works
of Kulik \cite{kulik} on SNS junctions and of Blonder et al. \cite{btk}
 on NIS junctions, who studied nonlinear transport within the framework 
of the BdG equations considering quasi one-dimensional models. On the other
hand, the scattering matrix technique was developed by Landauer and 
co-workers \cite{landauer,buttland} in the linear response regime,
resulting in the well known conductance formula for a normal metal. 
Lambert \cite{lambert} and Takane and Ebisawa \cite{takane,takane2} 
extended the approach to include superconducting segments, on
the basis of which Beenakker \cite{beenakker} derived a zero temperature, 
linear response conductance formula for the transport through NS junctions.
A few studies have been carried out recently \cite{marmokos,brouwer,claughton} 
combining 
the scattering matrix approach with the finite voltage transport model of 
Ref. \cite{btk}. They have limited themselves to the energy 
dependence of the scattering states to extract finite voltage properties of
the CVC in the subgap regime.
Here, we extend these works to voltages above the gap and additionally 
take into account the full voltage dependence of the transport problem. 
This is important within the context of the sign reversal symmetry of the
differential conductance which we discuss below. At the same time, we provide 
a common framework for the above studies, tracing them back to a single 
general formula.

In the present chapter, we derive the general expression for the current 
through 
a NS junction in the scattering matrix approach, valid for multiple channels, 
finite voltage, and nonzero temperature. We review the 
derivation of the current--voltage relation and express it in terms of a 
{\it spectral} conductance formula, thereby accounting for the full voltage
dependence of the transport problem. The reflection at 
the NS interface is made explicit using the Andreev approximation and 
a conductance formula is obtained expressing the result in terms of
the normal scattering matrix. We illustrate this
formula in the new regime of voltages above the gap and expose its 
connection to previously obtained limits. 
We describe the existence of resonances due to 
quasi-bound Andreev states, and show that they produce sharp conductance
peaks in both the single and multichannel junction. We explain the 
generic mechanism underlying the appearance of zero and finite bias 
anomalies in the ballistic two barrier system. Finally, we interpret
our results in connection to the recent experiments on zero and finite
bias anomalies in disordered NS junctions.

Chapter \ref{scattapproach} closely follows our paper published in 
\cite{leso}. We would like to point
out further works, both theoretical and experimental, which have been 
carried out since then \cite{hartog,raimondi,wilhelmns,lesoblatt}.\\

\begin{figure}[t]
\centerline{\psfig{figure=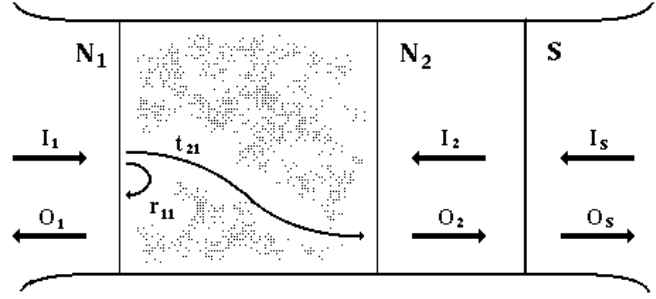,height=4.7cm}}
\caption{Schematic structure of a disordered NS junction. 
  Ballistic normal ($N_1$, $N_2$) and superconducting leads (S) are
  coupled to reservoirs at chemical potential $\mu-eV$ and $\mu$,
  respectively. Scattering is limited to the hatched region between
  leads $N_1$ and $N_2$.}
\label{nsfig1}
\end{figure}

We consider a normal--superconducting junction with quasi
one-dimensional, ballistic normal and superconducting leads, as shown
in Figure \ref{nsfig1}. The pair potential vanishes in the normal part, 
due to the absence of electron-electron interactions. Between the
normal lead and the NS interface, the electrons traverse a disordered
region, the transmission through and the reflection thereof are
described by a scattering matrix. 
Between the scattering region and the NS interface, a small
ballistic normal region serves to separate the scattering in the
normal part, which mixes all electron channels at a given energy,
from the scattering at the NS interface, where electron and hole
channels are mixed in the reflection process.
The coherent scattering in the disorder region of the normal metal is 
described by the $4N\times 4N$ scattering matrix,
\begin{eqnarray} \left(\begin{array}{c}O_1^e\\ O_1^h\\O_2^e\\
 O_2^h \end{array}\right) &=&  \left(\begin{array}{cccc} 
r_{11}\left(\epsilon\right) & 0 & t_{12}\left(\epsilon\right) & 0 \\ 
0 & r_{11}^*\left(-\epsilon\right) & 0 & t_{12}^*\left(-\epsilon\right)
 \\ t_{21}\left(\epsilon\right) & 0 & r_{22}\left(\epsilon\right) & 0 \\ 
0 & t_{21}^*\left(-\epsilon\right) & 0 & r_{22}^*\left(-\epsilon\right) 
\end{array}\right) \left(\begin{array}{c}I_1^e\\ I_1^h\\ I_2^e\\ I_2^h 
\end{array}\right) \nonumber \\
&=& \Large \left( \begin{array}{cc}\hat{r}_{11}
\left(\epsilon\right) & \hat{t}_{12}\left(\epsilon\right)\\ 
\hat{t}_{21}\left(\epsilon\right) & \hat{r}_{22}\left(\epsilon\right) 
\end{array}\right) \small \left(\begin{array}{c}I_1^e\\ I_1^h\\ I_2^e\\ 
I_2^h \end{array}\right), \label{scattmat} \end{eqnarray}
which we denote by ${\bf S}_N$. The matrix connects the incoming $N$
electron (hole) channels $I_i^e$ ($I_i^h$) on each side to the equal
energy outgoing channels $O_i^e$ ($O_i^h$) according to Figure \ref{nsfig1}
($i=1,2$). The $N$ channels represent the different transverse states, 
$r_{ii}$ and $t_{ij}$ are 
$N\times N$ reflection and transmission matrices for electron
channels, $\hat{r}_{ii}$ and $\hat{t}_{ij}$ the comprehensive
$2N\times 2N$ matrices including the complex conjugated reflection and
transmission amplitudes for holes \footnote{We denote the complex
conjugate of a matrix $m$ by $m^*$, the transposed matrix by
$m^{\top}$ and the adjoint by $m^{\dag}$.}. Following usual convention, 
we include the propagation in the ballistic region $N_2$ in the
scattering matrix. For states normalized to carry unit probability
current \cite{btk}, the continuity equation implies the unitarity of
the scattering matrix. By allowing for a dependence of the scattering matrices 
on voltage we can account for the full voltage dependence of the scattering 
problem. We can
thus describe the deformation of the states due to the space dependent 
potential including, e.g., the voltage dependent Schottky barrier at a 
semiconductor--superconductor (SmS) interface.

We define an analogous unitary scattering matrix for the NS interface 
${\bf S}_I$ (see Fig.\ \ref{nsfig1}) by
\begin{eqnarray} \small \left(\begin{array}{c}I_2^e\\ I_2^h\\O_S^e\\ 
O_S^h \end{array}\right) &=&  
\Large \left(\begin{array}{cc}\hat{r}_{I}\left(\epsilon\right) & 
\hat{t}'_{I}\left(\epsilon\right)\\  \hat{t}_{I}\left(\epsilon\right) & 
\hat{r}'_{I}\left(\epsilon\right)  \end{array}\right) \small \left(
\begin{array}{c}O_2^e\\ O_2^h\\ I_S^e\\ I_S^h \end{array}\right). 
\label{ascattmat} \end{eqnarray}
The $2N\times 2N$ matrices $\hat{r}_I$, $\hat{r}'_I$, $\hat{t}_I$, 
$\hat{t}'_I$ give the reflection and transmission amplitudes of the
states normalized to unit probability current in the normal and 
superconducting lead\footnote{Note that the connection the unit 
charge currents on either side of the interface will not produce 
unitarity, since quasiparticle charge current is not a conserved quantity in 
the superconductor, see also appendix \ref{bdgapp}}. 
In the evaluation of the current, we will consider the 
matrix (\ref{scattmat}) to be specified by an arbitrary model 
for the disorder, while the matrix (\ref{ascattmat}) will be made explicit 
below using the BdG equations. The total effect of all
scattering processes in the disorder region and at the NS interface can be
described by a unitary, global $4N\times 4N$ scattering matrix. 
We restrict ourselves to its sub-matrix
$\hat{r}$ describing the reflection into the normal region,
\begin{equation} \textstyle \left(\begin{array}{c}O_1^e\\ O_1^h \end{array}
\right)= \Large \hat{r}\left(\epsilon, V\right) \small \left(\begin{array}
{c}I_1^e\\ I_1^h \end{array}\right) = \left( \begin{array}{cc}r_{ee}
\left(\epsilon,V\right) & r_{eh}\left(\epsilon,V\right)\\ r_{he}\left(
\epsilon,V\right) & r_{hh}\left(\epsilon,V\right) \end{array} \right) 
\left(\begin{array}{c}I_1^e\\ I_1^h \end{array}\right) \label{globalR}. 
\end{equation}
$r_{ee}$, $r_{eh}$, $r_{he}$, and $r_{hh}$ are again $N\times N$
reflection matrices. $\hat{r}\left(\epsilon,V\right)$ can be expressed 
through the given scattering matrices (\ref{scattmat}) and
(\ref{ascattmat}),
\begin{equation} \hat{r}\left(\epsilon, V\right) =\hat{r}_{11}\left(
\epsilon\right)+\hat{t}_{12}\left(\epsilon\right)  \big[1-\hat{r}_{I}
\left(\epsilon\right)\hat{r}_{22}\left(\epsilon\right)\big]^{-1}  
\hat{r}_{I}\left(\epsilon\right) \hat{t}_{21}\left(\epsilon\right), 
\label{Rhat} \end{equation}
representing the sum over all scattering paths of an incident 
electron or hole excitation, multiply scattered between the disorder 
region and the NS interface. Apart from the direct reflection at the 
disorder region, the simplest process contributing consists of an 
excitation, which is first transmitted ($\hat{t}_{21}$) through the 
disorder region, reflected ($\hat{r}_{I}$) at the NS interface, and 
transmitted 
($\hat{t}_{12}$) back to the normal lead. All further paths result from
iterative scattering processes between the disorder region and the NS
interface.

We now derive the current in the normal lead, expressing it through 
the global reflection matrix $\hat{r}(\epsilon,V)$.
Applying a voltage $V$ (denoting the voltage in a two point measurement) on
the normal side has two consequences. First, the voltage induces an 
electrostatic potential drop over the disorder region in the NS junction, 
resulting in a deformation of scattering states. The coupling of incident 
and outgoing channels is thus voltage dependent in general, as described by
$\hat{r}\left(\epsilon, V\right)$. A stationary state
incident from the normal lead (of energy $\epsilon$ in channel $\nu$)
consists of the incident electron and the reflected electron- and hole-
states and carries the current ($e=|e|$),
\begin{equation} I_{\nu}\left(\epsilon, V\right) d\epsilon
= \frac{-e}{h}
  \left\{1-\sum_{\beta} \mid
    r_{ee}\left(\epsilon,V\right)_{\beta\nu}\mid^2 
+ \sum_{\beta} \mid
    r_{he}\left(\epsilon,V\right)_{\beta\nu}\mid^2 \right\} d\epsilon.
\end{equation} 
Second, the applied voltage shifts the chemical
potential of the reservoir attached to the normal lead by $-eV$ with
respect to the reservoir on the superconducting side. The deformation
of the states by itself produces no net current\footnote{This can be shown 
using the unitarity of the global scattering matrix including the 
Andreev process by an extension of the argument given for a normal 
junction.}. The net current flow results exclusively
from the difference in occupation of the (finite voltage) scattering
states incident from the left and right reservoirs. Writing the sum over 
channels as a trace, we obtain the current--voltage relation
\begin{equation} I= \int d\epsilon\, \frac{1}{e}\Big[f\left(\epsilon\right)-
f\left(\epsilon+eV\right)\Big]\, G_s\left(\epsilon,V\right), 
\label{current} \end{equation}
with the spectral conductance
\begin{equation} G_s\left(\epsilon,V\right)= \frac{2e^2}{h}\mbox{Tr}
\bigg[1-r_{ee}^{\dag}\left(\epsilon,V\right) r_{ee}\left(\epsilon,V
\right) + r_{he}^{\dag}\left(\epsilon,V\right) r_{he}\left(\epsilon,V
\right)\bigg]. \label{cond} \end{equation}
A factor two accounts for the spin degeneracy. The defined
{\it spectral} conductance $G_s\left(\epsilon,V\right)$ describes the 
current contribution of the incident scattering states at energy 
$\epsilon$, at a given voltage $V$  (by convention, 
the energy is measured with respect to the chemical potential in the
superconductor). Formulas (\ref{current}) and (\ref{cond}) imply the 
differential conductance
\begin{eqnarray} \frac{dI}{dV}\mid_V &=& -\int d\epsilon\, f'\left(\epsilon
+eV\right)\, G_s\left(\epsilon,V\right) \\ 
&&  +\int d\epsilon\, \frac{1}{e}\Big[f\left(\epsilon\right)-f\left(\epsilon
+eV\right)\Big]\, \frac{\partial G_s\left(\epsilon,V\right)}{\partial V},
\nonumber
\end{eqnarray}
with the expansion
$ dI/dV\mid_V = G_s(-eV,0) + 2V \partial_V 
G_s(\epsilon,V)\mid_{\epsilon=-eV,V=0} + \mbox{... }$
at zero temperature.
This differs from the differential conductance of Ref.~\cite{btk}, 
$dI/dV=G_s\left(-eV,0\right)$, by accounting for the change in the conductance
of the open channels with increasing voltage.

The expressions (\ref{Rhat}), (\ref{current}), and (\ref{cond})
determine the general form of the current--voltage relation of a
disordered NS junction, for an arbitrary scattering and pair 
potential at the NS interface. 

We close this section with a discussion of the symmetry of the CVC with 
respect to sign reversal of the applied voltage. In the subgap regime 
$e|V|<\Delta$, the incoming quasiparticle excitations may not enter the 
superconductor. The probability current of the states $|\epsilon|<\Delta$
is totally reflected and thus the global reflection matrix
$\hat{r}\left(\epsilon,V\right)$ of (\ref{globalR}) is unitary. The
unitarity produces the relations $r_{ee}^{\dag}r_{ee} + r_{he}^{\dag}r_{he}=1$
and $r_{ee} r_{ee}^{\dag} + r_{eh} r_{eh}^{\dag}=1$.
The symmetry of electron- and hole- type excitations in the BdG equations 
guarantees $r_{eh}\left(\epsilon,V\right)=-r_{he}^{*}\left(-\epsilon,V\right)$.
As a consequence, the subgap conductance takes the form
\begin{eqnarray} G_s\left(\epsilon,V\right)&=&\frac{4e^2}{h}\mbox{Tr}
\bigg[r_{he}^{\dag}\left(\epsilon,V\right) r_{he}\left(\epsilon,V
\right)\bigg] = \frac{4e^2}{h}\mbox{Tr}
\bigg[r_{eh}^{\dag}\left(\epsilon,V\right) r_{eh}\left(\epsilon,V
\right)\bigg] \nonumber \\
&=& \frac{4e^2}{h}\mbox{Tr}
\bigg[r_{he}^{\dag}\left(-\epsilon,V\right) r_{he}\left(-\epsilon,V
\right)\bigg] = G_s\left(-\epsilon,V\right). \end{eqnarray}
A subtle issue is that this symmetry does not yet imply a symmetry in 
the CVC under reversal of voltage \cite{lamblead}. The latter requires
that $G_s\left(\epsilon,V\right)=G_s\left(-\epsilon,-V\right)$, which
amounts to $G_s\left(\epsilon,V\right)$ being independent of voltage. 
Then we have $G_s\left(\epsilon\right)|_{\epsilon=-eV}=dI/dV|_V$ and the 
differential conductance is invariant under sign reversal of the voltage. 
Indeed, in recent experiments on SmS junctions \cite{magnee,sanquer}, 
an asymmetry in the CVC was found in the subgap regime, which can be 
understood on the basis of the above discussion taking into account the
voltage dependent Schottky barrier at the SmS interface. 
An explicit account of the voltage dependence of $G_s$ requires the
scattering matrix ${\bf S}_{N}$ to be determined in the applied electrostatic
potential. In principle, this task demands the self-consistent solution of the
scattering problem and the Poisson equation, see \cite{buttiker}. 
In the next section, we evaluate the spectral conductance (\ref{cond}) by 
using the Andreev approximation for the scattering at the NS interface.

\subsection{Spectral conductance in the Andreev approximation}

The stationary states in the
ballistic leads are solutions of the BdG equations \cite{deGennes} and
are of the plane wave type. A step function model for the pair
potential $\Delta\left(x\right)= \Delta_o e^{i\chi}\theta\left(x\right)$ 
is assumed, which neglects the suppression of the pair potential in the 
superconductor on the scale of a coherence length. The NS interface 
connects electrons and holes of the same channel with a reflection amplitude 
depending on the reduced chemical potential
$\mu_{\nu}=\mu-\hbar^2{\bf k}_{\bot}^2/2m$. In the limit
$\epsilon,\Delta \ll \mu_{\nu}$, the BdG equations are simplified by
linearizing the dispersion relation around the effective Fermi wave number
$k_{\nu}^{\left(0\right)}=\sqrt{2m\mu_{\nu}/\hbar}$. In this approximation 
made by Andreev, incoming electrons are purely reflected into
holes and vice versa. The reflection matrix at the NS interface is given
by
\begin{equation} \hat{r}_{I}\left(\epsilon\right)=  
\left(\begin{array}{cc} 0 & e^{-i\chi}\Gamma
\left(\epsilon\right)\\ e^{i\chi}\Gamma\left(\epsilon\right) & 0 
\end{array}\right), 
\label{ri}
\end{equation}
depending on the channel independent, scalar Andreev amplitudes 
\begin{equation} \Gamma\left(\epsilon\right)=\left\{\begin{array}{ll}
\frac{{\textstyle\epsilon-\mbox{sign}\left(\epsilon\right)
\sqrt{\epsilon^2-\Delta^2}}}{{\textstyle \Delta}} \,\sim \frac{{\textstyle 
\Delta}}{{\textstyle 2|\epsilon|}}, & |\epsilon| > \Delta, \\ 
\frac{{\textstyle \epsilon-i\sqrt{\Delta^2-\epsilon^2}}}{{\textstyle \Delta}}
 = \exp\left({\textstyle -i\arccos\frac{{\textstyle \epsilon}}{\Delta}}
\right) , & |\epsilon| < \Delta. 
\end{array} \right. \label{Gamma} \end{equation}
Inserting in (\ref{Rhat}) and (\ref{cond}), we 
obtain the multichannel spectral conductance formula 
\begin{eqnarray} G_s\left(\epsilon,V\right) = \frac{2e^2}{h}\left(1+\mid
\Gamma\left(\epsilon\right)\mid^2\right)   \mbox{Tr}\left\{t_{21}^{\dag}
\left(\epsilon\right)  
\left[1-\Gamma^*\left(\epsilon\right)^{2}r_{22}^{
\top}\left(-\epsilon\right)r_{22}^{\dag}\left(\epsilon\right)\right]^{-1}
\right.  \nonumber \\  
 \times \left. \left[1-\mid\Gamma\left(\epsilon
\right)\mid^2 r_{22}^{\top}\left(-\epsilon\right)r_{22}^*\left(-\epsilon
\right)\right]  
\left[1-\Gamma\left(\epsilon\right)^2 r_{22}\left(\epsilon
\right)r_{22}^*\left(-\epsilon\right)\right]^{-1}  t_{21}\left(\epsilon
\right)\right\}. \label{speccond} \end{eqnarray}
Combined with equation (\ref{current}), Eq.\ (\ref{speccond}) 
provides the finite 
voltage, finite temperature CVC of a disordered normal--superconducting
junction in the Andreev approximation. The spectral conductance
depends on the scattering matrices of the electrons at energies $\pm
\epsilon$ as a signature of the presence of Andreev reflection. The dependence 
of this formula on the phases of the reflection and transmission amplitudes
proves crucial in determining the resonance peaks in the conductance. 
The elementary process, which contributes to these phases is the propagation 
of an electron and a hole between the disorder region and the NS interface.

If no inter-channel mixing takes place, i.~e., the matrices $t_{ij}$ and
$r_{ii}$ are diagonal, the conductance $G_s\left(\epsilon,V\right)$ reduces 
to the quasi one-dimensional form
\begin{eqnarray} \small \lefteqn{ \frac{2e^2}{h}\sum_{\nu=1}^{N}
 \frac{\left(1+\mid\Gamma\left(\epsilon\right)\mid^2\right)\quad T_{\nu}
\left(\epsilon,V\right)\left[1-\mid\Gamma\left(\epsilon\right)\mid^2 
R_{\nu}\left(-\epsilon,V\right)\right]}{1+\mid\Gamma\left(\epsilon\right)
\mid^4 R_{\nu}\left(\epsilon,V\right) R_{\nu}\left(-\epsilon,V\right)- 2 Re
\left[\Gamma\left(\epsilon\right)^2 r_{\nu}\left(\epsilon,V\right) r_{\nu}^*
\left(-\epsilon,V\right)\right]}. }
\label{onechannel} 
\end{eqnarray}
\noindent
$r_{\nu}=\left(r_{22}\right)_{\nu\nu}$ is the reflection amplitude for a
state coming in from the right side of the scattering region (Fig. 
\ref{nsfig1}) and 
$R_{\nu}=|r_{\nu}|^2$ and $T_{\nu}=1-R_{\nu}$ denote the reflection and 
transmission probabilities of the $\nu$-th channel.

For voltages well {\it above} the gap, $|\epsilon|, V\, \gg\Delta$ (still 
assuming $|\epsilon|\ll \mu$), the Andreev reflection is strongly 
suppressed and drops according to 
$\Gamma\left(\epsilon,V\right)\sim \Delta/2|\epsilon|\,\to 0$. The spectral 
conductance (\ref{speccond}) asymptotically approaches the expression for
a normal junction,
\begin{equation} G_s\left(\epsilon,V\right)=\frac{2e^2}{h}\mbox{Tr}
\left[t_{21}^{\dag}\left(\epsilon,V\right)t_{21}\left(\epsilon,V\right)
\right]. \label{land} \end{equation}

At voltages {\it below} the gap, we make use of $\mid\Gamma\left(\epsilon
\right)\mid=1$ and $T_{\nu}=1-R_{\nu}$ and obtain the spectral conductance 
$G_s\left(\epsilon,V\right)$ in the form 
\begin{equation} \frac{4e^2}{h}\sum_{n=1}^{N} 
\frac{ T_{\nu}\left(\epsilon,V\right) T_{\nu}\left(-\epsilon,V\right)} 
{1+ R_{\nu}\left(\epsilon,V\right) R_{\nu}\left(-\epsilon,V\right)- 
2 Re\left[\Gamma \left(\epsilon\right)^2 r_{\nu}\left(\epsilon,V\right) 
r_{\nu}^*\left(-\epsilon,V\right)\right]}. \label{subonechannel} 
\end{equation}
The reflection and transmission coefficients at $\pm\epsilon$ are 
symmetrically involved in this formula, which results in the symmetry of 
the CVC discussed in the section above. In contrast, the spectral conductance 
(\ref{onechannel}) at voltages above the gap becomes increasingly asymmetric 
as it asymptotically approaches the Landauer expression. The sensitivity 
of (\ref{subonechannel}) to the phase of the reflection amplitudes 
allows for the distinction of zero and finite bias
peaks in the double barrier NS junction discussed below.

The linear response limit ($\epsilon,V\,\to 0$) of (\ref{subonechannel}) 
can be determined using $\Gamma\left(0\right)^2=1$ and $R_{\nu}=1-T_{\nu}$ 
and takes the form \cite{beenakker}
\begin{equation} G\left(0\right)=\frac{4e^2}{h}\sum_{n}\frac{T_{\nu}\left(0
\right)^2}{\left[2-T_{\nu}\left(0\right)\right]^2}. \label{been} 
\end{equation}

\subsection{Double barrier NINIS junction}

In this section we apply the above results to a double barrier NINIS junction, 
which allows to study the interplay between normal- and 
Andreev- levels in a Fabry--Perot type I$_1$NI$_2$ interferometer. 
First, we discuss the structure in the 
conductance of a single channel NI$_1$NI$_2$S junction, which we trace back to 
the presence of Andreev resonances.
Second, we present numerical results for a multichannel junction, showing
that the typical resonance structure of a single channel survives the 
summation over the channels. This stability is a peculiarity of the 
superconducting system, absent in a normal NI$_1$NI$_2$N double barrier 
junction. 

Since the channels separate in the double barrier problem, we can make use 
of the result (\ref{onechannel}) for the conductance $G_s$. 
$G_s$ depends on 
the phases $\varphi\left(\pm\epsilon\right)$ of the reflection amplitudes 
$r\left(\pm\epsilon\right)$ as well as on the complex amplitude 
$\Gamma\left(\epsilon\right)$ of the Andreev reflection. We use the notation
$T\left(\pm\epsilon\right)=T_{\pm}$, $R\left(\pm\epsilon\right)=R_{\pm}$, and
$r\left(\pm\epsilon\right)=\sqrt{R_{\pm}}e^{i\varphi\left(\pm\epsilon
\right)}$ for the reflection amplitude, the phase factors being determined 
by the potential barriers $I_1$ and $I_2$ and the propagation between them. 
We rewrite the amplitude of the Andreev reflection as
$\Gamma\left(\epsilon\right)=|\Gamma| e^{-i\vartheta\left(\epsilon\right)}$
with the phase $\vartheta\left(\epsilon\right)= \arccos\left(\epsilon/\Delta
\right)$ below the gap and vanishing above. The conductance 
$G_s\left(\epsilon\right)$ simplifies to
\begin{equation}  \!\!\!\!\! \frac{2\left(1+|\Gamma|^2\right)e^2}
{h} \frac{ T_+ \,\, \left(1-|\Gamma|^2 R_-\right)}{1 + |\Gamma|^4 R_+ R_-  
-2|\Gamma|^2\sqrt{R_+R_-} \cos\left[\varphi\left(\epsilon\right) -\varphi
\left(-\epsilon\right) -2\vartheta\left(\epsilon\right)\right] }, 
\label{advert} \end{equation}
\noindent
which is always less or equal to the universal value $4e^2/h$. Note that 
the Andreev reflection is suppressed above the gap ($|\Gamma| < 1$), while
the phase $\vartheta\left(\epsilon\right)$ vanishes.\\
Below the gap the conductance exhibits resonances, which is maximal if
$R_{\pm}$ are equal and the scattering phase difference of the electrons 
and holes compensates for the phase of the Andreev reflection, 
according to the resonance condition
\begin{equation} \cos\left[\varphi\left(\epsilon\right) -\varphi\left(
-\epsilon\right) -2\vartheta\left(\epsilon\right)\right]=1. \label{res} 
\end{equation}
 
Let us assume that the barrier at the NS interface vanishes, I$_2=0$, and 
denote the length of the interferometer by $d$. In the limit of a high 
potential barrier I$_1$, $R_+$ and $R_-$ are energy independent and
equal, and $\varphi\left(\pm\epsilon\right)=\pi + 2k_{\pm}d$. Using the wave number $k_{\pm}=mv_{\nu}\pm\epsilon/v_{\nu}$ 
($v_{\nu}$ is the Fermi velocity of channel $\nu$), 
the resonance condition (\ref{res}) yields the spectrum of Andreev levels,
\begin{equation} 
\epsilon_{n,\nu}=\frac{v_{\nu}}{2d}\left(n\pi+\arccos 
\frac{\epsilon_{n,\nu}}{\Delta} \right), 
\quad\quad\quad n=0,1, ... 
\label{alevel} 
\end{equation}
which predicts resonances in the conductance of a typical width proportional
to the transmission $T$ of the barrier. The phase $\vartheta\left(
\epsilon\right)$ varies between $-\pi/2$ and $0$ from $\epsilon=0$ to 
$\epsilon=\Delta$ and guarantees the existence of at least one 
Andreev resonance for arbitrarily small $d$. In the $d\to 0$ limit, 
the weight of this resonance lines up with the gap voltage and we recover 
the NIS junction, featuring a 
suppressed subgap conductivity and a peak in the differential conductance at 
the gap voltage. This peak can be understood in terms of the Andreev 
resonance which moves to the gap energy for $d\to 0$.

\begin{figure}[!t]
\psfig{figure=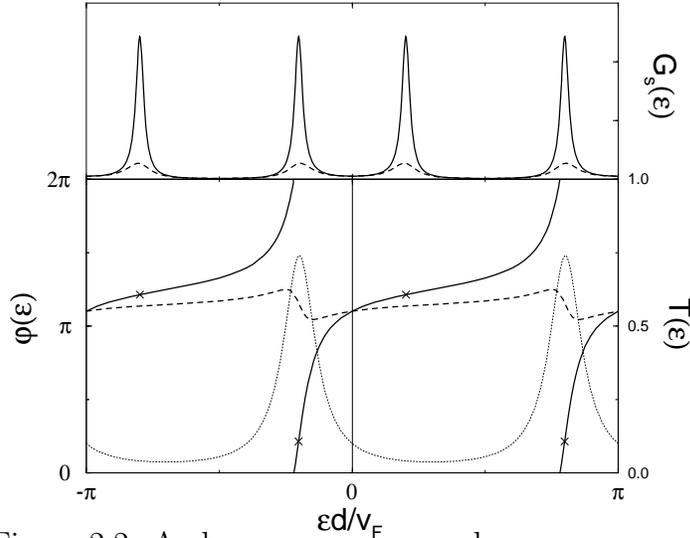,angle=-90,height=7.5cm}
\caption{Andreev resonances and resonance condition for the phase $\varphi$. 
Bottom: phase $\varphi\left(\epsilon\right)$ of the reflection amplitude versus
energy. The solid line represents the I$_1$NI$_2$ interferometer with barriers 
strengths $R_1=0.8$ and $R_2=0.5$, the 
dashed line stands for the inverse barrier sequence ($R_1=0.5$, $R_2=0.8$). 
The {\it Andreev} resonance condition for the phase is met for a pair of 
energies $\pm\epsilon_n$ with phase difference 
$\Delta\varphi\left(\epsilon_n\right)=\pi+2n\pi$.
This phase condition can be fulfilled only by the first barrier sequence 
($R_1>R_2$, solid line) at those energies indicated in the graph. The dotted 
line shows the transmission probability $T$ as determined by the 
{\it normal} resonances.
Top: conductance (arbitrary units) of the double barrier NS junction 
versus energy, the solid line again representing the barriers 
$R_1=0.8$, $R_2=0.5$, and the dashed line the barriers $R_1=0.5$, $R_2=0.8$.
Note the symmetry of the resonances with respect to $\epsilon=0$.}
\label{phasepl}
\end{figure}

We now turn to double barrier NI$_1$NI$_2$S junction with arbitrary 
barriers. Here $\varphi\left(\pm\epsilon\right)$ represent the phases for 
the reflection of electrons entering the double barrier scattering region 
from the right (superconducting side). The corresponding reflection 
amplitudes are given by
\begin{equation} r\left(\pm\epsilon\right)=r_2 + \frac{t_2^2 r_1 
e^{2ik_{\pm}d}}
{1-r_1 r_2 e^{2ik_{\pm}d}}, \label{rpm} \end{equation}
where $r_i, t_i$ are the amplitudes of the left ($i=1$) and the right ($i=2$) 
barrier. The phase of this reflection amplitude plays the major role in 
determining the structure of the conductance, as it controls the existence 
of resonances according to (\ref{res}). Let us fix the barrier I$_1$ and 
increase 
I$_2$ slowly, keeping their strengths I$_1>$ I$_2$. In this situation, the INI 
interferometer develops pronounced Andreev resonances. For $r_1\gg r_2$, the 
phase $\varphi\left(\pm\epsilon\right)$ of the reflection amplitude
$r\left(\pm\epsilon\right) \approx t_2^2r_1e^{2ik_{\pm}d}$ produces a linear 
energy dependence of the phase $\varphi\left(\epsilon\right)$, which changes by
$2\pi$ on the scale $v_F/d$ and results in (nearly) equidistant resonances,
 in accordance with (\ref{alevel}). As the strength of I$_2$ 
is increased, the resonances pair up as is 
illustrated in Figure \ref{phasepl}. The phase function 
$\varphi\left(\epsilon\right)$, as 
displayed in Figure \ref{phasepl} (solid line), can be used to determine the 
location of 
the resonances by finding those combinations of energies $\pm\epsilon$, which 
have a phase difference\footnote{The energy dependence of the Andreev 
reflection amplitude produces slight deviations close to the gap energy.} 
$\Delta\varphi\left(\epsilon\right)=\varphi\left(
\epsilon\right)-\varphi\left(-\epsilon\right)=\pi+2n\pi$. The period of 
$\Delta\varphi\left(\epsilon\right)$ reduced by half with respect to the 
period of $\varphi\left(\epsilon\right)$ accounts for the pairing of the 
resonances. 

When the strengths of the barriers become of the same order, I$_1\sim$ I$_2$, 
the spectral weight of the INI interferometer is shared by Andreev quasi-bound 
states of a mixed electron--hole character and  normal electron quasi-bound 
states. Due to the large gradient of the phase close 
to the normal resonances, see Fig.\ \ref{phasepl}, the Andreev resonances 
tend to be pinned to
normal resonances at either $+\epsilon$ or $-\epsilon$. 
While the Andreev bound 
states contribute to the current, normal bound states do not couple to the 
superconductor and thus do not participate in the charge transport. This is 
reflected by the subgap symmetry of the $G_s\left(\epsilon\right)$ under 
reversal of voltage which is is observed for all barrier strengths, see Fig.\ 
\ref{phasepl}.

As the barrier strength is increased further to I$_2>$ I$_1$, the Andreev 
resonances are weakened and eventually disappear. Although the normal 
resonances dominate in the INI interferometer in this regime, only the weak
Andreev resonances show up in the conductance, which thus exhibits only a 
weak subgap structure. The phase function 
$\varphi\left(\epsilon\right)$ in 
Figure \ref{phasepl} (dashed line) becomes nearly constant for $r_1 \ll r_2$, 
 see (\ref{rpm}), and the phase condition for resonance (\ref{res}) cannot be 
met.

A comparison of the double barrier systems NI$_1$NI$_2$S and 
NI$_2$NI$_1$S, i.e., with inverse sequences of the barriers I$_1$ and I$_2$, 
is given if Fig.\ \ref{phasepl}. The transmission 
$T\left(\epsilon\right)$  (dotted line) is identical for both cases.
Let us assume that I$_1 \gg$ I$_2$. 
In this first barrier sequence, we have a strong energy dependence of the 
phase $\varphi\left(\epsilon\right)$ (solid line in Fig.\ \ref{phasepl}), 
which implies the existence of Andreev type resonances at 
finite bias. The electrons entering the INI interferometer from the normal 
lead, are given enough time to build up an Andreev resonance and preferably 
leave into
the superconductor. For the inverse barrier sequence, the barrier I$_2$ at the
NS interface dominates. The weak energy dependence of the phase 
$\varphi\left(\epsilon\right)$ of reflection allows no sharp 
resonances to build up (dashed line in Fig.\ \ref{phasepl}). 
This reflects the fact that 
the electrons which enter the INI region leave through I$_1$ into the normal 
lead before an Andreev resonance can build up.
In summary, the spectral density in the INI interferometer changes radically
with the coupling strengths of the normal and superconducting leads. Normal
resonances dominate when the interferometer is coupled more strongly to the
normal lead, whereas Andreev resonances take over in weight when the coupling
to the superconductor is stronger. At any instant, however, only the Andreev 
states participate in the charge transport.

We turn to the numerical analysis of a multichannel NINIS junction. 
With Eq.\ (\ref{onechannel}) we extend the linear response study 
of Ref.\  \cite{melsen} to finite 
voltage. We investigate an NI$_1$NI$_2$S junction with two 
$\delta$-function 
barriers of typical strength $H=\int V(x) dx \approx \hbar v_F$ and 
corresponding reflection probability $R=H^2/\left(H^2+\hbar^2v_F^2\right)$ 
assuming values between $R=0.2$ and $R=1$. We vary the relative barrier 
strengths 
to cover the range between the two limits I$_1>$ I$_2$ and I$_1<$ I$_2$ 
discussed above. 
The distance between the barriers is chosen to be of the order of or larger 
than
the coherence length of the superconductor, providing the forward channel with
one to a few Andreev resonances. The number of resonances increases 
with the incidence angle of the channels. We chose leads with a cross section 
area of $\left(100/k_F\right)^2$, which amounts to about $800$ transverse 
channels. The ratio of the energy gap to the Fermi energy is assumed to be 
$\Delta/\epsilon_F=0.002$. Each channel features the typical conductance
structure of paired Andreev resonances exposed above. Their positions and 
widths 
depend on the ratio of the barrier strengths I$_1$ and I$_2$ as well as the 
longitudinal kinetic energy of the single channels. Remarkably, the overall 
conductance,
which is obtained through summation of single channel conductances, exhibits a 
characteristic subgap structure signaling the presence of Andreev resonances.
\begin{figure}[h]
\noindent
\makebox[8cm]{\psfig{figure=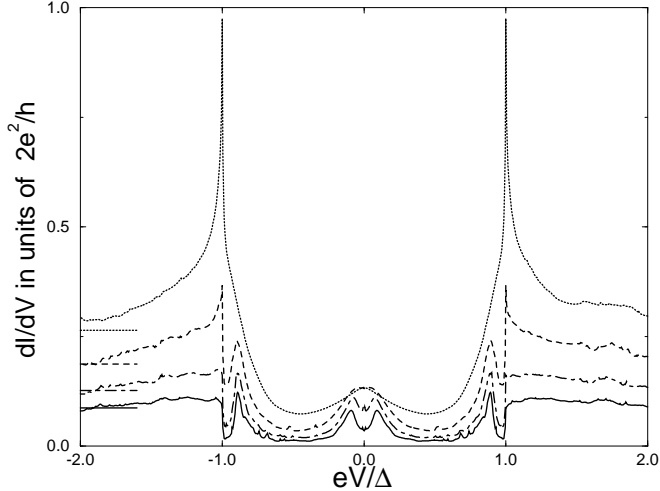,angle=-90,height=8cm}}
\vspace{1cm}
\caption{Differential conductance for a multichannel NINIS junction with 
interbarrier distance $d=2v_F/\Delta=2\pi\xi$. The average conductance per 
channel is plotted 
versus voltage at temperature $T=0$. The barrier strengths are chosen to 
be $R_1=0.2$ 
(dotted line), $0.5$ (dashed), $0.7$ (dot-dashed), and $0.8$ (solid) 
in the normal region, while $R_2=0.5$ at the NS interface. 
The normal state conductance, which is roughly 
independent over this voltage range, is indicated on the left. With increasing
barrier strength $R_1$ the zero bias anomaly develops into a finite bias 
anomaly as the Andreev resonance is formed for $R_1>R_2$.}
\label{nsfig4}
\end{figure}
\noindent
In contrast, the overall conductance of the corresponding NI$_1$NI$_2$N 
junction is practically constant, the normal resonances of the INI region 
having been averaged out.

The numerical study of a 3-dimensional NINS junction has shown that both the 
positions and the number of resonances in the overall conductance correspond to
those in the forward channel \cite{bagwell}\footnote{This is connected to 
the decrease of the transmission probability with increasing incidence 
angle, and the non-isotropic distribution of the channels over the 
incidence angles \cite{bagwell}, which is $\propto 
\sin\left(2\vartheta\right)$.}. In the NINIS junction, we do 
not find a direct correspondence of the resonances of the total conductance 
with the forward channel nor with any other specific channel, although a clear 
resonance structure still survives the summation over the channels.

\begin{figure}[h]
\noindent
\makebox[8cm]{\psfig{figure=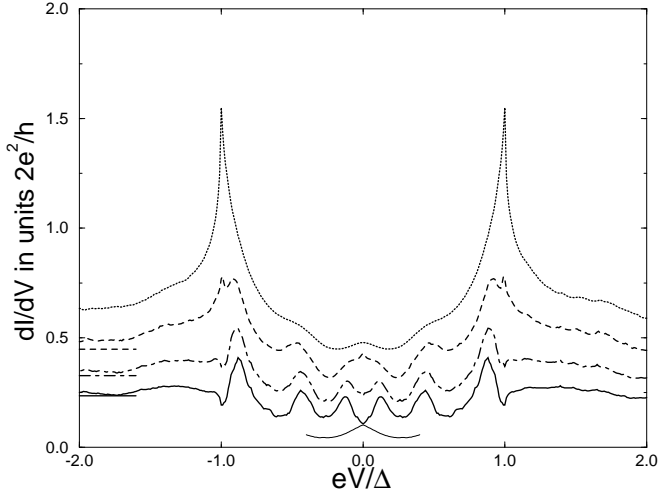,angle=-90,height=8cm}}
\caption{Differential conductance for a multichannel NINIS junction of 
width $d=4v_F/\Delta=4\pi\xi$. The average conductance per channel is plotted 
versus voltage for temperature $T=0$. The barrier strengths are $R_1=0.04$
(dotted line), $0.2$ (dashed), $0.4$ (dot-dashed), and $0.54$ (solid) 
in the normal region, and fixed at $R_2=0.2$ at the NS interface. 
The normal state conductivity is indicated on the left.
As the strength $R_1$ is increased, the zero bias anomaly
turn into an finite bias anomaly and several Andreev resonances appear.
For the last choice of barriers we have interchanged the barrier sequence: 
the conductance at zero voltage remains the same while 
changing from a zero bias minimum to a zero bias maximum (short solid line).}
\label{nsfig5}
\end{figure}

Let us concentrate on the conductance formula (\ref{subonechannel}), valid 
at subgap voltages, and on the properties of the CVC close to zero voltage. 
For I$_1>$ I$_2$, the denominator of (\ref{subonechannel}) changes rapidly 
with the strong energy dependence of the phase of 
$r_{\nu}\left(\epsilon\right)$ 
which is responsible for the appearance of conductance peaks at
finite voltage. The pronounced structure in the conductance survives the 
summation over the channels as displayed in Figures \ref{nsfig4} and 
\ref{nsfig5} (solid lines). 
The repulsion of the Andreev levels
around zero voltage produces a minimum in $dI/dV$ at zero voltage. For 
I$_1<$ I$_2$, the phase of the reflection amplitudes $r_{\nu}\left(\epsilon
\right)$ has negligible energy dependence and the numerator of 
(\ref{subonechannel}) dictates the features of the conductance. The expansion 
of the product $T_{\nu}\left(\epsilon\right)T_{\nu}\left(-\epsilon\right)=
T_{\nu}^2-\epsilon^2 T_{\nu}'^2$ about zero energy shows the existence 
of a zero bias maximum. 
The zero bias anomaly shows up as a characteristic property of the overall 
conductance, see Figures \ref{nsfig4} and \ref{nsfig5} (dotted lines). The 
zero bias maximum coincides 
with the maximum of the conductance product $G\left(\epsilon\right)G\left(-
\epsilon\right)$ of the corresponding NININ junction at zero energy. 
Figs.\ \ref{nsfig4} and \ref{nsfig5} illustrate the crossover from zero to 
finite bias anomalies for two different 
interbarrier distances $d$ as the strength of barrier I$_1$ is increased and
I$_2$ is kept fixed. For an interbarrier distance larger than the coherence 
length of the superconductor, several Andreev resonances show up (see Fig.\ 
\ref{nsfig5}). 
Note that the inversion of the barrier sequence transforms the zero
voltage conductance from a local minimum to a local maximum, while keeping the
same zero voltage conductance. 

The interference of multiple scattering processes between the scattering 
region and the NS interface thus produces an interesting structure in the 
differential conductance. Recently, experiments in disordered
NS and SmS junctions \cite{kastalsky,nguyen,nitta,bakker,magnee} have
concentrated on the observation of zero and finite bias anomalies. It has been
understood theoretically \cite{volkov,marmokos,yip,wees} that these features are
due to the interplay between the barrier at the NS interface and
disorder in the normal lead. At small disorder, the differential conductance 
exhibits a zero bias maximum, while at large enough disorder, a finite bias 
peak is expected \cite{yip} at a voltage of the order of the Thouless energy 
$E_c$ of the normal lead \cite{thouless}. 
This has recently been confirmed in an experiment \cite{sanquer,poirier}. 
Here, we have found the existence of the analog features in a ballistic double
barrier NS junction (in  addition, the double barrier junction shows higher 
harmonics in the resonances of the conductance). The ballistic point of view 
applied to the disordered NS junction thus explains the 
finite bias anomaly as a superposition of resonances due to 
quasi-bound Andreev states between the superconductor and the disorder.
This interpretation applies in the same way to recent experiments showing 
a low-temperature reentrance in the conductance 
\cite{petrashov,charlat}. The non-monotonicity is again due to the resonance 
structure in the spectral conductance $G(\epsilon)$, which now appears as 
a function of temperature.

\subsection{Discussion}

We have described the current--voltage characteristics 
of NS junctions by the scattering matrix approach and expressed 
it through the spectral conductance, which takes into account the explicit 
voltage dependence of the scattering problem. The spectral conductance 
is symmetric under sign reversal of voltage in the in the subgap regime.
We have derived Eq. (\ref{onechannel}) giving the spectral conductance 
in terms of the normal scattering amplitudes. This 
result has enabled us to carry out a study of a double barrier NINIS junction 
at finite voltage. As the ratio of the barrier strengths of the INI 
interferometer is changed, and we go over from normal to Andreev resonances, 
the conductance shows a cross-over from a zero bias to a finite bias 
peak.\\

\newpage

\section{Finite voltage shot noise in normal-metal -- superconductor junctions}
\label{shotnoise}

\markright{FINITE VOLTAGE SHOT NOISE IN ... }

\subsection{Introduction}

Shot noise arises from the current fluctuations in transport as a 
consequence of the discrete nature of the carriers, first predicted
by Schottky for the vacuum tube \cite{schottky}. In a coherent 
conductor the quantum fluctuations, which follow from the probabilistic
nature of the backscattering restricted by the Pauli principle, are 
reduced in comparison to the Schottky result \cite{lesovik,buttnoise}.
In a disordered normal-metal--superconductor (NS) junction, the shot noise 
is produced by the normal scattering processes as well as the imperfect 
Andreev reflection \cite{khlus,muzy,Jong,hessling}. The Andreev reflection 
introduces fluctuations proportional to the double electron charge, which may 
be iteratively increased to give fluctuations of several charge quanta in 
biased SNS junctions \cite{averimam}. For a review, we refer the reader
to \cite{deJong}.
In this chapter we study the shot noise at finite voltage in a disordered
NS junction as well as its  structure due to the iterative 
scattering processes between the disordered normal lead and the NS interface.
We derive a formula which expresses the differential shot noise of the
dirty NS junction in terms of the scattering matrix of the normal
lead and the (scalar) amplitude of Andreev reflection. In the systems of a 
single barrier normal-metal--insulator--normal-metal--superconductor 
(NINS) junction and a double barrier NINIS junction, we explain the existence 
of a non-trivial resonance structure. The results of chapter \ref{shotnoise}
have been published in \cite{alfnoise}. 

\begin{figure}[t]
\vspace{-3mm}
\centerline{\psfig{figure=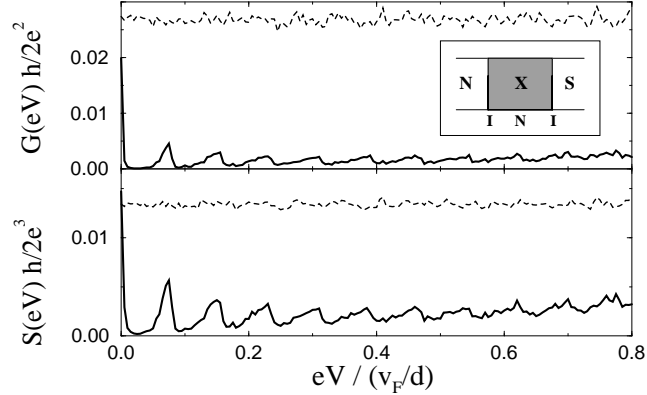,angle=-90,width=80mm}}
\vspace{-1.2cm}
\caption{Differential conductance (top) and shot noise (bottom) in a 
multichannel NINIS junction (see inset, symmetrical barriers of strength 
$\int dx\, V\left(x\right) = 
3\hbar v_F$, $T_1=T_2 \approx 0.05$, interbarrier distance 
$d=20\, v_F/\Delta$, $8\times 10^4$ channels). The average shot 
noise per channel [Eq.\ (\ref{diffnoise})] exhibits maxima at 
the resonances of the conductance (solid lines). Note the enhanced structure 
of the noise with respect to the conductance [Eq.\ (\ref{conduc})]. 
The dashed lines show the corresponding results for the normal state NININ 
junction.
Inset: schematic NXS junction, e.g., X $=$ INI. }
\label{fig1}
\end{figure}

We consider a disordered NXS junction, shown in the inset of Fig.\ \ref{fig1}, 
with an arbitrary elastic scattering 
region X in the normal lead, whose effect on the noise is to be determined.
We restrict ourselves to voltages below the superconducting gap $eV\ll \Delta$.
The low-frequency power spectrum of the current fluctuations is determined by 
the irreducible current--current correlator 
\be P =\int dt\, e^{i\omega t} \langle\langle 
I\left( t\right) I\left( 0\right) \rangle\rangle, \quad \quad   
\omega\to 0 
\label{power} \ee
($\langle\langle I\left(t\right) I\left(0\right) \rangle
\rangle  = \langle  \left(I\left(t\right)-\langle I\rangle\right) 
\linebreak[0] 
\left(I\left( 0\right)-\langle I\rangle \right) \rangle$ is the second 
cumulant in time). The time-dependence of the current operator is defined 
by 
\be I\left(t\right)=
\exp \left[i\left(H-\mu N\right)t\right]\, I \,\exp \left[-i\left(H-
\mu N\right)t\right].
\ee
In the mean field approximation for the Hamiltonian $H$
we do not account for the fluctuations of the order parameter in the S region. 
The effective Hamiltonian is diagonalized by a Bogoliubov 
transformation \cite{kuemmel}.
After integration over the cross-section, the net current operator can be 
expressed through the positive energy eigenfunctions,
\be I\left(t\right) = \frac{-e}{mi}\sum_{\epsilon_m,\epsilon_n>0} 
\int\! dydz  \big(u_m^*\hat{\partial_x} 
u_n \gamma_{m}^{\dag}\gamma_{n}  
- v_m^* \hat{\partial_x} v_n 
\gamma_{n}\gamma_{m}^{\dag} \big) 
e^{i\left(\epsilon_m-\epsilon_n\right)t} 
\label{currentop}
\ee
which may be evaluated in the normal lead. The operators $\gamma_{m}$ 
belonging to the wavefunctions $\left(u_m,v_m\right)$ annihilate the 
scattering states of the disordered NS junction 
(note $u\hat{\partial_x}v=u\partial_xv-v\partial_xu$). We note that
the symmetry of the BdG eigenfunctions $(u_n,v_n)$, $\epsilon_n$ with 
respect to sign reversal of energy to $(-v_n^*,u_n^*)$, $-\epsilon_n$ 
is the consequence of spin degeneracy in Nambu space, allowing the 
the negative energy states to be eliminated in favor of the positive
energy states, see Appendix \ref{bdgapp}.
\noindent 
\noindent
At voltages below the superconducting gap ($eV<\Delta$), 
the quasi-particles are injected from the normal reservoir only, and the 
wavefunctions depend on the $N\times N$ global reflection matrices 
$r_{ee}\left(\epsilon\right)$, $r_{he}\left(\epsilon\right)$, 
$r_{eh}\left(\epsilon\right)$, and $r_{hh}\left(\epsilon\right)$ of the 
disordered NS junction, see Eq.\ (\ref{globalR}). 
The scattering states indexed 
by $m=\left[^{e}_{h}, \mu,\epsilon\right]$ consist of an incident 
$e$(lectron) or $h$(ole) like quasi-particle in channel 
$\mu$ with energy $\epsilon$ superposed with the 
reflected electron and hole states. The occupation numbers are given by 
the Fermi-Dirac distribution $f_{e,h}=f\left(\epsilon \pm eV\right)$. 

The noise power (\ref{power}) is determined by the transitions 
induced by the current operator $I$ from initial states 
$|n\rangle = |n_n=1, n_m=0\rangle$ to intermediate states 
$|m\rangle = |n_n=0, n_m=1\rangle$ and back, which differ only by their 
occupation with respect to two single particle states with indices $n$ and $m$.
E.g., the transition between an incident electron 
$n = \left[e,\nu,\epsilon\right]$ and an incident hole 
state $m=\left[h,\mu,\epsilon\right]$ is produced by the
matrix element between the reflected electrons and holes of the two scattering 
states with respect to the current operator, 
$\langle m | I | n\rangle \propto f_{e} \left(1-f_{h}\right) 
 (r_{eh}^{\dag}r_{ee}-r_{hh}^{\dag}r_{he})_{\mu\nu}$.
Summed over the channels, the transitions contribute to the fluctuations 
with the weight 
$\sum_{\mu,\nu} |\langle m | I | n\rangle|^2 \propto  
f_{e}\left(1-f_{h}\right) \mbox{Tr}\{
(r_{ee}^{\dag}r_{eh}-r_{he}^{\dag}r_{hh}) \linebreak[0]
(r_{eh}^{\dag}r_{ee}-r_{hh}^{\dag} r_{he}) \} \linebreak[1] 
= f_{e}\left(1-f_{h}\right) \mbox{Tr}\{r_{he}^{\dag}r_{he}
(1-r_{he}^{\dag}r_{he})\}$.
Following the quantum mechanical formalism outlined, we obtain the 
low-frequency limit of the power spectrum \cite{datta,martin} valid for 
$T, eV < \Delta$,
\ba P = \frac{4e^2}{h} \int\limits_{0}^{\Delta} d\epsilon && \Bigg\{ 
\left[ f_{e}\left(1-f_{h}\right) + f_{h}\left(1-f_{e}\right) 
\right] 
\mbox{Tr}\Big[ r_{he}^{\dag}r_{he}\left(1-r_{he}^{\dag}r_{he}\right)
\Big]  \nonumber \\
&& +\left[ f_{e}\left(1-f_{e}\right) + f_{h}\left(1-f_{h}\right) \right] 
\mbox{Tr}\Big[ \left(r_{he}^{\dag}r_{he}\right)^2\Big] \Bigg\}. 
\label{pgeneral} 
\ea
\noindent
The first term describes the transitions between states of opposite current 
sign, while the second represents the transitions 
between states of equal current sign. At zero temperature, the second term 
is suppressed by the Pauli principle and the first term produces the 
shot noise  \cite{lesovik,lesolevi, levileso}. 
The thermal fluctuations describing the Johnson-Nyquist 
noise \cite{johnson,Nyquist} 
are due to both terms. Eq. (\ref{pgeneral}) is manifestly invariant under 
sign reversal of voltage, see Chp.\ \ref{scattapproach}.

\subsection{Interplay between normal scattering and Andreev reflection}

We express the electron-hole reflection matrix $r_{he}\left(\epsilon\right)$ 
in Eq.\ (\ref{pgeneral}) in terms of the normal reflection- and transmission 
matrices $r_{ii}$ and $t_{ij}$ of the scattering (X) region and the amplitude 
of Andreev reflection $\Gamma$ at the XS interface,
\be
r_{he}\left(\epsilon\right)=  t_{12}^*\left(-\epsilon\right)  
\left[1-\Gamma^2\left(\epsilon\right) r_{22}\left(\epsilon\right) 
r_{22}^*\left(-\epsilon\right) \right]^{-1} 
\Gamma\left(\epsilon\right)  t_{21}\left(-\epsilon\right),
\ee
see Eqs.\ (\ref{Rhat}) and (\ref{ri}), recalling
$\Gamma\left(\epsilon\right)=e^{i\vartheta\left(\epsilon\right)}$, 
$\vartheta\left(\epsilon\right)=\arccos\left(\epsilon/\Delta\right)$).
Inserting $r_{he}\left(\epsilon\right)$ in Eq.\ (\ref{pgeneral}) provides us
with a general multichannel expression for the shot noise in a disordered
NXS junction, explicit in the normal scattering matrix and the Andreev 
reflection amplitude.

Next we restrict ourselves to the situation of a junction with uniform
transverse cross-section, which permits the separation of channels and
will be used in the single and double barrier systems below.
We recall that the differential conductances of the channels 
$\nu=1 ... N$ 
are given by 
\be \frac{h}{4e^2} G_{\nu}\left(\epsilon\right)= \left(r_{he}^{\dag}r_{he}
\right)_{\nu\nu}, 
\label{conduc}
\ee
see Eq.\ (\ref{subonechannel}). 
At zero temperature, the shot noise (\ref{pgeneral}) 
exhibits the noise power 
\be 
P = \frac{1}{e} \int_0^{e|V|} d\epsilon\, \sum_{\nu} 
S_{\nu}\left(\epsilon\right),
\ee 
with the differential (low-frequency) shot noise of channel~$\nu$,
\ba \lefteqn{ \frac{h}{2e^3}S_{\nu}\left(\epsilon\right) \, = \, 4
\left(r_{he}^{\dag}r_{he} \left(1-r_{he}^{\dag}r_{he}\right)\right)_{\nu\nu} }
\nonumber \\ &=&
\frac{ 4 T_{\nu}\left(\epsilon\right) T_{\nu}\left(-\epsilon\right)
\left(R_{\nu}\left(\epsilon\right) + R_{\nu}\left(-\epsilon\right) -
2 {\rm Re}\left[\Gamma \left(\epsilon\right)^2 r_{\nu}\left(\epsilon\right) 
r_{\nu}^*\left(-\epsilon\right)\right]\right)} 
{\left(T_{\nu}\left(\epsilon\right) T_{\nu}\left(-\epsilon\right) + 
R_{\nu}\left(\epsilon\right) + R_{\nu}\left(-\epsilon\right) - 
2 {\rm Re}\left[\Gamma \left(\epsilon\right)^2 r_{\nu}\left(\epsilon\right) 
r_{\nu}^*\left(-\epsilon\right)\right]\right)^2}.
\label{diffnoise} 
\ea
$r_{\nu}\left(\epsilon\right)$ is the reflection amplitude of channel $\nu$ in 
the matrix $r_{22}\left(\epsilon\right)$ describing the reflection from the 
{\it right}-hand side of the X region, $R_{\nu}\left(
\epsilon\right)=|r_{\nu}\left(\epsilon\right)|^2$, and $T_{\nu}\left(\epsilon
\right)=1-R_{\nu}\left(\epsilon\right)$. 
The two energy dependencies of the reflection probabilities 
$R_{\nu}\left(\epsilon\right)$ and (the phases of) the reflection amplitudes
$r_{\nu}\left(\epsilon\right)$ translate into the voltage dependence 
of the shot noise. The term sensitive to the phases of the reflection 
amplitudes is due to the states multiply scattered between the disorder and the
NS interface. Note the similarity of Eq.\ (\ref{diffnoise}) to the
conductance formula (\ref{onechannel}).
In the limit $\epsilon\to 0$ ($\Gamma\to -i$) the 
phase dependencies drop out of Eq.\ (\ref{diffnoise}) and we recover 
the linear response result \cite{deJong},
\be 
\frac{h}{2e^2} S_{\nu}\left( 0\right)= \frac{16 T(0)^2 [1-T(0)]}{[2-T(0)]^4}.
\ee

\begin{figure}[tb]
\vspace{-5mm}
\centerline{\psfig{figure=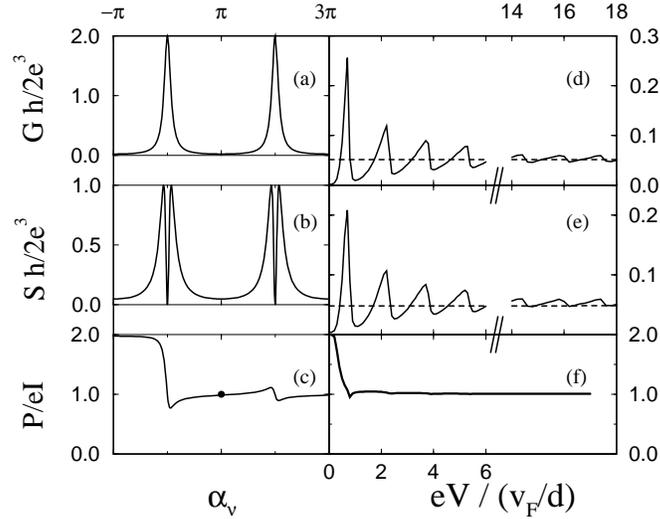,width=90mm,angle=-90}}
\caption[*]{Zero temperature conductance (a,d), differential shot noise (b,e), 
and finite voltage noise power to current ratio (c,f) for a NINS junction 
with a 
barrier of strength $\int dx\, V(x) = 3\hbar v_F$, average transmission 
$T=0.05$, 
$\epsilon_F=500\Delta$, $d=20\, v_F/\Delta$ 
($v_F/d \ll \Delta \ll \epsilon_F$). 
(a,b,c) follow from Eqs.\ (\ref{ninsnoise}) and (\ref{ninscond}) for one
channel. (d,e,f) from Eqs.\ (\ref{subonechannel}) and (\ref{diffnoise}) are 
averaged over 
$8\times10^4$ channels. Note the pronounced resonance structure in both 
conductance (d) 
and noise (e), which approaches the values for the corresponding NIN 
junction at large voltages (dashed lines). The noise to current ratio (c,f) 
decays to the classical Schottky value $P/eI=1$ above the Andreev minigap.}
\label{fig2}
\end{figure}

In order to describe the weak coupling of
a normal lead to an NS proximity sandwich, we consider a 
normal-metal--insulator--normal-metal--superconducting (NINS) 
junction with a potential 
barrier of low transmission ($T\ll 1$) placed at a distance $d$ away from the 
perfect NS interface. The NINS junction serves as a model system for 
a tunneling experiment from a metallic tip to a thin film 
NS layer structure, which has permitted the observation of the Rowell-McMillan 
oscillations \cite{rowell}. In this elementary model, the reflection 
amplitudes $r_{\nu}\left(\epsilon\right)=
\sqrt{R}e^{i\varphi_{\nu}\left(\epsilon\right)}$ 
have a roughly constant modulus $R$ and an energy dependent phase 
$\varphi_{\nu}\left(\epsilon\right)\approx 2\left(k_{\nu}+\epsilon/v_{\nu}
\right)d + \varphi_0$, accumulated during the propagation between the 
potential 
barrier and the NS interface ($k_{\nu}$, $v_{\nu}$ denote the Fermi wave 
number 
and velocity of channel $\nu$). Consequently the shot noise (\ref{diffnoise}) 
for the channel $\nu$,
\be \frac{h}{2e^3}S_{\nu}\left(\epsilon\right) =
\frac{ 8 T^2 R \left[1-\cos \alpha_{\nu}\left(\epsilon\right)
\right]}
{\left(T^2 + 2R \left[1-\cos \alpha_{\nu}\left(\epsilon\right)
\right] \right)^2}, \label{ninsnoise} 
\ee
depends only on the phase difference $\alpha_{\nu}(\epsilon) = 
\varphi_{\nu}\left(\epsilon\right)-\varphi_{\nu}\left(-\epsilon\right)
-2\vartheta\left(\epsilon\right)$ of electron and 
hole propagation through the X ($=$IN) region. The resonance structure of 
$S_{\nu}$ is intimately connected to the voltage (energy $\epsilon=eV$) 
dependence of the differential conductance (\ref{subonechannel}), 
\be \frac{h}{4e^2}G_{\nu}\left(\epsilon\right) = 
\frac{ T^2}{T^2 + 2R \left[1-\cos \alpha_{\nu}\left(\epsilon\right)
\right]}. \label{ninscond} 
\ee
Figures \ref{fig2}(a) and \ref{fig2}(b) show their generic dependence on 
$\alpha_{\nu}$. 
The minima of the denominator [$\cos \alpha_{\nu}=1 \, \Rightarrow \, 
\epsilon_{\nu,n}= v_{\nu}/2d (n\pi +\arccos \epsilon_{\nu,n}/\Delta)$] 
correspond to the resonances of the Andreev quasi-bound states. 
These conductance resonances are repelled from zero voltage, since the phase 
of the Andreev reflection $\vartheta\left(\epsilon\to 0\right)=\pi/2$ 
has to be compensated by the phase difference of electron and hole propagation 
$\varphi_{\nu}\left(\epsilon\right)-\varphi_{\nu}\left(-\epsilon\right)$.
The shot noise (\ref{ninsnoise}) vanishes at these resonances 
($S_{\nu}\left(\epsilon_{\nu,n}\right)=0$). The noise maxima are found from 
$\cos\alpha_{\nu}=\left(2R-T^2\right)/2R$ at energies doubly peaked close to
the resonances. The peak separation $\delta\epsilon \sim T v_F/d$ coincides 
with the width of the resonances. 

Interestingly, this non-trivial structure
survives in the multichannel NINS junction, whose results are shown in 
Figs.\ \ref{fig2}(d) and \ref{fig2}(e). The stability of the resonance 
structure is due to its pinning to the Fermi energy. 
In contrast, the resonance structure of a potential well 
in a normal conductor is washed out in the multichannel limit. As Figures 
\ref{fig2}(b,e) show, the narrow double peak structure in the noise 
is smeared out in the multichannel limit, and the noise $S$ takes a maximum
rather than a minimum at the resonance, retracing the shape of the conductance.

At large voltages ($eV\gg v_F/d$), the shot noise approaches a constant as a 
consequence of the dephasing between the channels. The channel average may be 
evaluated by averaging over the phase $\alpha_{\nu}$ in Eqs. 
(\ref{ninsnoise}) and (\ref{ninscond}) and we obtain the 
result ($\alpha_{\nu}$ ($1/N\sum_n \to 1/2\pi \int_0^{2\pi} d\alpha$),
\be \bar{S}=\frac{1}{2\pi}\int_0^{2\pi} d\alpha\, 
S_{\nu}\left(\alpha\right) = \frac{2e^3}{h} T, \quad\quad \bar{G} = 
\frac{2e^2}{h} T. 
\ee
It follows that both the shot noise and the conductance approach the normal 
state values at voltages $v_F/d \ll eV \ll \Delta$. This limiting behavior 
demonstrates that the NIN junction is effectively decoupled from the NS 
interface in the large voltage limit and dominates both noise and conductance 
due to its low transparency. 

Let us concentrate on the overall ratio of noise power to current 
$P/eI$ 
[$P=\int_0^{e|V|} d\epsilon\, S(\epsilon)$, $I=\int_0^{eV} d\epsilon\, 
G(\epsilon)$]. This ratio has already been used successfully in determining the
unit of the charge carriers in the Fractional Quantum Hall 
Effect \cite{glattli,picciotto}.
While at small voltages ($P/eI=2$) the noise carries the signature
of the Cooper pairs created by the Andreev reflection, at large voltages 
($v_F/d\ll eV \ll \Delta$) it decays to the normal state value ($P/eI=1$). 
Interestingly, this decay is immediate at the first Andreev resonance, as is 
seen from Figs.\ 2(c) and (f). In fact, in a single channel, the energy 
integration is equivalent to the integration over the phase $\alpha_{\nu}$, 
which at the midpoint between two resonances $\alpha_{\nu}$ has covered the 
period $2\pi$, thus producing $P/eI=1$ at $\alpha_{\nu}=\pi$, 
see Fig.\ 2(c). In the
transport through a normal lead weakly coupled to a NS sandwich, we thus 
find two characteristic regimes: At voltages below $eV < v_F/d$ the 
one-particle excitations in the NS structure are effectively 
gapped and we may only couple to the superconducting condensate which 
introduces the charge $2e$ in the noise power. At voltages
above the first Andreev resonance, we may couple directly to a one-particle
density of states in the normal layer and we loose the signature of the
superconductor at once.

\subsection{Double barrier junction}
 
In the study of normal versus superconducting junctions, it is of interest 
to investigate the impact the quantum coherence has on the 
macroscopic limit of transport (limit of an infinite number of channels) and, 
in particular, to compare the behavior of a specific scattering region 
connected to either a normal or a superconducting reservoir.
We consider an insulator--normal-metal--insulator (X$=$INI) interferometer 
which we connect to normal leads in a NININ junction and to a superconducting
lead in a NINIS junction. Besides the intrinsic interest in the double barrier 
system as a paradigm, now to be combined with a superconductor, 
it may also serve as a qualitative model for a disordered conductor, due to
its strong similarity in the transmission distribution \cite{doro,melsen}. 
As we consider voltages below the superconducting gap, we typically chose
the thickness of the interlayer to be $d>> \hbar v_F/\Delta$.

The resonance structure of the INI interferometer results in a 
bimodal distribution of its (overall) transmission probabilities $T$.  
For symmetric barriers it takes the form 
(infinite channel limit, $T_1=T_2\ll1$) \cite{deJong},
\be \rho\left(T\right)=\frac{1}{\pi}\frac{T_1}{2}\frac{1}{\sqrt{T^3
\left(1-T\right)}}, \quad \quad T \in \left[\frac{T_1^2}{\pi^2}, 1\right], 
\label{bimodal} \ee
which has its analog in a disordered conductor \cite{doro,melsen}. 
We recall that, in 
a NININ junction, the rich structure of this distribution has no impact on the 
macroscopic transport properties. The (linear) conductance and shot noise 
may be evaluated from the bimodal distribution (\ref{bimodal}), 
$\left(h/2e^2\right) G = \int dT\, \rho\left(T\right) T = T_1/2$, 
$\left(h/2e^3\right) S = \int dT\, \rho\left(T\right) T\left(1-T\right) = 
T_1/4$, and show no characteristics of coherent transport \cite{chen}. 
The conductance follows from the
series resistance of the two barriers, and an incoherent model of a double 
barrier junction gives the same suppressed noise $S/eG=1/2$ as a consequence 
of charge conservation \cite{buttbeen}.
When considering the INI interferometer in a NINIS junction instead, we find 
a non-trivial answer, both at small and large voltages. 
The {\it linear} response follows  
from Eqs. (\ref{subonechannel}), (\ref{diffnoise}), and (\ref{bimodal}),
\ba \frac{h}{2e^2} G\left(0\right)&=&\int dT\, \rho\left(T\right) \frac{2T^2}{
\left(2-T\right)^2} = \frac{T_1}{2} \frac{1}{\sqrt{2}}, 
\nonumber \\
\frac{h}{2e^3} S\left(0\right)& = & \int dT\, \rho\left(T\right) \frac{16T^2
\left(1-T\right)}{\left(2-T\right)^4} = \frac{T_1}{2} \frac{3}{4\sqrt{2}}.
\ea
The ratio of shot noise to conductance is \cite{deJong}  
$S\left(0\right)/eG\left(0\right)=3/4$. A comparison of the shot noise to 
conductance ratios in a single barrier junction (X$=$I or X$=$IN) and a
junction with disorder (X$=$D) is instructive, see Table 1. 
In the single barrier junction the distribution of transmissions is peaked at 
$T\ll 1$. The noise is due to the Schottky type fluctuations in 
nearly closed channels, and consequently the noise ratios $S_N/eG_N=1$ in the 
NIN and $S_S/eG_S=2$ in the NIS junction differ only by the charge 
quanta involved. In a junction with disorder, the noise-to-conductance ratio 
is also doubled for the NDS case, see Table 1. However, since in the 
bimodal distribution of the transmissions the noise is produced by the 
channels 
with intermediate transmission $0 < T < 1$ and the current by the open 
channels 
with $T \to 1$, the doubling is a non-generic and thus yet unexplained 
feature. This is demonstrated by the noise-to-conductance ratio of $3/4$ in 
the NINIS as compared to $1/2$ in the NININ junction. 
\begin{table}[ht]
\caption{Shot noise to conductance ratio $S/eG$ of a NXN 
compared to a NXS junction; single barrier X $=$ I, double barrier X $=$ INI, 
disorder X $=$ D. The results are valid for small transmissions $T\ll 1$, 
and many channels ($N \to \infty$) \cite{khlus,deJong,chen,buttbeen}. }
\vspace{0.5cm}
\centerline{
\begin{tabular}{l|ccc}
\hline\hline
      & single           & double barrier  & disorder \\ 
      & barrier $T\ll 1$ & $T_1=T_2\ll 1$ &  \\
\hline
$S_N/eG_N$ & 1     &1/2       & 1/3 \\ 
$S_S/eG_S$ & 2     &3/4       & 2/3 \\
\hline\hline
\end{tabular}
}
\end{table}

At {\it finite} voltage, the shot noise is described by Eq. (\ref{diffnoise}) 
with the reflection amplitudes 
$r_{\nu}\left(\pm\epsilon\right)= r_2 + t_2^2 r_1 
e^{2ik_{\nu}\left(\pm\epsilon\right)d} /\left( 1-r_1 r_2 e^{2ik_{\nu}
\left(\pm\epsilon\right)d} \right)$ of the double barrier interferometer. 
We have evaluated this expression for a symmetric double barrier junction 
($T_1=T_2=0.05$) and display the results in 
Fig.\ \ref{fig1}. At voltages of the order of the Andreev levels 
$eV\sim v_F/d$, we find a resonance structure independent of the number of 
channels. We observe again that the differential noise follows the 
resonance peaks of the conductance. At large voltages 
($eV\gg v_F/d$), the resonances disappear as the electrons 
and holes dephase, and we approach the regime where the 
conductance and shot noise become indistinguishable from the incoherent 
addition of the NIN and NIS junctions.

\subsection{Discussion}

In conclusion, we have expressed the differential shot noise in a disordered 
NXS junction in Eq.\ (\ref{diffnoise}) in terms of the normal and Andreev 
scattering amplitudes. We have described the resonance structure in the 
shot noise found at finite voltage in coherent transport. The 
robustness of the resonance structure in the multichannel limit is owed to 
its pinning to the Fermi energy. We have pointed out the possibility of a 
non-trivial normal versus superconducting noise ratio in the NINIS junction 
as as consequence of the bimodal distribution. 
Finally, we have found a rapid decay of double shot noise in a NINS junction 
above the Andreev minigap.

\chapter{Magnetic response of a normal-metal -- superconductor structure: 
Nonlocality}
\label{magresponse}

\section{Introduction}
\label{magintro}
\markboth{CHAPTER \ref{magresponse}.  MAGNETIC RESPONSE OF ...}{}

In this chapter we turn to the magnetic response of a normal metal 
in proximity to a superconductor. While the transport experiments
discussed in Chapter \ref{nonlinearity} probe the Andreev
quasi-particles in an out-of-equilibrium situation, which is
kept up by the voltage difference across the leads, the magnetic 
screening is a thermodynamic property of the condensate wavefunction.
Induced superconductivity, in fact, does not necessarily imply supercurrents.
Though the Andreev reflection may enhance the conductance 
in a normal-metal -- superconductor junction, the transport currents remain 
nevertheless dissipative. In the response to an applied magnetic field 
considered here, however, the Andreev quasi-particles produce diamagnetic 
supercurrents. The magnetism is a property of the $H-T$ phase diagram of the 
proximity system. 

Here we study the linear current response to an applied magnetic field. 
Combined with the self-consistent solution of the Maxwell equation this 
allows to 
describe the diamagnetic phase of the $H-T$ phase diagram. To be specific, 
we consider an normal-metal slab of thickness $d$ in contact with a 
bulk superconductor, as shown in Fig.\ \ref{fig:geometry}.
The pioneering works in the magnetic response of hybrid NS structures 
go back to the Orsay Group 
\cite{degennes:64,orsay}.  Using the Ginsburg-Landau (GL) equations 
for superconductivity, see \cite{orsay,geshkensokol}, they gained a good 
understanding of the rich phenomenology of these mostly dirty thin film 
systems. The normal metal in proximity showed an increasingly diamagnetic
response with decreasing temperature, eventually approaching a perfect 
Meissner-Ochsenfeld effect \cite{meissner}. Although they obtained the 
correct qualitative picture, the GL equations were
pushed beyond their strict domain of validity. The progress in the
fabrication technology opened the way for interesting experiments 
on increasingly clean NS proximity systems in the last two decades 
\cite{oda:80,mota:82,pobell:87,mota:89,mota:90,mota:94,oda:95}. Their 
quantitative understanding requires a more accurate microscopic 
description. 
Most importantly for this thesis was an astonishing finding
made by Mota and co-workers \cite{mota:90} in the low-temperature 
behavior of normal-metal coated superconducting cylinders. These Ag-Nb
cylinders were shown to exhibit a paramagnetic signal  
on top of the diamagnetic susceptibility.
Before a closer analysis of this fascinating experiment became possible 
--- we will return to it in Chapter \ref{paramagnetic} --- 
a quantitative analysis of their experimental data at higher 
temperature was needed for a characterization of the sample with respect
to the proximity contact and the impurity concentration. In fact, the theory
by \cite{orsay} could not give an quantitative description of the
usual diamagnetic behavior. The work described in this chapter has 
laid the basis for this understanding, tracing the screening behavior 
back to the non-locality of the current-field relations shown to be
typical for the proximity effect. 

A mile-stone in the theoretical development was the work by Zaikin 
\cite{zaikin} on the magnetic response of a clean normal-metal 
proximity layer. Applying the quasi-classical Green's function technique 
to this problem, he demonstrated the power of this approach, which has
been predominantly used since \cite{narikiyo:89,higashitani,bbs}. 
Section \ref{quasiclassprox} is devoted to the quasi-classical technique 
in the proximity effect. 
Zaikin found a non-local current-field dependence in this 
system, which exhibits a similarity to the Pippard equation \cite{pippard}. 
In section \ref{screening} we derive the general linear response kernel 
for a NS sandwich with an arbitrary impurity concentration. We find that the 
clean limit represents just one facet of the typical nonlocal constitutive 
relation 
in the proximity effect. The rich phenomenology is due to the self-consistency 
with the Maxwell equations required by the screening problem.
In section \ref{sensitivity}, we study 
the sensitivity of the nonlocality to the presence of impurities. The
superfluid density and the range of the current-field relations, both
diminishing with decreasing mean free path, are shown to affect the
screening ability in a contrary way and compete in the magnetic
response.

\section{Quasi-classical equations and proximity effect}
\label{quasiclassprox}

\subsection{Eilenberger equation}

The basic set of equations appropriate for describing spatially
inhomogeneous superconductors was developed by
Eilenberger \cite{eilenberger} and by Larkin and Ovchinnikov
\cite{larkin:68}. Starting from the widely used field theoretic
method for superconductivity, see \cite{agd,rickayzen,mahan}, 
by averaging over the oscillations on the scale of the Fermi
wavelength, a quasi-classical Green's function is obtained, which
retains the full information on the length scales of interest: 
the coherence length and the field penetration depth. The $2\times 2$ 
matrix Green's function $\hat{g}=\hat{g}_{\omega_n}({\bf x}, {\bf v}_F)$,
\be \hat{g}=\left( 
   \begin{array}[c]{cc}
     g&f\\
     f^\dag&-g
   \end{array} \right),
\label{green}
\ee
satisfies the Eilenberger equation\footnote{$e=|e|$, $\hbar=k_B=c=1$ 
throughout this thesis.}
\be  -\left({\bf v}_F\cdot {\bf \nabla}\right) 
\hat{g}_{\omega_n}\left({\bf x},{\bf v}_F\right)= 
\left[ \left(\omega_n+ie {\bf v}_F\cdot {\bf A}\left({\bf x}\right) \right) 
\hat{\tau}_3  + \hat{\sigma}_{\omega_n}({\bf x}), 
\hat{g}_{\omega_n}\left({\bf x},{\bf v}_F\right) \right], 
\label{eq:eilenberger}
\ee 
with the normalization condition $\hat{g}^2=1$. Here 
$g=g_{\omega_n}({\bf x},{\bf v}_F)$ 
is the normal- and $f=f_{\omega_n}({\bf x},{\bf v}_F)$, 
$f^{\dag}=f^{\dag}_{\omega_n}({\bf x,{\bf v}_F) }$ are the anomalous 
Green's functions, 
depending on the Matsubara frequency $\omega_n =(2n+1)\pi T$ ($n=0,\pm1,..$), 
the center-of-mass coordinate $\bf x$, and the direction of the Fermi
velocity ${\bf v}_F$ ($\hat{\tau}_i$ denote the Pauli matrices, 
$\left[\cdot,\cdot\right]$ is the commutator). The Eilenberger equation
allows to describe spatially inhomogeneous superconductors in equilibrium.
The self-energy $\hat{\sigma}=\hat{\sigma}^{(s)} + \hat{\sigma}^{(i)} + ...$ 
contains the off-diagonal pair potential
\be
\hat{\sigma}_{\omega_n}^{(s)}({\bf x}) =
\left(    \begin{array}[c]{cc}
     0               &   \Delta({\bf x}) \\
     \Delta({\bf x})^* &   0
   \end{array} \right),
\ee
as well as other contributions from impurity scattering or higher order 
electron-electron interactions. The quasi-classical technique represents a 
self-contained formalism, i.e, the self-energies can be expressed through the 
quasi-classical Green's functions themselves. Below we use the 
impurity scattering self-energy in the Born limit,
\be
\label{born}
\hat{\sigma}_{\omega_n}^{(i)}({\bf x}) = 
\frac{1}{2\tau} \langle \hat{g}_{\omega_n}({\bf x},{\bf v}_F) \rangle,
\ee
where $\tau$ the scattering time and $\langle ... \rangle$ denotes the
angular average $\int d\Omega/4\pi (...) $ over ${\bf v}_F$. 
The order parameter is self-consistently determined by
($N_0 = m k_F/\hbar^2\pi^2$) 
\be \Delta({\bf x}) = - V N_0 \pi T \sum_{\omega_n>0} 
\langle f_{\omega_n}({\bf x},{\bf v}_F) \rangle,
\label{eq:gap}
\ee
where $V<0$ describes the effective coupling constant for
the attractive electron-electron interaction. The systems of Eqs. 
(\ref{eq:eilenberger}) and (\ref{eq:gap}) are completed by the 
quasi-classical current expression,
\be 
{\bf j}({\bf x}) = 2ie N_0 \pi T \sum_{\omega_n>0} 
\langle {\bf v}_F g_{\omega_n}\left({\bf x},{\bf v}_F\right) \rangle.
\label{eq:current}
\ee
In the Gorkov formalism \cite{agd} the current is separated into a diamagnetic
and a paramagnetic contribution. Eq.\ (\ref{eq:current}) contains both 
contributions, even if it is formally similar to the paramagnetic current 
in terms of the Gorkov-type Green's function. This is due 
to the renormalization procedure used in the quasi-classical theory and
is explained in appendix \ref{quasiclassicaltheory}.

The Eilenberger equation is valid on all length scales $\zeta$ with 
$k_F\zeta \gg 1$, and for magnetic fields such that the Larmor radius 
$r_L=\hbar k_F c/eH$ exceeds the involved length scale $r_L \gg \zeta$ 
($r_L$ being the cyclotron radius of a particle with Fermi velocity).
The quasi-classical Green's functions need to be supplemented by a set of
boundary conditions \cite{zaitsev,kieselmann,hara}, 
because they are not continuous at the interfaces of different materials. 
For more details on the derivation, the symmetries and the validity of
the Eilenberger equations, we refer the reader to appendix 
\ref{quasiclassicaltheory} and several reviews on the subject 
\cite{serenerainer,shelankov,rammersmith,kopnin}. 

Let us illustrate the use of the quasi-classical equations for the  
simple case of a homogeneous superconductor. The Green's function 
takes the form
\be 
\hat{g}=\frac{1}{\Omega_n} 
\left(    \begin{array}[c]{cc}
     \omega_n               &   \Delta \\
     \Delta                 &   -\omega_n
   \end{array} \right),
\label{gsuper}
\ee
where $\Omega_n=\sqrt{\omega_n^2+\Delta^2}$ ($\omega_n>0$). 
Eq.\ (\ref{eq:gap}) yields the usual gap equation for the 
superconductor. Within a 
field ${\bf A}({\bf x})$ the solution of (\ref{eq:eilenberger}) is found
by substituting $\omega_n \to \omega_n + ie {\bf v}_F\cdot {\bf A}({\bf x})$ 
in (\ref{gsuper}), assuming a slow variation of the field on the scale 
$\lambda \gg \xi_0$ larger
than the superconducting coherence length $\xi_0=\hbar v_F/2\Delta$.
The current-field relation that follows from Eq.\ (\ref{eq:current}), 
\be 
{\bf j}({\bf x}) = - \frac{ n_s e^2}{mc} {\bf A}({\bf x}),
\label{eq:london}
\ee 
is the local London relation with the superconducting density
\begin{equation}
  n_s=\frac{k_{F}^3}{3\pi^2}\pi T \sum_{\omega_n>0}
    \frac{\Delta^2}{(\Delta^2 +\omega_{n}^2)^{3/2}},
\label{eq:ns}
\end{equation}
approaching the full electron density $n=k_F^3/3\pi^2$ for $T\to 0$.
In the limit of vanishing superconductivity $\Delta\to 0$ 
we recover the trivial Green's function for the normal metal 
$\hat{g}=\tau_3$, and the current expression vanishes.
In the following section we apply the quasi-classical formalism
to the proximity effect in a normal metal slab.

\subsection{Clean NS slab}

\begin{figure}[!tb]
\begin{center}
\leavevmode
\psfig{figure=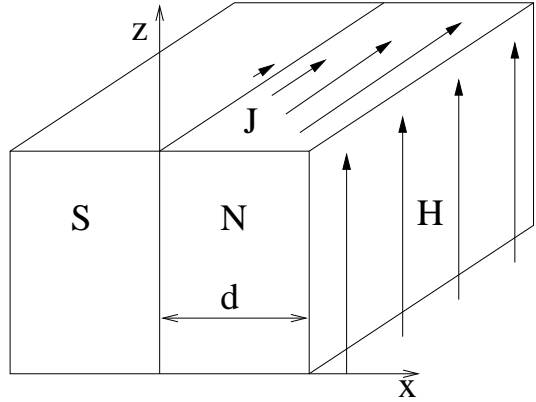,width=7cm}
\caption{Normal-metal layer (thickness $d$) in proximity to a bulk
superconductor. A magnetic field $\bf H$ parallel to the surface is
applied, driving screening current $\bf j$ along the surface. 
Current and fields are inhomogeneous in $x$-direction.}
\label{fig:geometry}
\end{center}
\end{figure}

In this chapter, we consider a normal-metal slab in contact with a 
bulk superconductor, see Fig.\ \ref{fig:geometry}. The normal layer 
($0<x<d$) is described within the free electron gas approximation.
We assume a perfect contact between the metal and the superconductor as well
as a specularly reflecting metal boundary. A magnetic field
$(0,0,H)$ is applied parallel to the metal surface, driving
screening currents $(0,j_y(x),0)$ along the surface. The magnetic 
induction is described by the vector potential $(0,A_y(x),0)$.
Taking a thickness $d \gg\xi_0$, we can neglect the 
self-consistency of the pair potential, and approximate the order 
parameter by a step function $\Delta(x)=\Delta\Theta(-x)$.

The relevant length scales in the system are given by the geometric
thickness $d$, the thermal coherence length $\xi_N(T)=\hbar v_F/2\pi T$, 
and the elastic mean free path $l=v_F \tau$. The thermal length
describes the distance over which quasi-particle excitations in a
typical energy interval given by the temperature $T$ are mutually 
dephased. The above length scales correspond to the respective energy 
scales $T_A=\hbar v_F/d$ of the Andreev levels, the temperature $T$ 
and the impurity self-energy $\hbar/\tau$, all present in 
the Eilenberger equation (\ref{eq:eilenberger}). In the following we 
first consider a clean NS slab, i.e. the electrons propagate ballistically 
between the NS interface and the metal-vacuum boundary ($\tau \to\infty$).
The Green's functions in the normal layer, in the low temperature
limit $T_A \ll \Delta$ of interest, take the form \cite{zaikin}
\be  \begin{array}{lll} 
g &=& \tanh \chi(d), \\
f &=& \frac{\exp\left[ {\rm sgn}(v_x) \chi(d-x)\right] }{\cosh \chi(d)} ,
\end{array}
\label{gfclean}
\ee
where (${\rm sgn} (v_x) = v_x/|v_x|$)
\be
\chi(x) = \frac{2\omega_n x}{|v_x|} + i \frac{2ev_y}{|v_x|} 
\int_0^x dx'\, A_y(x'),
\ee
($\omega_n >0$). The normal Green's function is constant as a signature
of the extent of the quasi-particle over the whole normal layer.
Carrying out the analytical continuation 
$\omega_n \to -iE+0$ and determining the poles of $g(E)$ yields the 
energies of the Andreev states bound to the normal layer,
\be 
E_n = \frac{|v_x|\pi}{2d} (n+\frac{1}{2}) - e\frac{v_y}{d}\int_0^d dx'\, 
A_y(x'), \quad\quad n=0,1,...\, .
\ee
We recover the energies of Eq.\ (\ref{alevel}) of the previous chapter
(in the limit $T_A \ll \Delta$), which are split in direction $\pm v_y$ 
and shifted by the vector potential. 
Note the non-local field dependence of the bound state energies.

From the anomalous Green's function we can derive the pair correlation 
function $F(x)=i\langle \hat{\Psi}_{\downarrow}(x) \hat{\Psi}_{\uparrow}(x) 
\rangle$ which satisfies $\Delta(x)= -V F(x)$,
\be 
F(x) = N_0 \pi T \sum_{\omega_n>0} \langle f_{\omega_n}(x,{\bf v}_F) \rangle.
\label{eq:F}
\ee
We consider the case in the absence of fields $A_y=0$ here. 
Neglecting the effect of the boundary at $x=d$ we find 
\be
F(x) \approx N_0 \frac{v_F}{x} e^{-x/\xi_N(T)}.
\ee
At high temperatures $T\gg T_A$, only the first Matsubara frequency 
$\omega_0=\pi T$ contributes to the sum (\ref{eq:F}), 
$F \sim \exp -x/\xi_N(T)$, while at zero temperature, the decay shows 
no length scale, $F \sim 1/x$.

\section{Magnetic screening}
\label{screening}

\subsection{Linear response kernel}

When studying the magnetic response, the task is to 
determine the current-field relation in the normal layer
in functional dependence of the vector potential j[A(x)] and
solve for the screening currents self-consistently with the
Maxwell equation
\be
-\partial^2_x A_y(x) = 4\pi j_y(x).
\label{ampere}
\ee
In this chapter we focus on the magnetic response linear in the
field. While the ground-state wavefunction experiences a weak perturbation 
that can be treated linearly, 
the screening of the magnetic induction is strong and requires the 
self-consistent solution of the Maxwell equation. Therefore 
the discussion of screening relies on the full knowledge of the 
{\it dispersive} response function $j(q)=K(q)A(q)$. 
While in a London superconductor the response 
$K(q) = K_0 = -c/4\pi\lambda_{\rm \scriptscriptstyle L}^2$ is local, 
implying a penetration depth $\lambda_{\rm \scriptscriptstyle eff}=
\lambda_{\rm \scriptscriptstyle L} \equiv (mc^2/4\pi n_s e^2)^{1/2}$, 
in a Pippard superconductor with $\xi_0>\lambda$ \cite{tinkham} with 
$K(q) \approx K_0 /(1+q\xi_0)$ the nonlocal response changes the
penetration depth to $\lambda_{\rm \scriptscriptstyle eff} = 
\sqrt[3]{\lambda_{\scriptstyle L}^2\xi_0} > 
\lambda_{\rm \scriptscriptstyle L}$. Here we derive the current response in 
functional dependence of the field for the NS slab geometry. We find that the 
current response in the proximity effect is nonlocal in general. 

The current functional linear in $A(x)$ takes the generic form
\begin{equation}
 \label{eq:functional}
 j_{y}(x) = -\int K(x,x') A_y(x')dx' 
\end{equation}
of a convolution with the response kernel $K(x,x')$. In the following we 
derive $K(x,x')$ for the NS sandwich shown in Fig.\ \ref{fig:geometry}, 
expressing it through the Green's function in absence of fields 
$\hat{g}_0$. Inserting the appropriate Green's function $\hat{g}_0$, 
we are able to describe the current in a normal metal 
layer with arbitrary impurity concentration, ranging from the clean 
to the dirty limit. To calculate the
linear diamagnetic response, we separate the Green's function into
a (real) zeroth order part and an (imaginary) first order part 
in $A(x)$ (we drop the Matsubara index $\omega_n$ in the following),
\be
\label{expand}
 \hat{g}({\bf x},{\bf v}_{F}) = \hat{g}_{0}(x,v_{x})+ i
 \hat{g}_{1}(x,v_{x},v_{y}) .
\ee
In the absence of external fields Eq.~(\ref{eq:eilenberger})
reduces to
\begin{eqnarray}
\label{eilenreal}
 -v_x \partial_x g_0(x,v_x) &=& \tilde{\Delta}(x) 
 \left[ f_0(x,v_x) - f_0^{\dag}(x,v_x) \right] \nonumber \\
 -{v}_{x}\partial_x f_{0}(x,v_{x})&=&
 2\tilde{\omega}(x) f_{0}(x,v_{x}) - 2\tilde{\Delta}(x) g_{0}(x,v_{x}) \\ 
 {v}_{x} \partial_x f^{\dag}_{0}(x,v_{x}) &=&
 2\tilde{\omega}(x)f_{0}^{\dag}(x,v_{x}) -
 2\tilde{\Delta}(x)g_{0}(x,v_{x})\; .\nonumber
\end{eqnarray}
We have introduced the effective frequency $\tilde{\omega}(x) = \omega
+ \langle g_{0}(x)\rangle /2\tau$ and pair potential
$\tilde{\Delta}(x)=\Delta(x)+\langle f_{0}(x)\rangle /2\tau$ as diagonal and 
off-diagonal potentials, respectively.
Equations (\ref{eilenreal}) imply that
$f_{0}(x,v_{x})=f_{0}^{\dag}(x,-v_{x})$, as $\tilde{\Delta}$ has been 
chosen real. We consider the solution of Eqs.\ (\ref{eilenreal}) as 
given in the following. The first-order parts obey the equations
\begin{eqnarray} \small 
\! -{v}_{x} \partial_x f_{1}(x,v_{x},v_{y})\! &=& \!
 2\tilde{\omega}(x) f_{1}(x,v_{x},v_{y}) -
 2\tilde{\Delta}(x) g_{1}(x,v_{x},v_{y}) \nonumber \\ 
&& +2ev_{y} A_y(x) f_{0}(x,v_{x}), 
 \label{eilenfirst1} \\ 
\! {v}_{x} \partial_x f^{\dag}_{1}(x,v_{x},v_{y}) \! &=& \!
 2\tilde{\omega}(x) f^{\dag}_{1}(x,v_{x},v_y) 
-2\tilde{\Delta}(x) g_{1}(x,v_{x},v_{y})
\nonumber \\ && +2ev_{y}A_y(x) f^{\dag}_{0}(x,v_{x}) ,
\label{eilenfirst2}
\end{eqnarray}
where $g_{1} = (f_{0} f^{\dag}_{1} + f_{1} f^{\dag}_{0})/ 2g_{0}$ 
follows from the normalization $\hat{g}^2=1$. Eqs.\ (\ref{eilenfirst1}) 
and (\ref{eilenfirst2}) can be mapped to two uncoupled Riccati equations 
and integrated out formally in terms of $g_0$, $f_0$, and $f_0^{\dag}$. 
Determining the current (\ref{eq:current}) we have extracted the 
kernel $K(x,x')$ appearing in (\ref{eq:functional}). The derivation of the 
results given here is deferred to Appendix \ref{kernel}, see also 
\cite{bbf}. Introducing a
'propagator'
\begin{equation}
\label{eq:defm}
 m(v_{x},x,x') = 
 \exp\left(\frac{2}{v_{x}}\int_{x}^{x'}
 \frac{\tilde{\Delta}(x'')}{
 f^{\dag}_{0}(x'',v_{x})} dx'' \right),
\end{equation}
with the properties $m(v_x,x,x') = m(v_x,x',x)^{-1}$ and 
$m(v_x,x,x'') m(v_x,x'',x') = m(v_x,x,x')$, the linear response 
kernel takes the form
\begin{eqnarray}
 \label{eq:kernel}
\lefteqn{ K(x,x')\! = \! e^{2} v_F N_0 \pi T \sum_{\omega_n>0}
 \int\limits_{0}^{v_{F}}du\frac{1\!-\!u^{2}/v_{F}^{2}}{u} 
 \left[1+g_{0}(x,u)\right] \left[ 1-g_{0}(x',u) \right] } \\
 \nonumber & &
 \bigg[\Theta(x-x^{\prime})m(u,x,x^{\prime})+
 \Theta(x^{\prime}-x)m(-u,x,x^{\prime})
 + m(-u,x,d)m(u,d,x^{\prime}) \bigg] .
\end{eqnarray}
The propagator (\ref{eq:defm}) determines the ratio of the current 
response at $x$ to the field at $x'$, 
thus containing the range of the current--field relations. 
The inverse decay length of the propagator is proportional to 
the off-diagonal part of the self-energy $\tilde{\Delta}$. 
The factor $1-g_0$ is a measure of the superfluid density, 
vanishing in the normal state $g_0=1$. The three summands in (\ref{eq:kernel}) 
stem from the quasi-particle trajectories connecting $x$ and $x'$, either 
directly or by one reflection at the normal-metal boundary.

For illustration we reproduce the current response of a half-infinite
superconductor. Setting $d=0$ the solution of the
Eilenberger equation (\ref{eilenreal}) takes the simple form
$g_0=\omega_n/\Omega_n$, $f_0=f^{\dag}_0=\Delta/\Omega_n$, where
$\Omega_n=(\Delta^2+\omega_n^2)^{1/2}$, see (\ref{gsuper}). 
Inserting in (\ref{eq:kernel}) we
obtain the linear-response kernel
\ba
 \label{kernelsuper} 
 K_{S}(x,x')  &=& e^{2} v_F N_0 \pi T
 \sum_{\omega_n>0} \frac{\Delta^{\!2}}{\Omega_n^2}
 \int\limits_{0}^{v_{F}} du \frac{1\!-\!u^{2}/v_{F}^{2}}{u} \\
&& \nonumber \left[ e^{-(2\Omega_n+ 1/\tau) 
 \frac{\scriptstyle|x-x'|}{\scriptstyle u}} +
 e^{(2\Omega_n+1/\tau) \frac{\scriptstyle x+x'}{\scriptstyle u}}
\right] ,
\ea
which describes the current response of a superconductor with arbitrary 
impurity concentration \cite{agd}, 
which here additionally includes the effect of the boundary. 
For fields varying rapidly
spatially we arrive at a non-local current-field relation of the
Pippard-type \cite{pippard}, while for slowly varying fields the kernel
can be integrated out in Eq.~(\ref{eq:functional}), producing the local 
result (\ref{eq:london}). We recall here certain generic features of this
kernel, which are of importance below. In a clean superconductor 
($1/\tau\ll\Delta$), the range is given by the superconducting coherence 
length $\xi_0$, while in a dirty superconductor
($\Delta \ll 1/\tau$) the range is given by the mean free path
$l=v_{F}\tau$, and is thus in both cases (nearly) temperature independent.
We discuss below how, in the proximity
effect, the range of the kernel varies from infinity to $l$ and
$\xi_N(T)$, exhibiting a strong temperature dependence, which leads to
non-trivial screening properties.

\subsection{Clean limit}

The linear response in the clean NS slab illustrates the nonlocality of
the current--field relation in the proximity effect. Inserting the clean
limit solution (\ref{gfclean}) (with $A=0$) into (\ref{eq:kernel}), we 
obtain straightforwardly \cite{zaikin},
\begin{equation}
 \label{eq:jclean}
 j_{y}=-\frac{1}{4\pi\lambda^{2}(T)\,d}
 \int\limits_{0}^{d}A(x)dx \; .
\end{equation}
The temperature-dependent penetration depth that is given
explicitly by
\begin{equation}
 \label{lambdaclean}
\frac{1}{\lambda_N^2(T)}=
 \left\{
 \begin{array}[c]{ll}\displaystyle
 \frac{4\pi e^{2}n}{mc^2} \equiv \frac{1}{\lambda_N^{2}}, &\; T=0
 \\[5mm]\displaystyle
 \frac{6T_A}{\lambda_N^{2} T} e^{-2 T/T_A}, &\; T\gg T_A.
 \end{array}\right.
\end{equation}
Here $n=k_F^3/3\pi^2$ is the normal electron density producing the 
London length typically $\lambda_N \ll d$. 
The propagator trivially becomes $m(v_x,x,x')=1$ implying an infinite 
kernel range. Interestingly, the thermal length $\xi_N(T)$ 
does not enter the kernel, as one might expect from the analogy to a 
clean superconductor with range $\xi_0$. The spatial extent of the 
current--field dependence is thus cut off by the geometry $0<x<d$.
Eq.\ (\ref{eq:jclean}) produces a constant current over the normal-metal 
layer, depending on the average vector potential.
Solving Maxwell's equation (\ref{ampere}) with the boundary
conditions $A_y(x\!=\!0)=0$ (perfect superconductor) and 
$\partial_x A_y(x\!=\!d)=H$ (applied magnetic field), we easily find
for $\lambda_N(T) \ll d$,
\ba
j_y(x) &\approx& -\frac{3H}{8\pi d}, \nonumber \\
B_z(x) &\approx& \frac{3H}{2d}x -\frac{H}{2},\\
A_y(x) &\approx& -\frac{H}{2}x + \frac{3H}{4d}x^2.  \nonumber 
\label{zaikinsolution}
\ea
The current $j\sim H/d$ screens the applied field on the geometric
scale $d$. The magnetic induction $B$ changes sign inside the normal metal 
approaching $B(x\! =\! 0)=-H/2$ at the NS interface: the field is overscreened.
While the screening currents minimize the average vector potential 
to 
\be
\frac{1}{d} \int_0^d A(x') dx' \sim Hd \frac{\lambda_N^2(T)}{d^2}, 
\label{aaverage}
\ee
the average magnetization 
\be 
4\pi M = \frac{A(d)}{d}-H \approx -\frac{3}{4} H, 
\ee
makes up only $3/4$ of the value for an ideal diamagnet. Thus the nonlocality 
results in a less efficient expulsion of the applied field, with overscreening
as its signature. Overscreening is observed at low temperature, 
as long as $\lambda_N(T) \ll d$, the 
diamagnetic susceptibility leveling off at $4\pi\chi = M/H \approx -3/4$.
At high temperatures the susceptibility is suppressed exponentially,
\be
4\pi \chi = - \frac{d^2}{4\lambda_N^2(T)} \sim e^{-2T/T_A},
\ee
see Fig.\ \ref{fig:rho1} (clean limit).

\section{Sensitivity to impurities}
\label{sensitivity}

\begin{figure}[tb]
 \begin{center}
 \leavevmode 
\psfig{figure=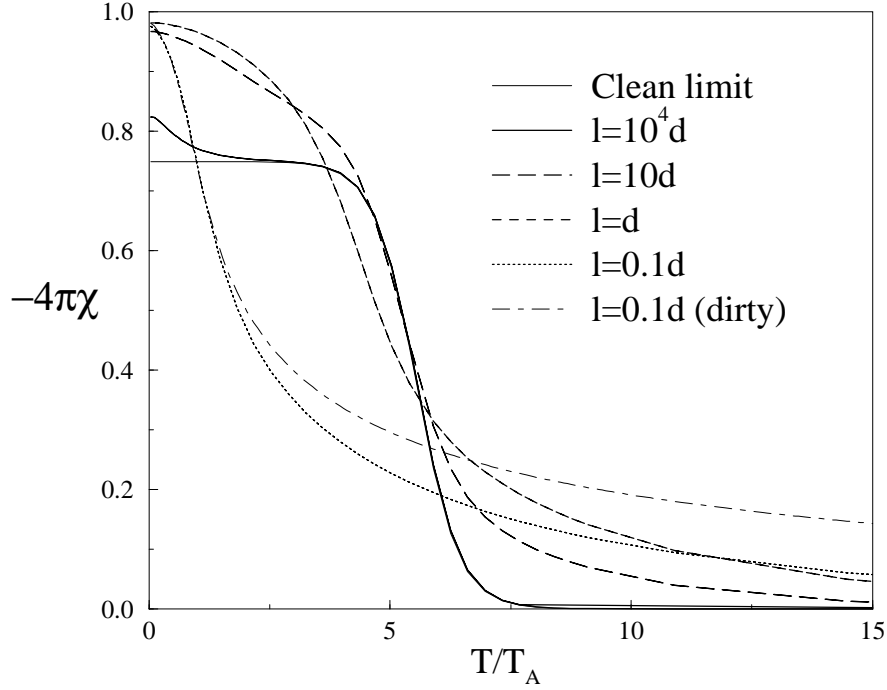,width=12cm}
 \caption[]{ 
Numerical results for diamagnetic susceptibility $4\pi \chi$ 
of the normal metal layer
($\lambda_N=0.003d$). The clean limit is indicated by a thin line
saturating at $4\pi\chi=-0.75$ for $T<5T_A$. The low temperature screening 
is enhanced by a finite  mean free path $l > d$ by reducing 
the kernel range from $\infty$ to $l$. 
Similarly, the high temperature signal is enhanced with decreasing mean
free path. As $l < d$, the 
superfluid (screening) density is reduced, suppressing the diamagnetic 
susceptibility at all temperatures. The dirty limit is shown to overestimate 
the screening at large temperature.}
 \label{fig:rho1} 
 \end{center}
\end{figure}

\begin{figure}[!tb]
 \begin{center}
\psfig{figure=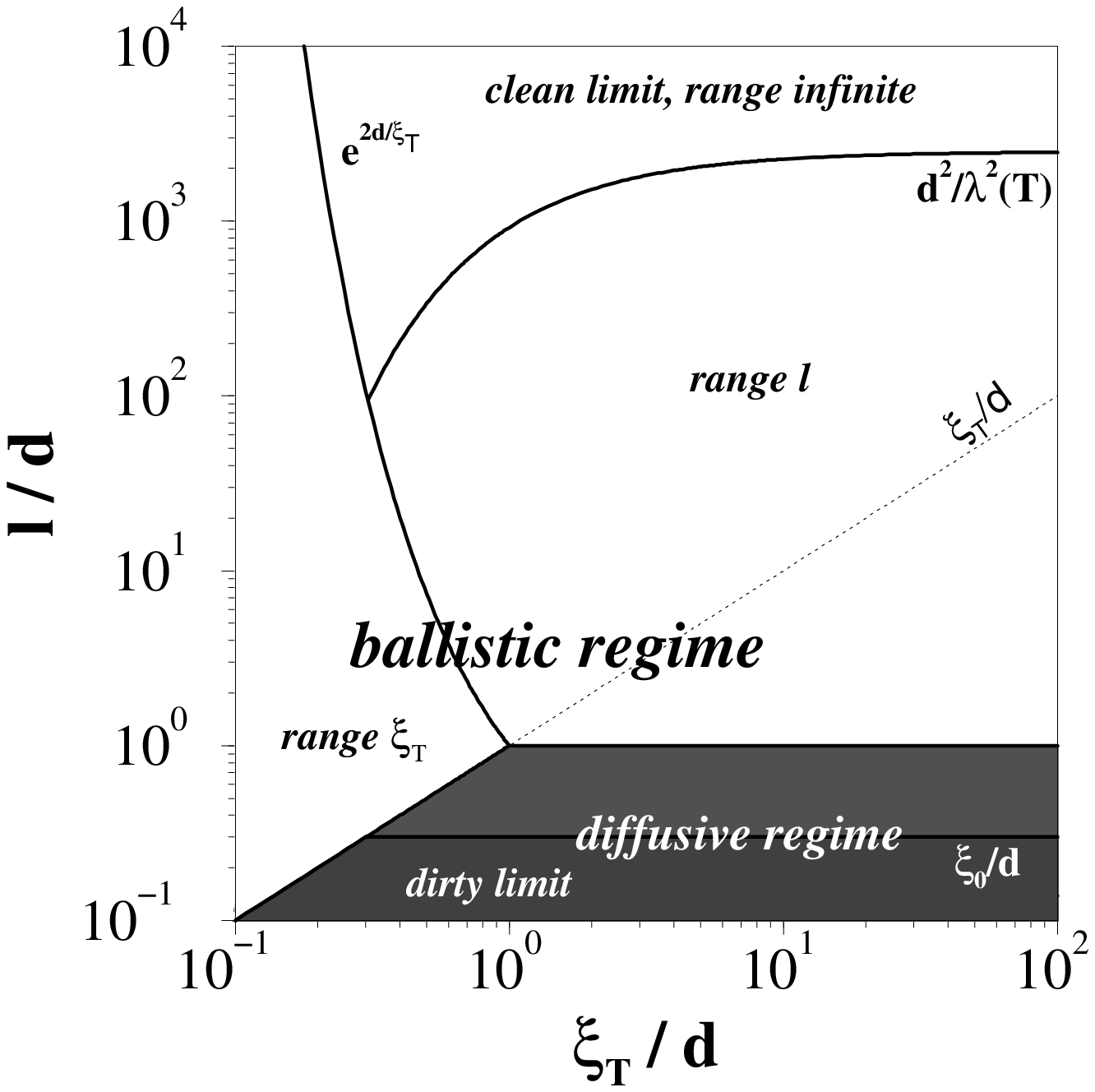,width=8cm} 
\caption{Dependence of the magnetic response on thermal length 
$\xi_N(T)=v_{F}/2\pi T$ and mean free path $l$. 
In the ballistic regime $l\gg {\rm min}\{d,\xi_N(T)\}$ 
we distinguish three regions: (a) the clean limit with infinite 
range of the kernel exhibiting a reduced diamagnetism (overscreening),
(b) the quasi-ballistic limit with finite range $\xi_N(T)$ increasing 
screening 
at large temperatures, (c) the ballistic limit where the finite range 
$l$ enhances the screening although $l\gg d$. 
In the diffusive regime $l\ll \xi_N(T),d$ the range of the kernel is given 
by $l$. The dirty limit with nearly isotropic Green's functions is 
restricted to $l\ll \xi_0, \xi_N(T), d$. Note that the current-field 
relations can still be local or non-local depending on the relative
size of penetration depth and mean free path.}
\label{fig:regime}
\end{center}
\end{figure}

For an arbitrary impurity concentration the screening problem has to 
be solved numerically. First the Green's function in the absence of fields 
$\hat{g}_0$ is determined, according to (\ref{eilenreal}). For the discussion 
of a stable algorithm therefore we refer the reader to \cite{belzigphd}. 
Second, the linear response kernel (\ref{eq:kernel}) is evaluated and solved
self-consistently for $A(x)$ with Maxwell's equation.
Depending on the relative size of the thermal length $\xi_N(T)$, the mean
free path $l$, and the thickness $d$ we distinguish various regimes of 
either ballistic or diffusive electron propagation which are 
shown in Fig.~\ref{fig:regime}. Note that the dividing line between 
ballistic and diffusive behavior is not given by the simple relation 
$\xi_N(T) =l$ 
(dotted line). Rather, the range of the current--field relation $\xi_N(T)$, 
$l$, or $\infty$ is needed for an accurate characterization of the magnetic
response. The numerical results for the magnetic susceptibility $4\pi \chi$ 
as a function of temperature are shown in Fig.~\ref{fig:rho1}. 
The different curves show the clean limit and
the mean free paths $l/d=10^{4},10,1,0.1$. Clearly, a finite
impurity concentration has a strong influence on the susceptibility,
even if $l>d$, and can either increase or decrease the diamagnetic
screening, depending on temperature. In the following, we explain Figs.\ 
\ref{fig:rho1} and \ref{fig:regime} in detail.

\subsection{Ballistic regime}

The ballistic regime is limited by $l\gg {\rm min}\{d,\xi_N(T)\}$, 
which ensures a ballistic
propagation of the electrons over the thickness or the thermal length
of the normal layer, respectively.
As a limiting case, we have already considered the clean limit 
($l\rightarrow \infty$).

An estimation using the zeroth order Eilenberger equation Eq.~(\ref{eilenreal})
easily shows that the clean-limit solution (\ref{gfclean}) remains valid 
in the absence of fields for
\begin{eqnarray} 
\label{eq:cleanvalid1}
l\gg d \exp(2d/\xi_N(T)), \quad & \mbox{if} &\quad \xi_N(T)\ll d, \\
\label{eq:cleanvalid2}
l\gg d, \quad & \mbox{if} & \quad \xi_N(T)\gg d ,
\end{eqnarray}
see Fig.\ \ref{fig:regime}. Note that this includes the region 
$d\ll l \ll \xi_N(T)$, the finite thickness preventing the small mean free 
path $l \ll \xi_N(T)$ from becoming effective.
In the evaluation of the kernel, the off-diagonal potential $\tilde{\Delta} 
= \langle f_0 \rangle/2\tau$ enters the propagator (\ref{eq:defm}), 
yielding the kernel (\ref{eq:kernel}), which approximately takes the form
\begin{equation}
 \label{kernelapprox}
 K(x,x^\prime)=\frac{1}{8\pi \lambda^2(T)d} 
 \left[e^{-\frac{|x-x^\prime|}{l}}+
 e^{-\frac{2d-x-x^\prime}{l}}\right] .
\end{equation}
Since $l\gg d$, the exponentials may be expanded to first order. As a
result, we obtain two contributions to the current 
$j=j_{\rm clean}+j_{\rm imp}$
\begin{eqnarray}
 \label{jyclean}
 j_{\rm{clean}}&=&
 \frac{-1}{4\pi \lambda^2(T)d}\int_0^d A(x)dx,\\
 \label{jyimp}
 j_{\rm{imp}}(x)&=&
 \frac{1}{8\pi \lambda^2(T)d}\int_0^d
 \frac{|x-x^\prime|+2d-x-x^\prime}{l}A(x^\prime)dx^\prime\; .
\end{eqnarray}
According to (\ref{aaverage}) solving the screening problem in the 
clean limit, we find $j_{\rm clean} \sim H/d$. Taking (\ref{jyimp}) as a 
small perturbation, we obtain 
$j_{\rm imp} \sim Hd^2/\lambda_N^2(T) l$, which becomes of the same order
of magnitude as $j_{\rm clean}$ if
\be
l \sim d \frac{d^2}{\lambda_N^2(T)}.
\ee
As long as $l \gg d^3/\lambda_N^2(T)$, the range of the kernel is 
effectively $\infty$ and the clean limit applies, see Fig.\ \ref{fig:regime}.
As $l < d^3/\lambda_N^2(T)$ --- which is the case for $l\gg d$ in general 
--- the mean free path $l$ enters the kernel range, as shown in Fig.\ 
\ref{fig:regime}. The diamagnetic susceptibility is enhanced at low 
temperature ($\xi_N(T)\gg d$), as Fig.\ \ref{fig:rho1} ($l=10^4 d$) 
illustrates. The screening currents are even sensitive to a mean free 
path $l \ll d$. The applied field is found to be screened on 
the effective penetration depth
\be
\lambda_{\rm eff} \equiv \sqrt[3]{\lambda^2(T)l} < d,
\label{lambdaeffclean}
\ee
in analogy to the effective penetration depth 
of a Pippard superconductor $\lambda_{\rm eff}=\sqrt[3]{\lambda_N^2(T)\xi_0}$, 
see e.g. \cite{tinkham}.

In the remaining part of the ballistic regime, 
see Fig.~\ref{fig:regime} (range $\xi_N(T)$), the full solution is not 
known analytically, but we may produce an approximate solution, which 
characterizes well the numerical results at high temperature 
($\xi_N(T) \ll d$). Limiting ourselves to high temperature with $\xi_N(T) \ll d$
allows us to consider the Green's function for the first Matsubara
frequency $\omega_0=\pi T$ only. We restrict ourselves to the forward
direction $v_{x}=+v_{F}$. From Eq.~(\ref{eilenreal}) we
find that $f_0(v_F) \approx 2 \exp(-x/\xi_N(T))$ remains unchanged as in
(\ref{gfclean}), and $f_0^{\dag}(v_F) \ll 1$ and $1-g_0(v_F) \ll 1$ obey the
approximate equations,
\begin{eqnarray}
 \label{eq:cleancorrection}
 \left(\frac{d}{dx}-\frac{1}{\xi_N(T)}\right)f_0^\dag(x,v_F)&=&-\frac1l\langle
 f_0(x)\rangle \\\nonumber 
\frac{d}{dx}(1-g_0(x,v_F))&=&-\frac1l\langle f_0(x)
 \rangle f_0(x,v_F)\;
\end{eqnarray}
with the solutions
\begin{eqnarray}
 \label{eq:clean_corr_solu_for_f}
 f_0^\dag(x,v_F)&=&\frac{\xi_N(T)}{2l}e^{\displaystyle-x/\xi_N(T)}\\
 \label{eq:clean_corr_solu_for_g}
 1-g_0(x,v_F)&=&\frac{\xi_N(T)}{2l}e^{\displaystyle-2x/\xi_N(T)}\; 
\end{eqnarray}
Here we have used $\langle f_0\rangle\approx f_0(v_F)/2$.
$f_0^{\dag}(v_F)$ and $1-g_0(v_F)$ are of order
$\xi_N(T)/l$ rather than exponentially suppressed $f_0^{\dag}\sim 
\exp -d/\xi_N(T)$, $1-g_0 \sim \exp -2d/\xi_N(T)$ as in
the clean limit (\ref{gfclean}). The correction to $g_0$ given by
Eq.~(\ref{eq:clean_corr_solu_for_g}) increases the superfluid
density in the vicinity of the superconductor via the factor
$1-g_0$ in the kernel. The range of the propagator is modified
by the correction (\ref{eq:clean_corr_solu_for_f}) to $f^\dag$,
leading to
\begin{equation}
 \label{eq:mcorr}
 m(v_F,x,x^\prime) \approx\exp\left(2\frac{x^\prime-x}{\xi_N(T)}\right)\; .
\end{equation}
The range of the kernel is now given by $\xi_N(T)$, which is
strongly temperature dependent, and the current flows in a layer 
of thickness $\xi_N(T)$ close to the NS interface,
\begin{equation}
 \label{eq:currcorr}
 j(x)\approx -\frac{1}{\lambda_N^2 l}\int_0^d dx^\prime
 e^{-2\frac{x^\prime+|x-x^\prime|}{\xi_N(T)}}A(x^\prime) .
\end{equation}
Thus in the high temperature limit of Fig.\ \ref{fig:rho1}, as the mean 
free path $l$ becomes smaller the diamagnetic susceptibility 
is enhanced. 

\subsection{Diffusive regime}

If impurity scattering dominates, as described by $\langle
g_0\rangle/\tau\gg\omega$ and $\langle f_0\rangle /\tau\gg \Delta$,
Eq.~(\ref{eilenreal}) can be reduced to the Usadel equation
\cite{usadel} for the isotropic part $\langle f_0(x)\rangle$.  Assuming
$T \ll\Delta$ the solution in the normal metal takes the form
\begin{equation}
 \label{eq:dirtysolution}
 \langle f_0(x) \rangle =
\cosh(\sqrt{\frac{2\omega}{D}}(d-x))/\cosh(\sqrt{\frac{2\omega}{D}}d)\;,
\end{equation}
where $D=v_{F}^2\tau/3$ is the diffusion constant.  Equation
(\ref{eq:dirtysolution}) shows that the important energy scale is the
Thouless energy $E_{c}=D/2\pi d^2$ \cite{thouless}. 
The $f$-function (\ref{eq:dirtysolution}) features the decay length 
$\xi_D(T)=(D/2\pi T)^{1/2}$ reflecting the diffusive
nature of the electron motion.

In the normal metal $l\ll\xi_N(T),d$ are necessary conditions for the
Usadel theory to be valid. However, the numerical solution of 
(\ref{eilenreal}) shows that the Usadel theory in the normal metal, 
which requires the Green's functions to be nearly isotropic, 
is only a good approximation for $l<\xi_0$: the dirty limit in 
Fig.\ \ref{fig:regime}. 

Using the fact that the zeroth-order Green's function is nearly
isotropic and varies on a scale $\xi_D(T)\gg l$, we find for the
kernel (\ref{eq:kernel})
\be
 \label{eq:kerneldirty}
 K(x,x')= \underbrace{\frac{\tau}{\lambda_N^2} 
 T\sum_{\omega_n>0} \langle f_0(x)\rangle^2}_{ \textstyle 
\frac{1}{4\pi \lambda^2(x,T)}} 
\frac{3}{4l} 
 \int\limits_{0}^{v_{F}} du \frac{v_{F}^{2}-u^{2}}{v_{F}^{3}u}
 \left[e^{-\frac{|x-x'|}{l u}} + e^{-\frac{ 2d-x-x'}{l u}} \right] .
\ee
The prefactor contains the full temperature dependence here and 
defines a space dependent penetration depth $\lambda(x,T)$. Qualitatively, 
the range of the kernel retains the same length scale $l$ as in the 
adjacent ballistic regime, while the screening density is 
suppressed by the impurities. In Fig.\ \ref{fig:rho1} we thus find that a 
small mean free path $l\ll d$ produces an overall reduction of the 
susceptibility.

For $\lambda(x,T)\gg l$ the vector potential may
be taken out of the integral in Eq.~(\ref{eq:functional}) and after 
integration over the kernel we recover the well-known local 
current-vector potential relation used in the Usadel theory\cite{usadel}, 
with penetration depth 
$\lambda(x,T)$. We note that as long as $\lambda(x,T) < l$, although 
the Green's functions are nearly isotropic, and in absence of
the field are given by the Usadel theory, the current response is
nevertheless nonlocal.
In Fig.\ \ref{fig:rho1} the magnetic susceptibility for $l=0.1d$ is compared 
to the local Usadel result (dirty limit), showing that the latter 
overestimates the screening at high temperature.

The high temperature ($T\gg E_c$) behavior is characterized by screening
limited to a layer of thickness $\xi_D(T)$ close to the superconductor,
and consequently the diamagnetic susceptibility is given by
\begin{equation}
 \label{eq:rhodirty1}
 4 \pi \chi \propto -\frac{\xi_{D}(T)}{d} \propto \frac{1}{\sqrt{T}} .
\end{equation}
This estimate is in agreement with the Ginzburg-Landau theory \cite{orsay} 
as well as the numerical Usadel results \cite{narikiyo:89}.

\section{Discussion}

There are two main differences in the observable properties of the 
induced screening between the clean and the dirty limit. First, the
zero temperature value of the susceptibility $-4\pi\chi$ saturates at $3/4$ 
in the clean limit, while it reaches unity in the dirty limit.
Second, the asymptotic behavior at high temperature differs: in the clean 
limit $\chi$ decays exponentially $\chi \propto \exp(-2 T/T_A)$, while in the 
dirty limit it follows an algebraic law $\chi \propto T^{-1/2}$.
The intermediate behavior illustrated in Fig.\ \ref{fig:rho1} is determined 
by the competition of the nonlocality range and the screening density.
With increasing disorder, a reduction of the range from $\infty$ to 
$\xi_N(T)$ or $l$ enhances the screening, while a reduction of the 
superfluid density weakens the diamagnetic currents. We have found 
several ballistic and diffusive regimes shown in Fig.\ \ref{fig:regime} 
with different characteristics. Remarkably, due to the absence of the 
small scale $\xi_0$ in the proximity effect, the nonlocal kernel 
(\ref{eq:kernel}) is crucial to the understanding of all the regimes.

Our results imply a non-monotonic dependence of the diamagnetic susceptibility 
on the mean free path. Starting from a clean sample, with decreasing 
purity first the range of the linear response kernel is augmented and
Meissner currents are enhanced, then the superfluid density is reduced 
and the screening currents are suppressed. 
Assuming a temperature-dependent scattering mechanism
with decreasing mean free path as a function of temperature, such as
electron-electron or electron-phonon interaction, we might speculate
to observe a non-monotonic (i.e. re-entrant) behavior of the
susceptibility (here the smallness of the scattering rate is
compensated by the high sensitivity of the non-local current-field
relation). However, as is evident from Eq.~(\ref{eq:defm}), the largest
off-diagonal self-energy ($\tilde{\Delta}$) which includes e.g.
impurity scattering will provide a (low-temperature) cutoff for this
behavior.

Our results allows for a quantitative agreement with the experiments. 
The fit of the experimental data for the diamagnetic susceptibility 
gives an independent determination of the mean free path in these samples, 
which has been compared to the results of the resistivity measurements \cite{mmb}. 
The magnetic susceptibility emerges as a sensitive indicator of the 
impurity concentration due to the nonlocality of the constitutive relation.

\chapter{Magnetic breakdown in a normal-metal -- superconductor 
proximity structure}
\label{breakdown}

\section{Introduction}
\markboth{CHAPTER \ref{breakdown}.  MAGNETIC BREAKDOWN IN ...}{}

In this Chapter we extend the study of magnetism from the linear response 
to the finite field regime. Recent experiments have demonstrated the 
non-trivial screening properties of hybrid normal-metal--superconductor 
structures, exhibiting a magnetic breakdown at finite fields  
\cite{oda:80,mota:82,pobell:87,visani}. The investigated 
samples have typical dimensions comparable to the 
thermal length $\xi_N(T)$ in the normal metal, 
attributing a key role to the quantum coherence of the electrons coupled to 
the macroscopic phase of the superconductor.

\begin{figure}
\noindent
\centerline{\psfig{figure=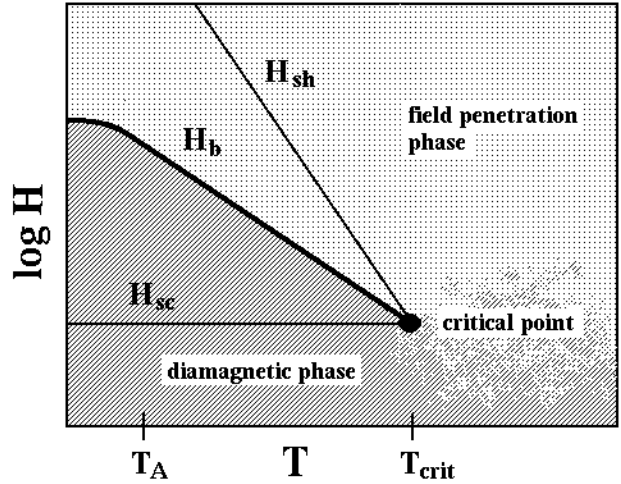,width=85mm}}
\caption[*]{$H-T$ phase diagram of the normal metal slab of thickness $d$. The 
breakdown field $H_b\left(T\right)\sim \Phi_{\circ}/\lambda_N\left(T\right)d$ 
marks the first order transition between the 
diamagnetic and the field penetration phase ($\lambda_N\left(T\right)$ denotes 
the penetration depth, $\Phi_{\circ}$ the superconducting flux unit). 
The critical point at the intersection of the spinodals 
$H_{sc}\left(T\right)\sim \Phi_{\circ}/d^2$ and 
$H_{sh}\left(T\right)\sim \Phi_{\circ}/\lambda_N^2\left(T\right)$  
separates the first order transition for $\lambda_N\left(T\right)\ll d$ 
from the continuous cross-over at large temperature 
($\lambda_N\left(T\right)\gg d$).}
\label{breakfig1}
\end{figure}

We consider a clean normal-metal slab of thickness $d$ in proximity 
with a superconductor, as shown in Fig.\ \ref{fig:geometry}. 
The self-consistent study of the finite field effect in a dirty NS sandwich 
within the framework of the Ginsburg-Landau (GL) equation
was carried out a long time ago by the Orsay group \cite{orsay}. 
Their work has provided the first understanding of the magnetic breakdown, 
which denotes the breakdown of the full magnetic flux expulsion in the 
normal layer, in analogy to the critical field of the superconductor. 
A quantitative agreement of the GL results  with recent experiments is not 
achieved, as is discussed in \cite{mota:89}. This is not surprising 
considering the role of nonlocality in the proximity effect, the local 
current-field relation of the GL theory applying only to the dirty limit.
The quasi-classical Green's function technique offers the possibility to 
describe the opposite clean limit using the Eilenberger 
equations \cite{eilenberger}, which we pursue here. 
In \cite{zaikin} the nonlinear response in the clean limit has been 
obtained along these lines. Using numerical methods, in \cite{bbs} 
the non-linear field regime of the screening problem has been investigated, 
and the two (meta)-stable solutions in both the clean and the 
dirty limit have been determined (the dirty limit extends the GL results
to low temperature using the Usadel equations \cite{usadel}). 
In this chapter we determine the $H$-$T$ 'phase' diagram shown in Fig.\ 
\ref{breakfig1} of the normal metal layer in the clean limit, where the 
bistable regime is particularly extended. In thermodynamic 
equilibrium, we find a magnetic breakdown at $H_b\left(T\right)$, which 
is a first order transition separating the phase of diamagnetic screening from 
the phase of magnetic field penetration. The results presented here have 
been published in \cite{alfbreak}.

Recent experiments \cite{mota:89,mota:94} have investigated the magnetic
response of metallic cylinders with a superconducting core, finding 
data, which was claimed to be characteristic for the ballistic 
limit \cite{mota:94}.
This has motivated us to derive the analytic dependence of the clean limit 
expression for the breakdown field $H_b$ on temperature $T$ and thickness $d$ 
of the normal layer and to compare it to the experiment. From the free energy
of the normal layer, which allows us to identify the two (meta)-stable states, 
we determine the spinodals, the 
thermodynamic breakdown field $H_b\left(T\right)$ and find the 
jumps in magnetization and entropy at the transition. 
Furthermore, we obtain a critical temperature which marks the upper limit of 
the bistable regime (see Fig.\ \ref{breakfig1}). Finally, we relate our 
results for the breakdown field to the signatures
of the nonlocality in the ballistic regime as they show up in the magnetic 
susceptibility $\chi$ and compare them with the experimental data. 
The following discussion is divided into four sections:  
the analysis of the constitutive relations (Sec.\ \ref{constitutive}), 
the solution of the magnetostatic problem (Sec.\ \ref{magnetostatic}), 
the determination of the breakdown field from the free energy (Sec.\ 
{thermodynamics}), and finally the comparison with the experiment
(Sec.\ \ref{exper}).

\section{Constitutive relations}
\label{constitutive}

The quasi-classical Green's function technique provides an appropriate 
description of a metal with nearly spherical Fermi surface. In a finite 
magnetic field, the vector potential 
${\bf A}\left({\bf x}\right)$ can be included as a phase factor along 
unperturbed trajectories, provided the dimensions of the normal metal are 
smaller than the Larmor radius ($r_L=\hbar k_Fc/eH$). 
In the ballistic limit, the quasi-classical $2\times 2$ matrix 
Green's function 
$\hat{g}_{\omega_n}\left({\bf x},{\bf v}_F\right)$ satisfies the Eilenberger 
equation
($e=|e|$, $\hbar=k_B=c=1$), 
\be  -\left({\bf v}_F\cdot {\bf \nabla}\right) 
\hat{g}_{\omega_n}\left({\bf x},{\bf v}_F\right)= 
\left[ \left(\omega_n+ie {\bf v}_F\cdot {\bf A}\left({\bf x}\right) \right) 
\hat{\tau}_3  + \Delta\left({\bf x}\right) \hat{\tau}_1, 
\hat{g}_{\omega_n}\left({\bf x},{\bf v}_F\right) \right], \label{eilen}
\ee 
discussed in Chapter \ref{magresponse}. We exclude impurity scattering 
here by assuming $\tau \to \infty$.

We consider a  normal metal slab of thickness $d$ on top of a bulk 
superconductor as shown in the inset of Fig.\ \ref{breakfig2}. 
The vector potential ${\bf A}=\left(0,A\left(x\right),0\right)$ describes a 
magnetic field ${\bf B}= \left(0,0,B\left(x\right)\right)$ applied parallel to 
the surface, which induces screening currents 
${\bf j}=\left(0,j\left(x\right),0\right)$. We make the usual  
idealizations in the description of the NS sandwich: The superconducting order 
parameter follows a step function
$\Delta\left(x\right)=\Delta \theta\left(-x\right)$ ($\Delta$ real), no 
attractive interactions being present in the normal layer. We assume a perfect 
NS interface as well as specular reflection at the normal-metal boundary. 

In the subsequent analysis we restrict our attention to the magnetic response 
of the normal layer. In the proximity effect, the macroscopic coherence of the 
superconducting condensate induces correlated electron-hole pairs in the 
normal layer through the process of Andreev reflection. The basic process 
consists of
an electron traveling forward and a hole traveling backward along a 
quasi-classical trajectory as shown in Fig.\ \ref{breakfig2} (at discrete 
energies, bound Andreev states are found along these trajectories).
In the presence of a magnetic 
field, the area enclosed by the trajectory (see Fig.\ \ref{breakfig2}) is 
threaded by the flux 
\be \Phi\left(a,\vartheta,\varphi\right)= \oint {\bf A}\left({\bf x}\right) 
\cdot d{\bf x} = 2\tan\vartheta \cos\varphi \int_0^d A\left(x\right) dx,
\ee
which can be expressed through the integral $a=\int_0^d A\left(x\right) dx$ 
times a geometric factor due to the inclination of the trajectory (the 
spherical angles $\vartheta$ and $\varphi$ parameterize the direction of the 
trajectory with respect to the $x$-axis). 
The current carried along a trajectory depends on the phase factor 
$\Phi\left(a,\vartheta,\varphi\right)/\Phi_{\circ}$ acquired by the 
propagation 
of both the electron and the hole along the Andreev loop, and we arrive 
at an intrinsic non-local current--field dependence $j\left(a\right)$.
The total current is determined by the sum over the currents along the 
quasi-classical trajectories, see (\ref{eq:current}), from which we 
obtain the current expression for finite fields \cite{zaikin},
\be 
j\left(a\right) = \int_0^{\pi/2} d\vartheta \int_0^{\pi/2} d\varphi \, 
j\left( \vartheta, \varphi, \Phi\left(a,\vartheta,\varphi\right)  \right), 
\label{jybreak} \ee
where ($\alpha_n=2\omega_nd/v_F\cos\vartheta$)
\ba && j\left(\vartheta,\varphi, \Phi\left(a,\vartheta,\varphi\right) \right)
=-2e v_F N_0 T\sum_{\omega_n>0} \sin^2\vartheta \cos\varphi 
\label{jgeneral} \\ && \times
\frac{\Delta^2 \sin 2\pi\Phi/\Phi_{\circ}}{\left(\omega_n \cosh\alpha_n 
+ \sqrt{\omega_n^2+\Delta^2}\sinh\alpha_n\right)^2
+\Delta^2\cos^2\pi\Phi/\Phi_{\circ}}. \nonumber 
\ea
The induced currents 
for each trajectory depend only on the flux $\Phi$ modulo the 
superconducting flux quantum $\Phi_{\circ}=\pi\hbar c/e$, reflecting
gauge invariance. At small fields ($a/\Phi_{\circ}\ll 1$),
the current response is diamagnetic for all trajectories and the proximity 
effect produces screening currents in the normal metal. 
As the field increases to $a/\Phi_{\circ} \sim 1$, some of the more extended 
trajectories produce paramagnetic currents, since the reduced flux 
$\Phi \in \left[-\Phi_{\circ}/2,\Phi_{\circ}/2\right]$ they enclose becomes 
negative, and the net diamagnetic current response is reduced. 
As we reach large fields ($a/\Phi_{\circ} \gg 1$), the Andreev loops become 
mutually dephased due to a uniform distribution of the reduced flux. The 
associated currents are randomly
dia- or paramagnetic and the net current vanishes. Note that the proximity 
effect, i.e., the existence of the Andreev levels is not 
destroyed in this limit, leading to a finite kinetic energy of the 
currents induced by the magnetic field. 

\section{Magnetostatics}
\label{magnetostatic}
 
Owing to the independence of $j$ on $x$, the Maxwell equation 
$-\partial_x^2A\left(x\right)=4\pi j$ and the constitutive equation 
(\ref{jybreak}) 
combined with the boundary conditions $A\left(x\! =\! 0\right)=0$ and 
$\partial_xA\left(x\! =\! d\right)=H$ can be given a formal solution. 
We arrive at a parabolic dependence for 
\be
A\left(x\right) = Hx + 4\pi j\left(a\right) x\left(d-\frac{x}{2}\right),
\ee 
parameterized by $a=\int_0^d A\left(x\right) dx$, which in turn is 
determined through the self-consistency condition
\be  a=\frac{Hd^2}{2} + \frac{4\pi}{3} j\left(a\right) d^3. 
\label{selfconsistency} 
\ee
The total magnetization $\cal M$ (per unit surface) is defined by 
\be 
4\pi{\cal M}= \int_0^d dx\, \left(\partial_xA\left(x\right)-H\right) 
= 2\pi j\left(a\right) d^2. \label{magnet}
\ee
Eq.\ (\ref{selfconsistency}) contains the essential physics of the
problem: For small fields ($a\to 0$), the current $j \sim -H/ d$ 
linearly suppresses the magnetic induction to the $B\left(0\right)\to -H/2$ 
at the NS boundary (overscreening). The current is given by the
linear response expression in this limit,
\be j\left(a/\Phi_{\circ}\ll 1\right) \approx 
-\frac{1}{4\pi\lambda_N^2\left(T\right)d} a, \label{jlinear} 
\ee
which depends on penetration depth $\lambda_N\left(T\right) \ll d$, 
see (Eq. (\ref{lambda})). When inserted back into 
(\ref{selfconsistency}), 
the vector potential is found to be strongly suppressed to 
$a \sim H \lambda^2_N\left(T\right)$, and we obtain a consistent 
diamagnetic solution (i.e., $a/\Phi_{\circ}\ll 1$) for fields up to 
$H < \Phi_{\circ}/\lambda^2_N$. 
At large fields, the current vanishes ($j\to 0$) 
and the magnetic field penetrates the normal layer. From Eq.\ 
(\ref{selfconsistency}) we find $a\approx Hd^2/2$, consequently this metallic 
behavior is expected down to magnetic fields $H > \Phi_{\circ}/d^2$, as 
follows from the condition $a/\Phi_{\circ} \gg 1$ for the Andreev levels 
to be dephased.
With $\Phi_{\circ}/d^2 \ll \Phi_{\circ}/\lambda^2_N$ the diamagnetic and 
field penetration solution coexist in the regime $\Phi_{\circ}/d^2 < H < 
 \Phi_{\circ}/\lambda^2_N$. These simple estimates for the limits of the 
bistable regime elucidate the numerical data given in \cite{bbs}.

\section{Thermodynamics} 
\label{thermodynamics}

In the phase diagram of Fig.\ \ref{breakfig1} the upper and lower bounds of 
the bistable regime found from the above mean-field analysis are identified 
with the spinodals of the transition, the super-cooled field 
$H_{sc}\sim \Phi_o/d^2$ and the super-heated field 
$H_{sh} \sim \Phi_o/\lambda_N^2\left(T\right)$. In the
thermodynamic equilibrium, a magnetic breakdown occurs at an intermediate 
field, connecting the diamagnetic regime to the field penetration regime by a 
first order transition. In the following, we determine this breakdown field 
and the associated entropy and magnetization jump from the free energy.

The energy (per unit surface) of the currents $j\left(x\right) = 
-\delta F/\delta A\left(x\right)$ is obtained via an integration over the 
non-linear current expression, 
\ba F\left(a\right)&=& -\int_0^a j\left(a'\right) da' \nonumber \\
&=& \hbar v_F N_0 T\sum_{\omega_n>0}\int_0^{\pi/2} d\vartheta 
\int_0^{\pi/2} d\varphi \sin\vartheta \cos\vartheta \label{fofa} \\
&& \!\!\!\!\!\! \!\!\!\!\!\!\!\!\!\!\!\!\! \log\frac{\left(\omega_n\cosh
\alpha_n+\sqrt{\omega_n^2+\Delta^2}\sinh\alpha_n\right)^2 +\Delta^2}
{\left(\omega_n\cosh\alpha_n+\sqrt{\omega_n^2+\Delta^2}\sinh\alpha_n\right)^2 
+ \Delta^2\cos^2\pi\Phi/\Phi_{\circ}}. \!\!\!\! \nonumber
\ea
$F\left(a\right)$ describes the difference in free energy between the metal 
layer under proximity and in the normal state. $F\left(a\right)$ is a 
monotonous and strictly positive function, reflecting the absence of 
condensation energy in the normal layer, and expresses the cost of the 
induced proximity effect lying in the kinetic energy of the currents induced 
by the vector potential. 
The free energy ${\cal F}\left(T,{\cal M}\right)$ is constructed by adding the 
electro-magnetic field energy and subtracting the vacuum field 
contribution, 
\be {\cal F}\left(T,{\cal M}\right)= F\left(a\right) + \int_0^d dx\,
\left(\frac{\left(\partial_x A\left(x\right)\right)^2}{8\pi}-\frac{H^2}{8\pi}
\right) . \label{free}
\ee
We do not include the condensation energy and the kinetic energy of the 
screening currents in the superconductor. The field dependent term of the 
condensation energy might in fact be of the order of the free energy in the 
normal 
layer and would be expected to produce numerical corrections in the results, 
which are not accounted for by our idealized choice of the order parameter
$\Delta\left(x\right)=\Delta \theta\left(-x\right)$. The kinetic energy of 
the screening currents $\sim H^2\lambda$ may be neglected.

\begin{figure}[tb]
\noindent
\centerline{\psfig{figure=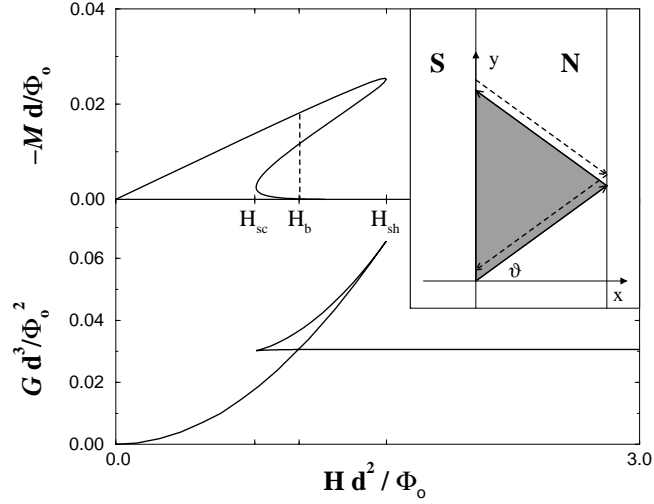,angle=-90,width=82mm}}
\vspace{-0.5cm}
\caption[*]{Magnetization ${\cal M}\left(H\right)$ and free energy 
${\cal G}\left(T,H\right)$ at a temperature $T=1.5\,T_A$ ($T_A=v_F/2\pi d$). 
The representation is universal
in the thickness $d$. The small and large field branches represent 
(meta)-stable solutions describing the diamagnetic and field penetration
phases, which overlap in the field interval $H_{sc}<H<H_{sh}$. The Maxwell 
construction determines the first order transition between the phases at the 
breakdown field $H_b$ (dashed line). 
Inset: Cross-section of the normal-metal slab in contact with the bulk 
superconductor. The flux enclosed by the quasi-classical electron-hole 
trajectory (Andreev loop) enters as a phase factor in the Green's function.}
\label{breakfig2}
\end{figure}

After a Legendre transformation, we obtain the Gibb's free energy 
\ba {\cal G}\left(T,H\right)&=& F\left(T,{\cal M}\right) - {\cal M}H \nonumber 
\\ &=& F\left(a\right)+\int_0^d dx\, \frac{\left(\partial_x 
A\left(x\right)-H\right)^2} {8\pi}. \label{freeenergy}
\ea
The field term in Eq.\ (\ref{freeenergy}) describes the work necessary to 
expel the magnetic field. The extrema of the free energy $\cal G$ 
with respect to $a$ reproduce the equation of state (\ref{selfconsistency}). 
Fig.\ \ref{breakfig2} shows the free energy 
${\cal G}\left(H\right)$ as obtained from the parameterization of $\cal G$ 
and $H$ through $a$. The breakdown field $H_b\left(T\right)$ is
determined by the intersection of the free energies $\cal G$ of the two
(meta)-stable solutions. We note that this procedure is equivalent to 
the Maxwell construction in the magnetization curve 
${\cal M}=-\partial{\cal G}/\partial H$ of Fig.\ \ref{breakfig2}.

In the following, we consider the free energy (\ref{fofa}) in the two 
temperature limits $T=0$ and $T_A\ll T \leq \Delta$ and obtain
($T_A=\hbar v_F/2\pi d$),
\ba F_{T=0}\left(a\right)\! &\! =\! &\! \frac{1}{2} v_F N_0 T_A \int_0^{\pi/2} 
\! d \vartheta 
\int_0^{\pi/2} \! d\varphi \sin\vartheta \cos^2\vartheta 
\left\{ \arctan\left[\tan\pi\Phi/\Phi_{\circ} 
\right]\right\}^2\! , \label{FT0} \\ \!\!\!\!\!\!\!\!\!
F_{T\gg T_A}\left(a\right)\! &\! =\! & \! 4 v_F N_0 T \gamma^2
\left(T,\Delta\right) 
\int_0^{\pi/2} d\vartheta \int_0^{\pi/2} d\varphi \sin\vartheta \cos\vartheta
\nonumber \\ && \exp\left(-\frac{2T}{T_A \cos\vartheta}\right) 
\sin^2\pi\Phi/\Phi_{\circ}. \label{FT>}
\ea 
The finite value of the superconducting gap $\Delta$ is accounted for by 
the dimensionless parameter 
$\gamma\left(T,\Delta\right)=\Delta/\left(\sqrt{\Delta^2+
\left(\pi T\right)^2}+\pi T\right) < 1$.

The free energies of the two (meta)-stable states can be approximated 
by their asymptotic forms in the limits $a\to 0$ and $a\to \infty$. 
In the diamagnetic regime, the expansion in $a/\Phi_{\circ}$ up to quadratic 
order of (\ref{FT0}) and (\ref{FT>}) provide the result 
\be F\left(a\right) \approx   \frac{a^2}{8\pi\lambda_N^2\left(T\right)d}. 
\label{flinear} \ee
Eq.\ (\ref{flinear}) is valid both in the low and high temperature limits 
using the penetration depth $1/\lambda_N^2\left(0\right)\equiv 1/\lambda_N^2 
= \left(4\pi n e^2/m\right)$ at $T=0$ and 
\be \frac{1}{\lambda_N^2\left(T\right)} =  \frac{1}
{\lambda_N^2}\gamma^2\left(T,\Delta\right) 
\frac{6T_A}{T} e^{-2T/T_A} \label{lambda}
\ee
for $T\gg T_A$. Note that the derivative $j = -\partial F/\partial a$
applied to Eq.\ (\ref{flinear}) produces the linear response constitutive 
relation of Eq.\ (\ref{jlinear}).
The Gibb's free energy follows from Eqs.\ (\ref{free}) and (\ref{flinear}), 
using the solution of the Maxwell equations, 
\be {\cal G}\left(a\ll \Phi_{\circ}\right)\approx  
\frac{3}{32\pi}H^2d.  \label{f1}
\ee
Eq.\ (\ref{f1}) is dominated by the magnetization work necessary to expel the 
field, which is parametrically larger (by $\left(d/\lambda_N\left(T\right) 
\right)^2$) than the kinetic energy of the currents. We note that
the scaling in the free energy (\ref{f1}), ${\cal G} \sim H^2d$, 
is parametrically independent of the nonlocality range of the constitutive 
relation (\ref{jgeneral}). 
 
In the field penetration regime we approximate the free energy
by its asymptotic value at $a\to\infty$. In this limit we replace the
strongly oscillating functions of $\Phi$ in (\ref{FT0}) and (\ref{FT>}) by
their average value $\langle\left(\arctan\tan\Phi\right)^2\rangle=\pi^2/12$ 
and $\langle\sin^2\Phi\rangle=1/2$ and obtain
\ba  {\cal G}_{T=0}\left(a\gg \Phi_{\circ} \right)&\approx &
\frac{1}{384\pi}\frac{\Phi_o^2}{\lambda_N^2d}, \nonumber \\
{\cal G}_{T\gg T_A}\left(a \gg \Phi_{\circ}\right)&\approx 
&\frac{3}{16\pi^3} \gamma^2\left(T,\Delta\right)    
\frac{\Phi_o^2} {\lambda_N^2d} e^{-2T/T_A}. \label{f2}
\ea
The magnetization energy vanishes in this limit. The corrections to the
free energy (\ref{f2}) are of relative order $\left(\Phi_{\circ}/a\right)^2$.

The magnetic breakdown field $H_b\left(T\right)$ is determined by the 
intersection of the two asymptotics of the free energy ${\cal G}$ given by 
Eqs. (\ref{f1}) and (\ref{f2}),
\ba H_b\left(T=0\right)&\approx & \frac{1}{6} \frac{\Phi_o}{\lambda_N d}, 
\label{b0}\\
    H_b\left(T\gg T_A\right)&\approx & \frac{\sqrt{2}}{\pi} 
    \gamma\left(T,\Delta\right)  \frac{\Phi_o}{\lambda_N d} 
   e^{-d/\xi_N\left(T\right)}.
\label{breakdownfield} \ea
We note three important features of this result: The temperature 
dependence is a simple exponential with the exponent 
$d/\xi\left(T\right)=T/T_A$, where $\xi_N\left(T\right)=v_F/2\pi T$ denotes 
the thermal length. The amplitude of the breakdown field scales 
inversely proportional to the thickness of the normal layer, $H_b \sim 1/d$. 
In the limit $T\to 0$ the magnetic 
breakdown field saturates to a value which is suppressed by the factor 
$\pi/6\sqrt{2}\approx 0.37$ as compared to the extrapolation of the 
high temperature result. The formula for the breakdown field $H_b \sim 
\Phi_o/\lambda_N(T)d$ is similar to the critical field $H_c \sim 
\Phi_o/\lambda \xi_0$ of a Type I superconductor \cite{blatter:review}, 
the geometric scale $d$ replacing the superconducting coherence length $\xi_0$.

We arrive at the $H-T$ phase diagram shown in Fig.\ \ref{breakfig1}. 
The first order transition between the diamagnetic and the field penetration 
regime takes place between the spinodals $H_{sc}\sim \Phi_o/d^2 < 
H_b\left(T\right) < H_{sh}\sim \Phi_o/\lambda_N\left(T\right)^2$ which 
delimit the (meta)-stable regime. Their intersection marks the critical 
temperature 
\be T_{crit}\approx T_A \log \left(d/\lambda_N\right), \label{tcrit}
\ee
where 
$\lambda_N\left(T_{crit}\right) \approx d$. Below $T_{crit}$ the penetration 
depth is small, $\lambda_N\left(T_{crit}\right) < d$, and we observe 
a first order transition. Above the critical point $T_{crit}$, 
where $\lambda_N\left(T_{crit}\right) > d$, a continuous and reversible 
cross-over between the diamagnetic and field penetration regime is expected. 

The latent heat (at $T\gg T_A$) of the transition follows 
from Eqs.\ (\ref{f1}), (\ref{f2}), 
and (\ref{breakdownfield}) using $S=-\partial{\cal G}/\partial T$,
\be T\Delta S  \approx  \frac{3}{16\pi} \frac{T}{T_A} H_b^2\left(T\right)d, 
\label{latent}
\ee
and is related to the magnetization jump
\be 4\pi \Delta {\cal M} \approx  \frac{3}{4} H_b\left(T\right)d \label{Mjump}
\ee
via the Clausius-Clapeyron equation.

In the derivation of the breakdown field we have used the asymptotic 
expansions of the free energies in $a/\Phi_{\circ}$ and $\Phi_{\circ}/a$, 
respectively. Their quality at the transition point is determined by the 
range of overlap between the diamagnetic and the field penetration regimes 
in Fig.\ \ref{breakfig2}, which
is governed by the parameter $\lambda_N\left(T\right)/d$. In the diamagnetic
phase, the corrections are of the order of $\left(a/\Phi_{\circ}\right)^2\sim
\left(H_b\lambda^2_N\left(T\right)/\Phi_{\circ}\right)^2\sim
\left(\lambda^2_N\left(T\right)/d\right)^2$, and similarly in the 
field penetration regime. The expansion thus breaks down at 
$\lambda_N\left(T\right)\approx d$, which is the critical point of the 
transition line.
We note that the total magnetization changes from its diamagnetic value 
${\cal M} \sim H_b d$ to the strongly suppressed value 
${\cal M} \sim H_b d \left(\lambda_N\left(T\right)/ d\right)^2$ at the 
transition, reflecting its strong first order character.

\section{Experiment}
\label{exper}

The breakdown field has been measured in fairly clean Ag-Nb cylinders 
recently \cite{mota:89,mota:94}. Let us compare our results from the 
clean limit theory with the experimental data. We neglect 
the difference in geometry, cylindrical for the sample and planar in the 
theoretical model.
\begin{figure}[htb]
\noindent
\centerline{\psfig{figure=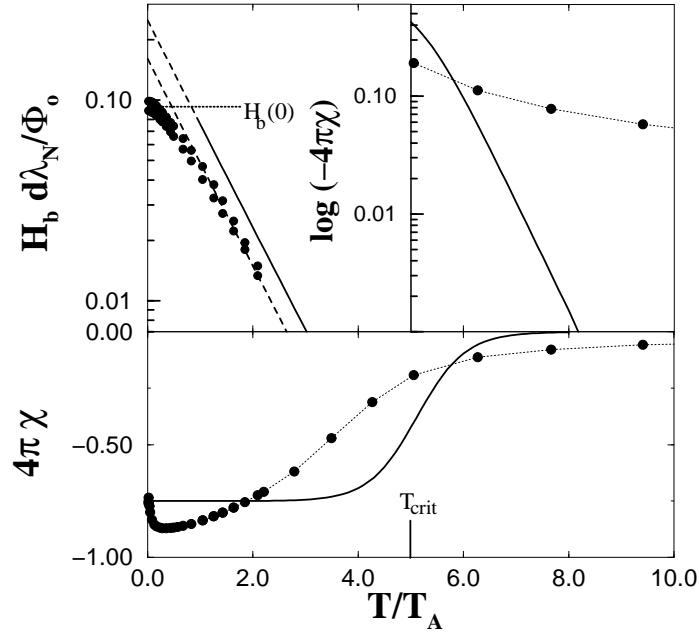,angle=-90,width=82mm}}
\vspace{-1cm}
\caption[*]{Breakdown field $H_b\left(T\right)$ and linear 
susceptibility $4\pi\chi$ from theory and experiment (the analysis 
applies to a Ag-Nb sample of thickness $d=5.5\mu\mbox{m}$). 
Theory: results of Eqs.\ (\ref{breakdownfield}) and (\ref{suscept}) shown as 
solid lines; 
$H_b\left(T\right)$ is rescaled to fit the zero temperature value 
$H_b\left(0\right)$ (horizontal line) to the experiment. Experiment: data 
shown as solid dots, the dotted line is a guide to the eye\cite{mota:94}. 
Note that the logarithmic slope of the breakdown field is reproduced 
precisely (dashed line), while the one of the susceptibility is much 
smaller than expected.}
\label{breakfig3}
\end{figure}

In Fig.\ \ref{breakfig3} we show the two data sets for the breakdown field 
data obtained on heating and cooling a sample of thickness $d=5.5\mu\mbox{m}$
exhibiting hysteresis. We note that the theoretical values of the super-cooled 
and super-heated fields $H_{sc}$ and $H_{sh}$ are not reached in the
experiments, as the phase transition is smoothed by the breakdown of 
local (Andreev) current loops, see Fig.\ \ref{breakfig2}. 
The data saturates at low temperatures, in qualitative agreement with our 
theoretical analysis. Given the electron 
density in Ag, $n=5.8\cdot 10^{22}\mbox{cm}^{-3}$ 
($\Rightarrow \lambda_N=2.2\cdot 10^{-2} \mu\mbox{m}$) 
and $d=5.5 \mu \mbox{m}$, the breakdown field is determined by 
Eq.\ (\ref{b0}) and (\ref{breakdownfield}). Due to the idealization of our 
model, which assumes a step
function for the order parameter, as well as the difference between the planar 
and cylindrical geometry, we expect a numerical factor correcting the 
amplitude of $H_b$. Making use of the scaling factor $\approx 0.56$ in Eqs. 
(\ref{b0}) and (\ref{breakdownfield}), we calibrate the theoretical result to 
fit the zero temperature value $H_b\left(0\right)$, as shown in Fig.\ 
\ref{breakfig3}. The theoretical prediction for the high temperature 
behavior then follows from Eq.\ (\ref{breakdownfield}) and
is shown as a solid line in Fig.\ \ref{breakfig3}. 
Eq.\ (\ref{breakdownfield}) accurately reproduces the logarithmic slope 
$-1/T_A$ of the experimental data, thus correctly tracing the signature 
of the Andreev levels. 
The amplitude of $H_b\left(T\right)$ deviates from the data 
by the constant ratio $\approx 0.64$, which can be attributed to the presence 
of a barrier at the NS interface, see below.

A further agreement between theory and experiment is found in the 
scaling of the breakdown field with sample thickness $d$, which was reported 
to be $\propto 1/d$, in accordance with Eq.\ (\ref{breakdownfield}) 
(the experimental study involved 
$10$ samples \cite{mota:94} with thicknesses ranging from 
$d=2.9\mu\mbox{m}$ to $d=28\mu\mbox{m}$ 
\footnote{We acknowledge unpublished data provided by B.\ M\"uller-Allinger and 
A.\ Mota, ETH Z\"urich.}). Similarly, the critical temperature 
determined in the experiment\cite{mota:89} exhibits the same scaling 
$\propto 1/d$, in agreement with Eq.\ (\ref{tcrit}) ($T_A\propto 1/d$).

For comparison, we cite the dirty limit result, which follows from the GL 
equations 
\cite{orsay},
\be H_D\left(T\right)\approx 1.9 \frac{\Phi_{\circ}}
{\lambda_D \xi_D} 
\exp \left(-d/\xi_D\right), \label{dirtyHb} 
\ee
where $\xi_D= \sqrt{v_F l /6\pi T}$ and $\lambda_D$ is a fitting 
parameter for the penetration depth at the NS interface. Both the 
exponential dependence on $d/\xi_D\propto \sqrt{T}$ and the independence
of the overall amplitude on $d$ clearly deviate from the experimental data.

The good agreement between the clean limit theory and experiment for the 
breakdown field does not trivially generalize to other physical 
quantities, however. In particular, the linear susceptibility 
$\chi={\cal M}/H$ sensitively depends on the non-locality of the constitutive 
relation $j\left(a\right)$.
From Eqs.\ (\ref{magnet}) and (\ref{jlinear}) we obtain the susceptibility
\be 4\pi \chi = \frac{4\pi{\cal M}}{H} =  
-\frac{3}{4}\,\frac{1}{1+3\lambda_N^2\left(T\right)/d^2}, 
\label{suscept} \ee
which exhibits a temperature dependence with the following characteristics: 
$4\pi \chi$  decays exponentially $\propto 1/\lambda^2_N\left(T\right)$ at 
large temperatures, twice as fast as the breakdown field. The susceptibility 
takes half its maximal value at $3\lambda_N^2 \left(T_{1/2}\right)/d^2 
\sim 1$, which roughly coincides with the critical temperature $T_{crit}$.
The logarithmic derivative at $T=T_{1/2}$ is predicted to be 
$\chi'\left(T_{1/2}\right)/\chi\left(T_{1/2}\right) = 1/T_A$.
Below the critical point, the susceptibility saturates as the penetration
depth decreases below the sample thickness ($\lambda_N\left(T\right) < d$). 
Due to the non-locality, the penetration depth drops out of the 
expression for $4\pi\chi \approx -3/4$ and 
we are in the regime of over-screening. 
In Fig.\ \ref{breakfig3} we show the linear susceptibility according to 
the clean limit predictions (\ref{suscept}) (there is no fitting parameter).
The experimental data fails to show the typical saturation of
the susceptibility expected below the critical temperature. At low temperature 
the experimental value clearly exceeds the maximal diamagnetic value $-3/4$ 
found in the clean limit. The decay at large 
temperature is slower than the decay of the breakdown field, while 
Eq.\ (\ref{suscept}) predicts a decay with twice the logarithmic slope, 
see Fig.\ \ref{breakfig3}. This discrepancy finds a natural explanation in 
the different sensitivity of $H_b(T)$ and $\chi(T)$ to the degree of 
nonlocality in the constitutive relation, as we discuss in the following 
section.

Let us consider the influence of an insulating barrier at the NS interface.
The consequences of a finite reflectivity at the NS interface on the linear 
current response has been analyzed in \cite{higashitani}. 
Their results allow for the reflection coefficient $R$ to be  
included in the penetration depth $\lambda_N\left(T\right)$ 
by redefining the factor 
$\gamma_R\left(T_A\ll T\ll \Delta\right)=\left(1-R\right)/\left(1+R\right)$,
in Eq.\ (\ref{lambda}); $\lambda_N\left(0\right)\equiv \lambda_N$ remains 
unchanged. Inserting the modified penetration depth into 
Eq.\ (\ref{suscept}) we 
obtain the linear susceptibility. The additional factor $\gamma$ does not 
change the characteristic shape of the susceptibility (saturation, 
logarithmic slope at $T_{1/2}$, exponential decay), 
but only lowers the position of the half-value of $\chi$ to 
$T_{1/2}\approx \log\left[d\left(1-R\right)/\lambda_N\left(1+R\right)\right]$. 
The finite reflection does not remedy the qualitative discrepancy between the 
susceptibility in theory and experiment, in consistency with the above 
considerations.
Considering the structure of the equations we may expect the dependence on the
reflection $R$ to enter in a similar fashion into the breakdown field 
$H_b\left(T\right)$. 
Eq.\ (\ref{suscept}) inserted in Eq.\ (\ref{breakdownfield}) gives the high 
temperature behavior, while the zero temperature result of Eq.\ (\ref{b0})
remains unchanged. We fit the breakdown field data by using first an overall 
scaling factor needed to adjust $H_b\left(0\right)$ and second, a finite 
reflectivity, which only enters at high temperatures. The fit of the 
high temperature behavior provides us with an estimate of the reflectivity 
$R\approx 0.21$ of the NS interface, see Fig.\ \ref{breakfig3} (dashed line). 

\section{Discussion}

We have calculated the clean limit expression for the breakdown 
field separating the diamagnetic phase and the field penetration
phase by a first order transition. We have determined the spinodals, the 
critical temperature as well as the latent heat of the transition.
 
In Chapter \ref{magresponse} we found that the solution of the 
screening problem crucially depends on the nonlocality range. The high 
sensitivity to impurities is a consequence of the self-consistency problem 
with the Maxwell equations. In the ballistic regime $l>d$ the linear 
screening was found to be enhanced as compared to the clean limit, at low 
temperature due to the kernel range $l$ and at high temperature due to the 
range $\xi_N(T)$. Here the clean limit result for the breakdown field 
has been found to be in good agreement with the experimental data on fairly 
clean samples ($l \succeq d$). We resolve this apparent discrepancy by 
showing the the self-consistent screening problem drops out of the 
derivation of the breakdown field:
The breakdown field $H_b \sim \Phi_0/\lambda_N(T)d$ 
is obtained from matching the magnetization energy ${\cal G} \sim H_b^2d$ 
in the diamagnetic phase with the kinetic energy in the penetrating field 
${\cal G} \sim \Phi_0^2/\lambda_N^2(T)$. The magnetization energy overshadows 
the kinetic energy of the currents in the diamagnetic phase, which 
contain the solution of the screening problem. Thus the breakdown field 
is remarkably stable towards a finite impurity concentration. A correction 
to the breakdown field of order $d/l$ is expected from the kinetic term in the 
field penetration phase at temperatures $T\sim T_A$. We note that
the high temperature corrections, which are relevant beyond $T_{1/2}$ 
hardly affect the breakdown field, as $T_c$ and $T_{1/2}$ coincide.
The inclusion of a finite 
reflection at the NS interface permits an accurate fit of the breakdown 
field and gives an estimate for the quality of the NS interface. 

\chapter{Paramagnetic instability of Andreev electrons}
\label{paramagnetic}

\section{Introduction}
\markboth{CHAPTER \ref{paramagnetic}.  PARAMAGNETIC INSTABILITY OF ...}{}

A normal metal in contact with a superconductor exhibits the
phenomenon of proximity --- the superconductor exports its coherent
state across the interface into the normal metal. On a microscopic
level, this phenomenon is described through the Andreev reflection
of the normal-metal quasi-particles at the 
NS interface, converting normal- to
supercurrent. Proximity superconductivity exhibits a rich
phenomenology and has attracted considerable interest
recently \cite{curacao}.  A particularly puzzling finding is the
ultra-low-temperature reentrance observed in
normal-metal coated superconducting cylinders  \cite{mota:90}, where,
contrary to expectation, the fully diamagnetic cylinder develops a
paramagnetic response at low temperatures.  Recently, it has been
speculated that some novel kind of persistent current states circling 
the cylinder might be responsible for this phenomenon \cite{bi}, 
but closer inspection
of the experimentally measurable quantities reveals that the
predicted effect is by orders of magnitude too small \cite{fgb}. 
In this chapter, we demonstrate that the presence of a repulsive
electron-electron interaction in the normal metal naturally leads to 
the appearance of a paramagnetic instability at very low temperature, 
offering a possible explanation of the reentrance effect in the NS 
cylinders.

For simplicity, we consider a clean normal-metal slab of
thickness $d$ ($0<x<d$), in perfect contact with a bulk, conventional
superconductor. The proximity effect is mediated by the Andreev
reflection at the interface with the superconductor, which binds the
quasi-particles states to the normal layer for $E<\Delta_S$.  In the
usual free electron gas description of the normal metal, the Andreev
bound states are found at $E_n = \hbar v_x (2n+1)\pi /4d$ ($n=0, 1,
..$; $v_x=v_F\cos\vartheta$) producing a linear suppression of the
DOS \cite{james,saintjames} 
\be N(E) \sim N_0 Ed/\hbar v_F, 
\ee 
close to the Fermi level
$E=0$ ($N_0=mk_F/\hbar^2\pi^2$). In the following we assume that the
electron-electron interaction in the normal layer, which follows from
the delicate balance between the phonon-mediated- and the
Coulomb-interaction, is repulsive. As a consequence, a finite order
parameter $\Delta(x)$ is induced in the metal, opposite in sign as
compared to $\Delta_S$ in the superconductor, see Ref.\ 
 \cite{degennes:64}.  The NS junctions then behaves like a Josephson
junction with a phase difference $\pi$, trapping quasi-particle states
at the Fermi energy close to the NS interface. The local density of
states $N(E,x)$ exhibits a peak at zero energy on top of the Andreev
density of states, as shown in Fig.\ \ref{dosfig}.  This peak involves
a macroscopic number of states with density $n_p\sim k_F^2/d$, which
in the following we call the $\pi$-states.

The change in the DOS crucially affects the response of the proximity
metal.  The linear current response $j[A]$ can be divided into two
contributions $j=j_{\rm dia}+j_{\rm para}$, the diamagnetic current
$j_{\rm dia}=-(e^2n/mc)\, A$ giving the rigid response of the bulk
density $n=k_F^3/3\pi^2$ and the paramagnetic current $j_{\rm para}$
following from the deformation of the wavefunction at the Fermi
surface \cite{schrieffer}, 
\be j_{\rm para}= \frac{e^2n}{mc}\, A \int
dE \left(-{ \frac{\partial f} {\partial E}}\right) \frac{N(E)}{N_0} 
\label{paraestimate} 
\ee 
for slowly varying fields $A$ ($f$ is the Fermi occupation
number).  While in a bulk superconductor the paramagnetic current is
quenched by the energy gap at low temperatures producing a net
diamagnetic response, the paramagnetic current of a bulk normal metal
cancels the diamagnetic current exactly. In the non-interacting metal
under proximity, the linear density of states suppression $N(E)\propto
E$ is still sufficient to suppress the paramagnetic current at zero
temperature \cite{zaikin}. Including a repulsive interaction places the
system in the opposite limit: The sharp DOS peak at the Fermi level
produces a paramagnetic signal which {\it over}-compensates the
diamagnetic response. Such a paramagnetic response naturally leads to
an instability: The free energy $\delta F = -c j \delta A <0$ can be
lowered via a non-zero magnetic induction induced by spontaneous
currents along the NS interface. The interface currents are associated
with an orbital magnetization $M(T)$ producing a low-temperature
reentrance in the magnetic susceptibility.

\section{Zero energy bound states}

In the following we present a quantitative analysis of the
paramagnetic instability induced by the $\pi$-states. The magnetic
induction $B_z(x)$ parallel to the surface is described by the vector
potential $A_y(x)$ which drives the currents $j_y(x)$. The
electron-electron interaction in the superconductor is accounted for
by an effective coupling constant $V_S<0$ and similarly $V_N>0$ in the
normal metal, see also \cite{zhouspivak,nazarov}. Two
self-consistency problems have to be solved: First, we evaluate the
order parameter $\Delta(x)$ accounting for the different coupling
constants in the superconductor and the normal metal, and obtain the
local DOS $N(E,x)$.  Second, we determine the current functional
$j[A]$ which we solve together
with Maxwell's equation to find the spontaneous interface currents.

\begin{figure}[tb]
\psfig{figure=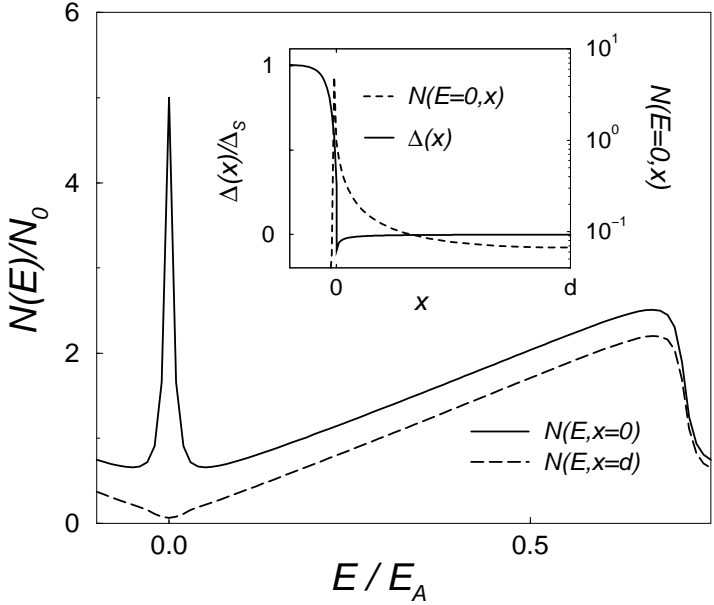,width=11cm,angle=-90}
\vspace{0.3cm}
\caption{Local DOS $N(E,x)$ at the NS interface $x=0$ and 
at the metal boundary $x=d$ ($E_A=\hbar v_F/d$), 
as it follows from the self-consistent
solution of the Eilenberger equation, Eqs.\ 
(\ref{eilenberger}) and (\ref{gapeq}), for a thickness  
$d=10 \hbar v_F/\Delta_S$ and the coupling constants $V_S=-0.3$ and 
$V_N=0.1$. Inset: Spatial dependence of the order parameter $\Delta(x)$ 
and local DOS $N(E=0,x)$ at the peak energy (S: $x<0$, N: $x>0$).}
\label{dosfig}
\end{figure} 

We use the quasi-classical description following from the Eilenberger
equation, see (\ref{eq:eilenberger}), 
\be -v_x \partial_x \hat{g} =\left[
  \{\omega_n +ie v_y A_y(x)\} \hat{\tau}_3 + \Delta(x)
  \hat{\tau}_1, \hat{g} \right],
\label{eilenberger}
\ee
where the 2$\times$2 matrix $\hat{g}$ have been defined in Chp.\ 
\ref{magintro}. Eq.\ (\ref{eilenberger}) is completed
by the self-consistency relation for the pair potential 
($\langle .. \rangle$ is the angular average),
\be \Delta(x) = - V N_0 \pi T \sum_{\omega_n>0} \langle f_{\omega_n}(x,v_x) 
\rangle.
\label{gapeq}
\ee
The self-consistent numerical solution
of Eqs.\  (\ref{eilenberger}) and (\ref{gapeq}) is shown in the inset of
Fig.\ \ref{dosfig}. 
The course of the order parameter in the normal layer is asymptotically 
given by $\Delta(x) \sim -V_N N_0 \hbar v_F/x$, as expected from the 
$f$-function in the non-interacting case $V_N=0$ \cite{falk}. 
$\Delta(x)$ decays from a value $\sim -|V_N/V_S| \Delta_S$ at 
the NS interface, to $\sim  -V_N N_0 \hbar v_F/d$ at the outer
boundary. Close to the NS interface, the local DOS
\be N(E,x)=N_0 {\rm Re} [\langle g_{-iE+\delta}(x,v_x)\rangle ] ,
\ee 
exhibits a pronounced peak at zero energy, as shown in Fig.\ \ref{dosfig}.
At the outer metal boundary $x=d$ the DOS is suppressed linearly, as in
the noninteracting case \cite{james}. Such a pseudogap is typical in 
proximity induced superconductivity \cite{hara,frahm}. We note that a 
peak structure as a consequence of interactions could be related to one 
found for a Luttinger liquid in proximity with a superconductor 
\cite{faziohekking}.

In order to proceed with analytical results, we approximate the order 
parameter by a step function,
\[ \Delta(x) = \left\{ \begin{array}{ll} \Delta_S, & x<0, \\ 
-\Delta_N, \quad\quad& 0<x<d, 
\end{array} \right. 
\]
where $\Delta_N \propto V_N$ enters as a parameter. The Green's function 
in the normal layer $x>0$ can be determined exactly and takes the form
\be g_{\omega_n} (x,v_x) = 
\frac{\omega_n \sinh \left[ \chi(d)-\gamma\right] + 
\Delta_N \cosh \left[ \chi(d-x)\right] }
{\Omega_n \cosh \left[\chi(d)-\gamma\right]},
\label{green_finite_d}
\ee
where $\chi(x) = 2 \Omega_n x / v_x$, $\Omega_n^2 = \Delta_N^2 + 
\omega_n^2$, and $\tanh \gamma=\Delta_N/\Omega_n$ 
(we consider the limit $\Delta_S \gg \Delta_N, T $).
The second term in (\ref{green_finite_d}) describes the $\pi$-states 
at the NS interface.  
The poles of the Green's function at $\omega_n \to -iE+0$ yield the 
bound state energies. For $E> \Delta_N$ 
the bound states given by
\be \sqrt{E^2-\Delta_N^2} = \frac{\hbar v_x}{2d} \left(n\pi 
   +\arccos\frac{E}{\Delta_N}\right), \quad\quad  (n=0,1,...),
\ee
down-shifted by $\delta E_n \approx - 2\Delta_N/(2n+1)\pi$ 
with respect to the Andreev states of the free electron gas.
Below the gap $E<\Delta_N$ we find the $\pi$-states at 
\be E = \Delta_N / \cosh \frac{2\sqrt{\Delta_N^2-E^2}d}{v_x} 
\sim \Delta_N {\rm e}^{-2\Delta_N d/ v_x},  
\label{energy_zbs} 
\ee
exponentially close to Fermi energy. All trajectories with 
$v_x=v_F\cos\vartheta \ll \Delta_N d/$ possess a bound state 
at $E\approx 0$, thus producing the macroscopic weight of the zero 
energy DOS peak: For $\Delta_N > v_F/d$ the number of $\pi$-states 
per unit surface $N_{\rm surf}$ is equal to the number of transverse 
levels $N_{\rm surf} \sim k_F^2$, while for $\Delta_N < v_F/d$ 
it is reduced to $N_{\rm surf} \sim k_F^2 (\Delta_N d/v_F)^2$ 
via the reduction of the available solid angle $\cos\vartheta < 
\Delta_N d/v_F$.

\section{Spontaneous currents and magnetization}

We derive the current-field relations at low temperatures, assuming 
$T \ll v_F/d, \Delta_N$. This implies a thermal length 
$\xi_N(T) = \hbar v_F/2\pi T$ larger than the thickness $d$ and no 
thermal smearing on the scale $\Delta_N$. Only the trajectories with
$\cos\vartheta < \Delta_N d/ v_F$ contribute to the current 
at low temperatures. We describe them in the limit 
$v_x/\Delta_N d \to 0$ by
\be g_{\omega_n}(x,v_x) = \frac{\omega_n}{\Omega_n} + \frac{\omega_n \Delta_N}
{\Omega_n\left(\Omega_n-\Delta_N\right)} {\rm e}^{-\chi(x)}. 
\label{green_infinite_d}
\ee
The current in the presence of a slowly varying vector potential $A$, follows 
from Eq.\ (\ref{green_infinite_d}) after replacing $\omega_n$ by 
$\omega_n + iev_y A$ and inserting it into the quasi-classical current 
expression, see (\ref{eq:current}). In addition to the diamagnetic 
current 
\be j_{\rm dia}  = -(c/4\pi \lambda_N^2)\, A,
\label{jtdia}
\ee 
[$\lambda_N = (mc^2/4\pi n e^2)^{-1/2}$ denotes the London length], we
obtain the paramagnetic current 
\be
j_{\rm para} \approx  \frac{c}{4\pi \lambda_N^2} {\rm e}^{-x/\alpha\xi_N^0}
    \alpha \frac{3 \Phi_0 }{2\pi \xi_N^0}\arctan \frac{e v_F A}{\pi T}, 
\label{jtpara} 
\ee
in the 
limit $ev_F A/c, T \ll \Delta_N$. Here, $\Phi_0=\pi\hbar c/e$ denotes the 
flux quantum, and the coherence length in the normal metal  
\be 
\xi_N^0 = \hbar v_F/2\Delta_N,
\ee
gives the extent of the $\pi$-states. Under the assumption $\Delta_N> v_F/d$ 
we set $\alpha=1$. At temperatures $T \gg ev_F A/c$ the paramagnetic current 
$j_{\rm para}\sim (1/\lambda_N^2) (\Delta_N/T)  A$ is 
linear in $A$ and $\propto 1/T$, a signature of the thermally smeared zero 
energy DOS peak, and competes with the diamagnetic current on the scale 
$\xi_N^0$. 
At $T\to 0$, Eq.\ (\ref{jtpara}) is nonlinear in the field and generates the 
spontaneous paramagnetic current. This paramagnetic interface current results 
from the energy splitting of the $\pi$-states in the field, 
$E\approx \pm e v_F A$, allowing the system to gain energy by shifting the 
DOS below the Fermi surface \footnote{A similar mechanism producing 
spontaneous currents is discussed by Honerkamp {\it et al.} \cite{honerkamp}}.
For $\Delta_N < \hbar v_F/d$, the paramagnetic current is reduced 
by the factor $\alpha=(\Delta_N d/\hbar v_F)<1$ in Eq.\ (\ref{jtpara}).
The surface current $I=\int j dx \sim \alpha^2 c\Phi_0/\lambda_N^2$ is in 
agreement with the current estimate $I_{\pi} \sim N_{\rm surf} e v_F$ based on 
the number of $\pi$-states at zero energy. Eq.\ (\ref{jtpara}) thus 
always produces a net paramagnetic response at low temperature and 
fields.

\begin{figure}[tb]
\psfig{figure=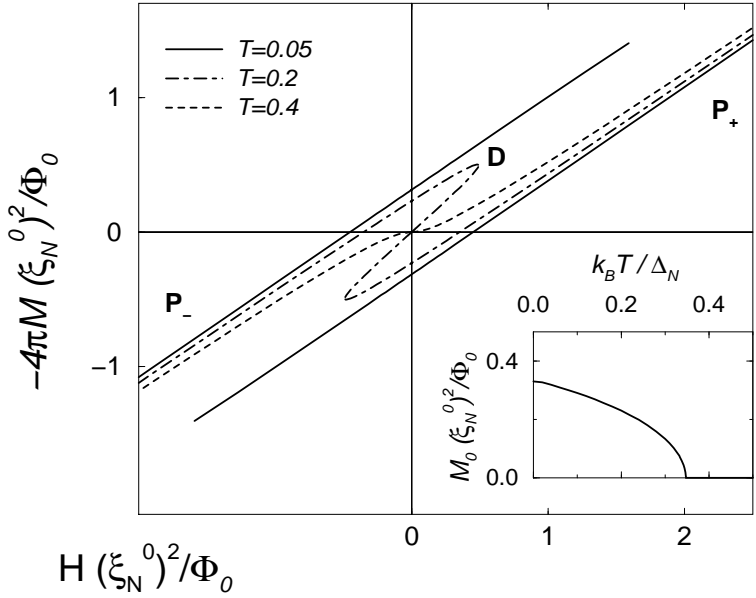,width=11cm,angle=-90}
\vspace{0.5cm}
\caption{Magnetization $M(T,H)$ curve at various temperatures (in units of 
$\Delta_N$), for $\lambda_N=0.3d$ and $\xi_N^0=d$.
The two meta-stable branches $P_{\pm}$ exhibit a spontaneous magnetization  
in zero field, the diamagnetic branch $D$ is unstable. 
Inset: Zero field magnetization $M_0(T)$.}
\label{magfig}
\end{figure} 

The evaluation of the induced magnetization requires the self-consistent 
solution of Maxwell's equation $-\partial_x^2 A(x)= 4\pi j(x)/c$ together 
with the current functional $j[A(x)]$. The solution of the screening problem 
requires the full dispersive relation between $j(q)$ and $A(q)$. According 
to the study of Chp.\ \ref{magresponse}, the coherence length in the normal 
metal $\xi_0$ gives the nonlocality range, as $\Delta_N$ is the off-diagonal 
self-energy in (\ref{eq:eilenberger}). Eq.\ (\ref{jtpara}) represents the 
long wavelength limit $q\to 0$. For simplicity, here we use 
Eq.\ (\ref{jtpara}) under the assumption of local response. Including 
the nonlocality in the constitutive relations does not alter the 
spontaneous interface currents qualitatively. 

The magnetization curves $M(T,H)$, which follow from Eq.\ 
(\ref{jtpara}) are shown in Fig.\ \ref{magfig}. Approaching from large
fields, Fig.\ \ref{magfig} shows two paramagnetic branches $P_{\pm}$
with a linear diamagnetic slope exhibiting a spontaneous magnetization
in zero field. They result from the superposition of the paramagnetic
magnetization $M_0(T)$ and the Meissner response to the applied field
$H$.  As the field is decreased (increased) past $H=0$, the branch
$P_+$ ($P_-$) becomes meta-stable. The spontaneous magnetization
$M_0(T)$ appears below a second order transition point $T_c^M$ and
saturates at low temperatures, as shown in the inset of Fig.\ 
\ref{magfig}.  The magnetization curve includes a diamagnetic branch
$D$, which arises from the competition between the paramagnetic
instability and the thermal smearing and is thermodynamically
unstable.

In the following we give a semi-quantitative analysis of the
magnetization $M = \int dx \,M(x)/d$, first at zero temperature and
field [$M_0$], proceeding to finite temperatures [$M_0(T)$], and
finally including an applied magnetic field $H$ [$M(T,H)$]. The
boundary conditions are given by $A(0)=0$ and $\partial_x A(d)=H$. We
concentrate on the most relevant limit where $\xi_N^0, d \gg \lambda_N$.
At $T=0$, according to Eq.\ (\ref{jtpara}), the paramagnetic interface
current $j\sim \alpha \Phi_0/\lambda_N^2\xi_N^0$ remains unscreened
until being matched by $j_{\rm dia}\sim -A/\lambda_N^2$, producing a
vector potential $A\sim \alpha \Phi_0/\xi_N^0$ on the scale $\lambda_N$.
The vector potential $A$ saturates beyond $\lambda_N$, as the para- and
diamagnetic currents cancel each other. Assuming that the $\pi$-states
extend up to the outer metal surface ($\alpha<1$), the induced
magnetization $M = A(d)/4\pi d$ is given by 
\be M_0 \sim \alpha
\frac{\Phi_0}{\xi_N^0 d} \sim \frac{\Phi_0}{(\xi_N^0)^2}.
\label{m0}
\ee
We note that although the spontaneous currents increase as
$\Delta_N > \hbar v_F/d$ ($\alpha=1$) they are screened exponentially 
beyond the extent of the $\pi$-states in this limit, giving a magnetization 
$M_0 \sim (\Phi_0/\xi_N^0d) \exp[-(d-\xi_N^0)/\lambda_N]$. 
We assume $\alpha < 1$ in the following.

At finite temperature, the spontaneous magnetization is suppressed by
the factor $\arctan (e v_F A / \pi  T)$, which itself depends
on the magnetization via $A \sim M d$, implying the implicit equation
\be \frac{M_0(T)}{M_0} \sim \arctan \frac{M_0(T) \alpha \Delta_N}
{M_0  T}.
\label{condM}
\ee 
The spontaneous magnetization appears below a second order
transition at $T_c^M \sim \alpha \Delta_N$,
saturating at low temperatures, as shown in the inset of Fig.\ 
\ref{magfig}. The transition temperature is equal in magnitude to the
energy splitting of the DOS peak $E \sim e v_F A/c \sim \alpha
\Delta_N$.
 
Under an applied magnetic field $H$, the Meissner current $j_{dia}$  
screens both the spontaneous interface current and the applied field.
At zero temperature we deal with a linear problem and the magnetization 
is given by the superposition $M(H) = M_0 + \chi H$ 
of the spontaneous magnetic moment $M_0$ and the Meissner response 
$\chi H$. As the temperature increases, $M_0(T)$ decreases and 
the meta-stable regime shrinks. At $T>T_c^M$ the spontaneous magnetization 
in zero field has disappeared, the signature of the paramagnetic currents 
remains, however, reducing the diamagnetic susceptibility $\chi$ at small
fields. At large temperature $T\!\gg\! T_c^M$ we recover the pure Meissner 
response.

Note that the two meta-stable branches $P_+$ and $P_-$ in the
magnetization curve, see Fig.\ \ref{magfig}, imply a first order
transition with changing field at $H=0$. The first order transition is
similar to the magnetic breakdown occurring in the same system at large
fields between the fully diamagnetic phase and a field penetration
phase \cite{alfbreak}.  The rotation of the magnetic moments to the
energetically more favorable polarization will show the hysteretic
behavior typical for a first order transition. The transition from
$P_+$ to $P_-$ implies a paramagnetic slope in the thermodynamic
$dc$-magnetization $\langle M(T,H) \rangle$, which will link the
meta-stable solutions $P_{\pm}$ in Fig.\ \ref{magfig} and cross the origin at
$M(H=0)=0$. In summary, we find that on approaching $T_c^M$ from above, the 
diamagnetic susceptibility $\chi_{dc}=\langle M(T,H)\rangle/H$ is reduced, 
exhibiting a low-temperature reentrance. Below $T_c^M$, the 
spontaneous interface currents produce a net paramagnetic susceptibility 
$\chi_{dc}$.

\section{Discussion}

In the following we discuss our results in the context of the
experiments by Mota and co-workers, who have measured the magnetic
response of normal-metal coated superconducting cylinders at low
temperatures \cite{mota:90,mota:89}. In the previous chapters 
\ref{magresponse} and \ref{breakdown} we have established a
quantitative understanding of the magnetic response of these samples
at higher temperatures, studying the screening at small impurity 
concentration 
and the magnetic breakdown at finite field. The samples are typically 
characterized by a mean free path $l\sim d$ and an interface transparency 
of order unity. The Nb-Ag and Nb-Cu cylinders show an anomalous paramagnetic 
signal in the magnetic
response in the low-temperature -- low-field corner of the $H-T$ phase
diagram \cite{mota:90,mota:94}. A direct comparison with our theory
requires the magnetization curve $\langle M(T,H) \rangle$ which has
not yet been measured. The observed $dc$-susceptibility $\chi_{dc}(T)$
as a function of temperature shows an increase at low
temperature \cite{frassanito}. The measured $ac$-susceptibilities
$\chi_{ac}(T)$ and $\chi_{ac}(H)$ exhibit a reentrance both as
a function of temperature and field \cite{mota:90}. The reentrance is
accompanied by an out-of-phase response signaling dissipation and by
hysteresis in the field dependence. These features are in qualitative
agreement with our results for the magnetization curve.  We find that 
theory and experiment agree in order of magnitude for $\alpha^2 \sim
0.1$, implying a transition temperature $T_c^M \sim 100 {\rm mK}$ and a
spontaneous magnetization $M_0 \sim 1 {\rm G}$.
A more quantitative comparison with experiment requires a
self-consistent treatment of the spontaneous currents with the pair
potential, accounting for the nonlocality of the current-field
relation and its sensitivity to disorder. 

In conclusion, we have demonstrated that the inclusion of a finite
electron-electron repulsion in a proximity coupled normal-metal layer
naturally produces spontaneous interface currents leading to a
paramagnetic reentrance in the magnetic response: The sign change in
the coupling across the NS interface leads to the trapping of
$\pi$-states at the Fermi energy.  The frustrated NS junction relaxes
through the generation of spontaneous interface currents, inducing a
paramagnetic moment. The spontaneous magnetization implies a first order 
transition in zero field in the low-temperature sector of the $H-T$ phase 
diagram. A non-trivial issue remains the requirement that
the electron-electron interaction be repulsive at the low energy
scales involved.  Interesting consequences of this assumption have
been discussed in the context of the proximity effect \cite{degennes:64}
and most recently in relation to the low temperature transport in
mesoscopic NS structures \cite{petrashov,nazarov}.  In fact, the noble
metal coatings used in the experiments of Mota and co-workers
\cite{mota:90} appear to be the most plausible candidates for a
repulsive electron-electron interaction. Turning the argument around,
in the light of our findings the experimental observation of a
paramagnetic reentrance can be taken as an indication of the presence
of a repulsive interaction in these materials.

\chapter{Nonlocality in Josephson junctions: Anomalous current -- flux 
periodicity}
\label{nonlocalsns}

\section{Introduction}
\markboth{CHAPTER \ref{nonlocalsns}.  NONLOCALITY IN JOSEPHSON ...}{}

The equilibrium transport in Josephson junctions is caused by the quantum 
interference of the two overlapping reservoir wavefunctions 
\cite{bib:Josephson}. The supercurrent is driven by the nonlocal phase 
difference between the superconducting leads. This is of particular 
interest in superconductor--normal-metal--superconductor (SNS) junctions, 
where the thickness of the interlayer can be much larger than the 
superconducting coherence length $\xi_0$, without suppressing the Josephson 
current. The transport relies on the coherence of the quasi-particle 
population between the superconductors, on the mesoscopic scale of the
thermal and phase coherence lengths of the normal metal \cite{likharev}.

In early studies on SNS junctions \cite{kulik} the supercurrents were 
already expressed in terms of the quasi-particles which are bound to 
the interlayer due to the Andreev reflection. 
These bound states consist of a forward propagating electron and a back 
propagating hole and carry the current of the double electron charge $2e$, 
which is converted from an to supercurrent at the NS interfaces. 
Recently, there has been a renewed interest in those states as they 
can be brought out of equilibrium in a controlled fashion 
\cite{otbk,averin:95,gorelik,argaman}. Due to the advances in 
nanofabrication technology, which have achieved a good coupling of a 
two-dimensional electron gas (2DEG) to superconducting reservoirs 
\cite{taka1,taka2}, ballistic properties of these quasi-particles like the
quantization of supercurrent \cite{houten} have been observed. 

In the present chapter we determine the critical-current--flux relation 
in a ballistic SNS junction. The current-carrying 
quasi-particles traversing the weak link are sensitive to the Aharonov-Bohm 
phase, producing a nonlocal dependence of the current density on the 
magnetic induction in the junction and the superconducting phase difference. 
As a consequence, the nonlocality and the finite size junction 
produce an anomalous doubling of the critical-current--flux periodicity 
$I_c(\Phi)$. 

\begin{figure}[tb]
  \centerline{
  \psfig{file=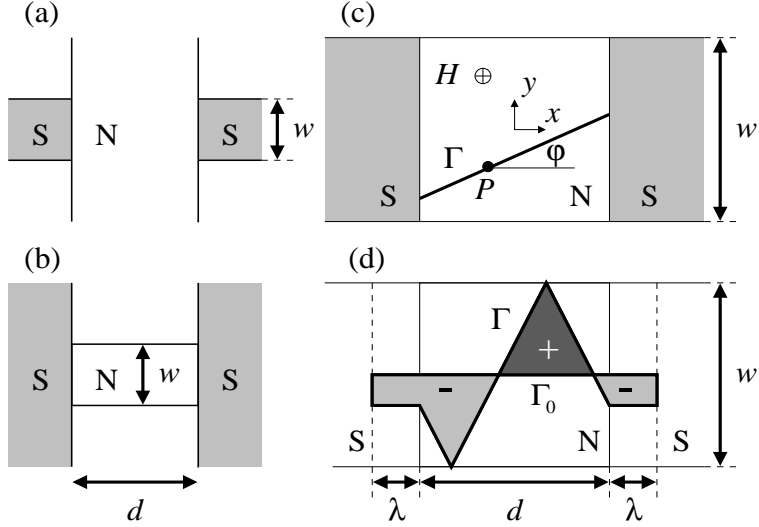,width=10cm,height=7cm}}
  \vspace{4mm} \caption{(a) Junction with a point-contact (open) geometry as
  discussed in [10], where $w$ is the width of the two
  \emph{super}conductors. (b) The junction studied here has a strip geometry
  with $w$ the width of the \emph{normal} conductor. (c) The magnetic field
  $H$ is applied in the $z$-direction and the coordinate system is chosen
  symmetric with respect to the junction center. The current density in the
  point $P$ involves contributions from all trajectories $\Gamma$
  parameterized by the angle $\varphi$. (d) The phase difference $\gamma$
  along the trajectory $\Gamma$ can be expressed through the enclosed flux
  $\phi$; using the trajectory $\Gamma_{0}$ as our reference, the flux 
  through the areas above (below) $\Gamma_{0}$ contributes with a positive 
  (negative) sign.}  \label{f:SNSJunction}
\end{figure}

The motivation for our work is drawn from an 
experiment on S-2DEG-S junctions \cite{bib:Heida} of width $w$ comparable 
to length $d$, where the a $2\Phi_0$ periodicity was found instead of the 
usual $\Phi_0$. A first attempt to explain this finding is due to 
\cite{bib:Zagoskin}; considering the point-contact geometry of Fig.~1(a) with
open boundary condition in the metal, they indeed recover a $2\Phi_{0}$
periodicity for a geometric ratio $w/d\rightarrow 0$. 
However, the experiment in \cite{bib:Heida} is carried out in the strip 
geometry of Fig.~1(b) and involves dimensions $w\sim d$ of the same order. 
Here we determine the critical current $I_{c}$ as a function of flux, 
taking proper account of the reflecting boundaries in the normal-metal 
characteristic for the strip geometry of Fig.~\ref{f:SNSJunction}(b).
We find that the periodicity of the critical current changes from $\Phi_{0}$
to $2\Phi_{0}$ as the flux through the junction increases, i.e., as a 
function of field. At low temperatures
the crossover to the $2\Phi_{0}$ periodic current appears at a flux
$\sim\Phi_{0} w/d$, thus explaining the result of Heida {\it et al.}
\cite{bib:Heida}, who found a $2\Phi_{0}$ periodic pattern for all fields in
devices with $w/d\sim 1$. The results of this chapter have been accepted  
for publication \cite{lfb}.

\section{Current density pattern}

In our study, we neglect the screening of the applied fields by the 
supercurrent in the normal-metal layer, which is justified for a
small critical current $j_c$. The scale over which the fields can be 
screened is given by the Josephson length 
\be
\lambda_J \sim \sqrt{c\Phi_0/j_c},
\ee
where $\tilde{d}=d+2\lambda_S$ is the effective length of the junction, extended
by the penetration depth $\lambda_S$ in the superconductors. We assume that 
the Josephson vortex distance \cite{tinkham}, 
\be
a_0 = \Phi_0/\tilde{d},
\ee
is the smaller length scale, $a_0 \ll \lambda_J$. Expressing the critical 
current by the screening density $n\propto 1/\lambda_N(T)^2$ introduced in 
Chp.\ \ref{magresponse}, $j_c \sim c\Phi_0/\lambda_N^2(T)d$, we find that 
the field regime of interest lies above the 'critical field' of the normal 
interlayer, $H\gg \Phi_0/\lambda_N(T)d$. 
We consider the SNS-junction sketched in Fig.~\ref{f:SNSJunction}. In the
quasi-classical Green's function technique, the current density 
(\ref{eq:current}) in a point 
$P$ results from the contributions over all quasi-particle {\it trajectories} 
connecting the two NS interfaces through $P$. In a junction of infinite 
width, the trajectories involve no reflection at the boundaries. 
In the case of a finite junction, boundary conditions at the 
normal-metal--vacuum boundary have to be applied, which we idealize through 
the assumption of specular reflections. 
Furthermore, we adopt the usual approximations: perfect Andreev
reflections at the SN-interfaces and a coherence length $\xi_{0}$ in the two
superconductors with $\xi_{0}\ll d$, allowing for a step-like approximation 
of the order parameter~$\Delta$ \cite{kulik}. The
quasi-classical Green function is calculated by matching the partial 
solutions in N and S at the interfaces. For the current density 
$\mathbf{j}$, we arrive at a generalization of the results given by 
Svidzinskii and co-workers \cite{bib:Svidz}. 
For finite temperatures with $d\gg\xi_{\scriptscriptstyle{N}}$, the current 
$\mathbf{j}$ takes the form,
\begin{eqnarray}
  \frac{{\bf j}(x,y)}{j_{c,0}}=-\frac{6}{\pi} \sqrt{ \frac{2}{\pi}}
  \int_{-\pi/2}^{\pi/2}\!\!\!d\varphi\,\hat{\mathbf{p}}\,
  \sin(\gamma)
  \frac{d}{\sqrt{\xi_{\scriptscriptstyle{N}} l(\varphi)}}
  \exp\left(-\frac{l(\varphi)}{\xi_{\scriptscriptstyle{N}}}\right),
  \label{eq:CurrentDensHighTemp}
\end{eqnarray}
while in the low temperature limit, $d\ll\xi_{\scriptscriptstyle{N}}$,
\begin{eqnarray}
  \frac{{\bf j}(x,y)}{j_{c,0}}=\frac{4}{\pi^{2}}\sum_{k=1}^{\infty}
  \frac{(-1)^{k}}{k}\!\int_{-\pi/2}^{\pi/2}\!\!\!
  d\varphi\;\hat{\mathbf{p}}\sin(k\gamma)\frac{d}{l(\varphi)},
  \label{eq:CurrentDensLowTempST}
\end{eqnarray}
where $\hat{\mathbf{p}}=(\cos(\varphi),\sin(\varphi),0)$,
$l(\varphi)=d/\cos(\varphi)$ is the length of a trajectory with slope
$\varphi$, and the zero temperature critical current density is
\begin{equation}
  j_{c,0}=\frac{ne^{2}}{mc}\frac{\Phi_{0}}{2d}.  \label{eq:cd}
\end{equation}
In (\ref{eq:cd}), $n$ denotes the electron density in the normal metal, 
and we have assumed $T\ll \Delta$ as usual.
While in the low temperature limit all harmonics $\sin(k\gamma)$
($k=1,2,\ldots$) contribute to the current density \cite{bib:Ishii}, at 
finite temperatures, the thermal smearing of the Andreev levels leads to a 
suppression of
the higher harmonics $\propto\exp(-kd/\xi_{\scriptscriptstyle{N}})$ and only
the first term $\propto\sin(\gamma)$ survives. An individual trajectory
contributes with a weight $\propto\exp(-l/\xi_{\scriptscriptstyle{N}})$ at
finite- and $\propto d/l$ in the low temperature limit. In a wide junction,
$\gamma$ takes the form\footnote{
In the discussion of the point-contact geometry, Barzykin and Zagoskin
make use of the above results, but limit the integration
in (\ref{eq:CurrentDensHighTemp}) and (\ref{eq:CurrentDensLowTempST}) to 
those trajectories connecting the superconducting contacts \cite{bib:Zagoskin}.
} \cite{bib:Svidz},
\begin{equation}
  \gamma(x,y;\varphi)=\gamma_{0}-\frac{2\pi}{\Phi_{0}}H\tilde{d}
  \left[y-x\tan(\varphi)\right].  \label{eq:GammaInf}
\end{equation}
In the more general result derived here, $\gamma$ is given by the gauge 
invariant phase difference
\begin{equation}
  \gamma(x,y;\varphi)=\Delta\varphi-\frac{2\pi}{\Phi_{0}} \int_{\Gamma}{\bf
  A}\cdot d{\bf s}, \label{eq:GaugeInvPhd}
\end{equation}
where $\Delta\varphi$ denotes the phase difference between the two
superconductors and $\Gamma$ is the path which goes through the point $(x,y)$
with slope $\varphi$. Combining the current expressions
(\ref{eq:CurrentDensHighTemp}) or (\ref{eq:CurrentDensLowTempST}) and
(\ref{eq:GaugeInvPhd}) with the Maxwell equation $\nabla^{2}{\bf A}=-4\pi{\bf
j}[{\bf A},\Delta\varphi]/c$, we obtain the transverse vector-potential ${\bf
A}$ allowing to solve the full screening problem. The solution of the screening 
problem has only been carried out for a tunnel junction so far \cite{bib:Scalapino}.

In the following, we neglect screening and concentrate on junctions with the
strip geometry of Fig.~\ref{f:SNSJunction}(b), including the (reflecting)
trajectories $\Gamma$ in (\ref{eq:CurrentDensHighTemp}) and
(\ref{eq:CurrentDensLowTempST}). We express the gauge invariant phase
difference (\ref{eq:GaugeInvPhd}) in terms of the flux $\phi$ enclosed by
$\Gamma$ and the reference path $\Gamma_{0}$ and obtain,
\begin{equation}
  \gamma(x,y;\varphi)=\gamma_{0}-\frac{2\pi\phi(x,y;\varphi)}{\Phi_{0}},
\end{equation}
where for negligible screening $\phi(x,y;\varphi)=H S(x,y;\varphi)$
and $S$ is the properly weighted enclosed area, see
Fig.~\ref{f:SNSJunction}(d). Note the analogy of this result to the 
Aharonov-Bohm phase picked up by an Andreev loop in a SN structure, see
Fig.\ \ref{breakfig2}. The surface $S$ is calculated as a
function of the number of reflections the trajectory $\Gamma$ undergoes (in
the following called the `order' of the trajectory). The point-contact
geometry of Fig.~\ref{f:SNSJunction}(a) then is described by the order-zero
trajectories alone \cite{bib:Zagoskin}, while in the strip geometry of
Fig.~\ref{f:SNSJunction}(b), higher orders have to be included.

\begin{figure}[tb]
  \centerline{\psfig{file=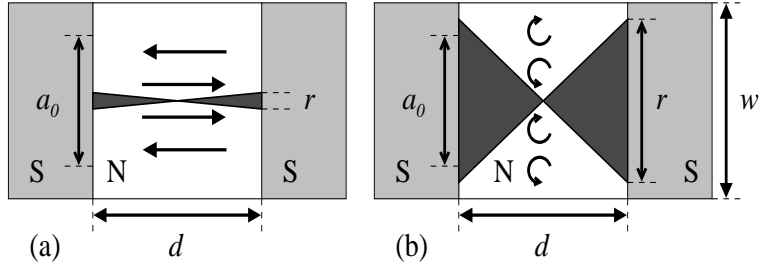,width=10cm,height=3.5cm}}
  \vspace{4mm} \caption{The `bow ties' of width $r$ display the ensemble of
  trajectories contributing to the current density in a given point. The
  arrows indicate the current flow. (a) For weak nonlocality,
  $r<a_{\scriptscriptstyle{0}}$, the current flows straight through the
  junction. (b) The strong nonlocality, $r>a_{\scriptscriptstyle{0}}$, leads
  to the formation of vortex-like domains of circular flow.}
  \label{f:nlrange}
\end{figure}
\begin{figure}[tb]
\psfig{file=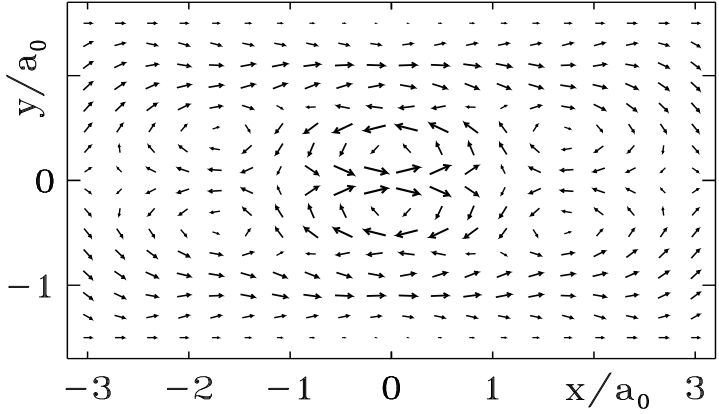,width=10.5cm,height=6cm}
    \caption{The current density for $d/\xi_{\scriptscriptstyle{N}}=3$,
    $r/a_{\scriptscriptstyle{0}}\approx 3.5$, and $\Phi/\Phi_{0}=3$. Apart
    from the vortex-like domains in the middle produced by order-zero
    trajectories, additional circular flow is set up by the order-one
    trajectories.}  \label{f:cdnl}
\end{figure}

The geometrical pattern in the current density $\mathbf{j}$ depends strongly
on the sample dimensions $d$ and $w$, the normal metal coherence length
$\xi_{\scriptscriptstyle{N}}$, and the applied field $H$. At finite
temperature, the current density in $P$ draws its weight from trajectories
with $\varphi<\sqrt{\xi_{\scriptscriptstyle{N}}/d}$, allowing us to introduce
the transverse nonlocality range $r=\sqrt{\xi_{\scriptscriptstyle{N}} d}$ (in
the low temperature limit, $\varphi\sim 1$ and we define $r=d$). This range 
of nonlocality has to be compared to the scale
$a_{\scriptscriptstyle{0}}=\Phi_{0}/H\tilde{d}$ of transverse variations in
$\mathbf{j}$ (see Fig.~\ref{f:nlrange}): For \emph{weak} nonlocality,
$r<a_{\scriptscriptstyle{0}}$, the flow is uniform along $x$ with amplitude
$j_{c}$ and changes direction on a distance $a_{\scriptscriptstyle{0}}/2$
along the $y$-axis. This contrasts with the \emph{strongly} nonlocal case
$r>a_{\scriptscriptstyle{0}}$ which is found with increasing field, 
where the current density forms domains of left-and right-going circular flow. 
While the local case is similar to that in a
tunnel junction, the pattern in the \emph{non}local situation reminds of the
usual vortex structure in a superconductor, see
Fig. \ref{f:cdnl}. For finite temperatures with
$d\gg\xi_{\scriptscriptstyle{N}}$, the current density of the order-zero
trajectories is given by
\begin{eqnarray}
  \frac{j_{x}(x,y)}{j_{c,0}}&=&-\frac{12}{\pi}\sin\left(\gamma_{0}-\frac{2\pi
  y}{a_{\scriptscriptstyle{0}}}\right) \exp\left[-\alpha(x)\right],
  \nonumber\\ \frac{j_{y}(x,y)}{j_{c,0}}&=&-\frac{12}{\pi}
  \cos\left(\gamma_{0}-\frac{2\pi
  y}{a_{\scriptscriptstyle{0}}}\right)
  \frac{\xi_{\scriptscriptstyle{N}}}{d}\frac{2\pi
  x}{a_{\scriptscriptstyle{0}}}\exp\left[-\alpha(x)\right],
\end{eqnarray}
where
\begin{equation}
  \alpha(x)= \sqrt{\left(\frac{d}{\xi_{\scriptscriptstyle{N}}}\right)^{2}
  +\left(\frac{2\pi x}{a_{\scriptscriptstyle{0}}}\right)^{2}}.
\end{equation}
For weak nonlocality, $a_{\scriptscriptstyle{0}}<r$, the exponent remains
approximately constant in the normal part, $\alpha(x)\approx\alpha(0)$,
leading to a uniform current flow, while for strong nonlocality,
$a_{\scriptscriptstyle{0}}>r$, $\alpha(x)$ grows as $x$ approaches the
interfaces, $\alpha(\pm d/2)\gg\alpha(0)$, such that the current concentrates
in the middle of the junction. For $a_{\scriptscriptstyle{0}}>r$, the higher
order trajectories lead to a refinement of the current pattern, see
Fig. \ref{f:cdnl}. Similar results are obtained in the low temperature limit, 
see \cite{ledermann}.

\section{Critical current -- flux dependence}

\begin{figure}[tb]
  \psfig{file=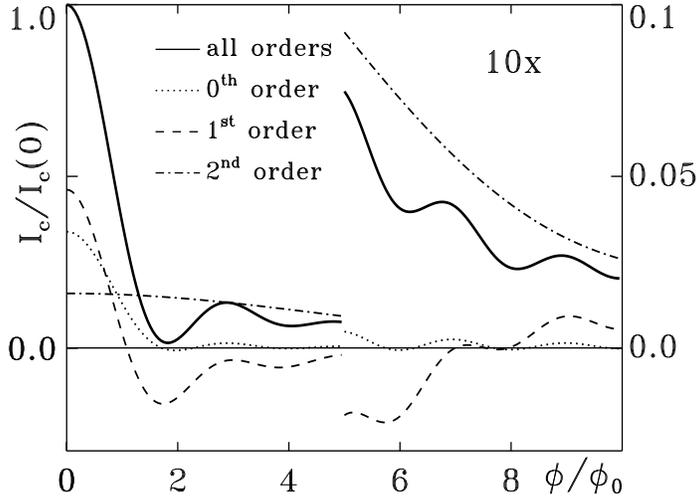,width=9cm,height=7.2cm}
  \caption{The critical current for $d/\xi_{\scriptscriptstyle{N}}=5$ and
  $w/d=1/3$. The solid curve shows the full critical current and the dashed
  curves are the contributions from the orders 0, 1, and 2. The orders 0 
  and 1
  oscillate with periodicity $2\Phi_{0}$, while the second order decreases
  monotonically, remaining always positive. The current pattern produced by
  the orders 0 and 1 is lifted by the order 2 contributions, and the critical
  current attains the periodicity $2\Phi_{0}$.}  \label{f:CritCurrentLift}
\end{figure}

The ratio $r/a_{\scriptscriptstyle{0}}$ and its associated characteristic
current pattern manifest themselves in the (pseudo)-periodicity of the
critical current,
\begin{equation}
  I_{c}(\Phi)=\max_{\gamma_{0}}\int_{-w/2}^{w/2}dy\;
  j_{x}(0,y;\gamma_{0},\Phi),
\end{equation}
versus flux $\Phi=H\tilde{d}w$ in the junction. In the case of weak
nonlocality, $r<a_{\scriptscriptstyle{0}}$, the relevant contribution to the
critical current comes from the order-zero trajectories resulting in a
$\Phi_{0}$ periodicity. For strong nonlocality, $r>a_{\scriptscriptstyle{0}}$,
higher orders are relevant and \emph{lift} the order-zero result as shown in
Fig.~\ref{f:CritCurrentLift} --- the periodicity of the critical current
changes to $2\Phi_{0}$. For a critical current $j_c < \Phi_0/cd$ 
the cross-over $a_0\sim r$ lies within the negligible screening regime. 
To be specific, we discuss in detail the orders 0, 1,
and 2 for the case of finite temperatures with
$d\gg\xi_{\scriptscriptstyle{N}}$ (the qualitative arguments for
$d\ll\xi_{\scriptscriptstyle{N}}$ are similar).

For $w>r>a_{\scriptscriptstyle{0}}$, the critical current due to the
order-zero trajectories takes the form
\begin{equation}
  I_{c}^{(0)}(\Phi)=-\frac{\sqrt{2}I_{c,T}}{\sqrt{\pi}}\frac{w}{r}
  \frac{\cos(\pi\Phi/\Phi_{0})}{(\pi\Phi/\Phi_{0})^{2}},
\label{icofphi}
\end{equation}
where $I_{c,T}=w (12/\pi) j_{c,0}\exp[-d/\xi_N(T)]$. 
For the first-order trajectories, we numerically
find a $2\Phi_{0}$ (pseudo-) periodic contribution as well. Both components
vanish with field $\propto 1/\Phi^2$. The second- and all following 
even-order
trajectories exhibit a large current amplitude of order $j_{c}$ on a scale
$a_{\scriptscriptstyle{0}}\propto 1/\Phi$ in the junction center $(0,0)$, a
consequence of the $\varphi$-independence of the gauge invariant phase
difference $\gamma$ along trajectories through $(0,0)$. Their contribution
scales $\propto 1/\Phi$ and therefore dominates over the zeroth- and
first-order terms at large enough fields --- as the strongly nonlocal limit
with $a_{\scriptscriptstyle{0}}<r$ is reached, the periodicity changes 
to $2\Phi_{0}$.  Samples with a small width $w<r$ are always in the strongly
nonlocal limit and their current pattern is $2\Phi_{0}$ periodic throughout
the entire field axis. At low temperatures, the condition $w<r$ transforms
into the geometric requirement $w < d$. 
The same arguments apply for the $r>w>a_0$, where the zero-order current 
is given by \cite{bib:Heida}
\begin{equation}
  I_{c}^{(0)}(\Phi)=-\frac{I_{c,T}}{2\sqrt{\pi}}\frac{w}{r}
  \left( \frac{\sin(\pi\Phi/2\Phi_{0})}{\pi\Phi/2\Phi_{0}}\right)^{2}.
\label{jcofphizago}
\end{equation}
While Eq.\ (\ref{jcofphizago}) trivially implies a $2\Phi_0$ periodicity when 
taking the absolute value, more generally Eq.\ (\ref{icofphi}) only produces 
a $2\Phi_0$ periodicity when accounting for the lifting by the higher order 
contributions. 
We give a complete classification of the regimes of 
simple or double flux periodicity, depending on the nonlocality range 
$r$, and the Josephson vortex distance $a_0$ in Table~\ref{t:ptable}.
\begin{figure}[tb]
  \psfig{file=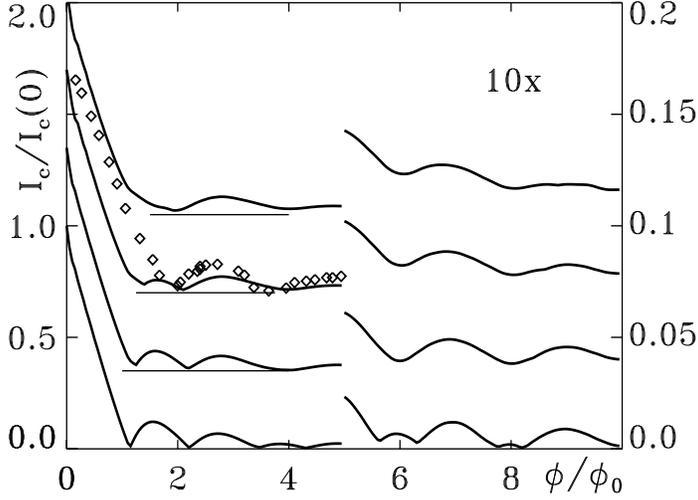,width=9cm,height=7.2cm}
  \caption{The critical current in the low temperature limit,
  $d\ll\xi_{\scriptscriptstyle{N}}$, for $w/d=0.8$, 0.9, 1.1, 1.5 (from top 
  to
  bottom). Successive plots have been offset by 0.35. Measured data ([9],
  diamonds) is shown for the case $w/d=0.9$.}  
\label{f:MaxScLT}
\end{figure}
\begin{table}[b]
\caption{The periodicity of the critical current is controlled by the 
 parameters $w/r$, and $r/a_{\scriptscriptstyle{0}}$. The table has to
 be read as a flow chart, starting at the top row and selecting the proper
 condition proceeding down the rows. The nonlocality range is 
 given by $r=\sqrt{\xi_N(T)d}$ for $\xi_N(T)\ll d$ and $r= d$ 
 for $\xi_N(T)\gg d$.
}  
\begin{center}\begin{tabular}{|c|c|c|c|}\hline\hline
    ratio & \multicolumn{3}{c|}{value}\\\hline
    $w/r$     & \multicolumn{2}{c|}{$>1$} & $<1$  \\
    $r/a_0$  & $<1$ & $>1$ &  \\\hline
    period    & $\Phi_0$ & \multicolumn{2}{c|}{$2\Phi_0$} 
\\\hline\hline
\end{tabular}\end{center}
\label{t:ptable}
\end{table}

\section{Discussion}

We have demonstrated that the current density and the critical
current in a clean SNS-junction with strip geometry depend on the
ratio $r/a_{\scriptscriptstyle{0}}$ between the nonlocality range $r$ and the
vortex distance $a_{\scriptscriptstyle{0}}$. The period of the critical
current changes with increasing field from a $\Phi_0$ periodicity 
for $a_0>r$ to a $2\Phi_0$ pseudo-periodicity for strong nonlocality $a_0<r$.

Recently, Heida {\it et al.} \cite{bib:Heida} observed such a $2\Phi_{0}$
periodicity in strip like ($w\sim d$) S-2DEG-S junctions made from Nb
electrodes in contact with InAs operating at low temperatures $T = 0.1$ K.
As the total flux through the junction is difficult to
determine in the experiment, Heida {\it et al.} had to infer their 
$2\Phi_{0}$
periodic structure from a fit on four samples with different ratios $w/d$
ranging from 0.9 to 2.2. In Fig.~\ref{f:MaxScLT}, we present the results of
our numerical calculations for the strip geometry, where we have properly 
taken into account the finite penetration depth of the flux into the 
superconducting banks. While geometries with $w/d<1$ clearly exhibit a 
$2\Phi_{0}$ periodicity
throughout the entire field region, a $\Phi_{0}$-component starts to develop
at low fields in wide junctions. The comparison with the data of Heida 
{\it et al.} ($w/d = 0.9$) gives a satisfactory description of the 
pseudo-periodic
structure. 

For further experimental studies on wider junctions, we predict a crossover 
from a $\Phi_{0}$- to a $2\Phi_{0}$-periodicity with increasing fields.
The cross-over is accompanied by the detachment of the current pattern 
form the superconducting leads, which form a vortex structure shown in
Fig.\ \ref{f:cdnl}.

\chapter{Quiet SDS Josephson junctions for quantum computing}
\label{quietsds}

\section{Introduction}
\markboth{CHAPTER \ref{quietsds}.  QUIET SDS JOSEPHSON JUNCTIONS ...}{}

Quantum computers take advantage of the inherent parallelism of the
quantum state propagation, allowing them to outperform classical
computers in a qualitative manner. Although the concept of quantum
computation has been introduced quite a while ago \cite{Feynman},
wide spread interest has developed only recently when specific
algorithms exploiting the character of coherent state propagation 
have been proposed \cite{Ekert}. Here we deal with the
device aspect of quantum computers, which is flourishing in the wake of
the recent successes achieved on the algorithmic side.  Two
conflicting difficulties have to be faced by all hardware
implementations of quantum computation: while the computer
must be scalable and controllable, the device should be almost 
completely detached from the environment during operation 
in order to maximize phase coherence. The most advanced 
propositions are based on trapped ions \cite{Cirac-Zoller,Monroe}, 
photons in cavities \cite{Turchette}, NMR spectroscopy of
molecules \cite{Gershenfeld}, and various solid state implementations
based on electrons trapped in quantum dots \cite{Loss,burkard}, the Coulomb 
blockade in superconducting junction arrays \cite{schoenqc1,schoenqc2,Averin}, 
or the flux dynamics in Superconducting Quantum Interference Devices 
(SQUIDs) \cite{Bosco}. Nano\-structured solid state quantum gates 
offer the attractive feature of large scale integrability, once 
the limitations due to decoherence can be overcome \cite{Haroche}.

In this chapter we propose a new device concept for a (quantum) logic gate
exploiting the unusual symmetry properties of unconventional
superconductors. The basic idea is sketched in Fig.\ 1: 
connecting the positive (100) and negative (010) lobes of a 
$d$-wave superconductor with a $s$-wave material produces 
the famous $\pi$-loop with a current carrying ground state
characteristic of $d$-wave symmetry \cite{vanHarlingen,sigristrice}.
We make use of an alternative geometry and match the $s$-wave
superconductors (S) to the (110) boundaries of the $d$-wave (D)
material. As a consequence, the usual Josephson coupling
$\propto (1 - \cos \phi)$ vanishes due to symmetry reasons and we arrive
at a bistable device, where the leading term in the coupling takes 
the form $E_d \cos 2\phi$ with minima at $\phi = \pm \pi/2$ 
($\phi$ denotes the gauge invariant phase drop across the junction). 
In our design we need the minima at the positions $\phi = 0,\pi$ --- 
the necessary shift is achieved by going over to an asymmetric SDS' 
junction with a large DS' coupling, see Fig.\ 1. The static DS' 
junction shifts the minima of the active SD junction by the desired 
amount $\phi = \pm \pi/2$. A similar double-periodic junction has 
recently been realized by combining two $d$-wave superconductors 
oriented at a $45^\circ$ angle \cite{Ilichev}.
\begin{figure} [bt]
\centerline{\psfig{file=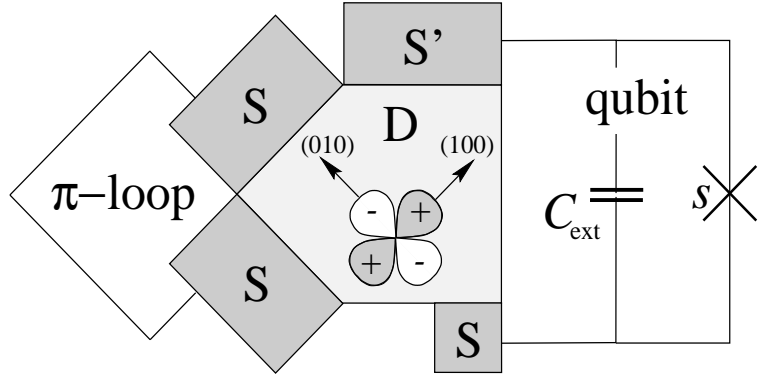,width=10cm,height=5cm}}
\vspace{4mm}
\caption{Geometrical arrangements between $s$-wave and $d$-wave
superconductors producing a $\pi$-loop (used in the phase sensitive
experiment by Wollman {\it et al.} [1]) and a qubit, the basic
building block of a quantum computer.}
\end{figure}
The ground states of our SDS' junction are degenerate and carry 
no current, while still being distinguishable from one another: 
e.g., connecting the junction to a large inductance loop, 
the $\pi$ state is easily identified through the induced current. 
It is this double-periodicity and the associated degeneracy in 
the ground state of the SDS' junction which we want to exploit 
here for quantum computation: combining the SDS' junction, 
a capacitor, and a conventional $s$-wave junction into a SDS' 
SQUID loop, we construct a bistable element which satisfies 
all the requirements for a qubit, the basic building block of 
a quantum computer. This work has been submitted for publication \cite{sds}.

\section{Device functionality}

In the following we give a detailed account of the operational features 
of our device. Consider a small-inductance ($L$) SQUID loop with $I_{\rm
\scriptscriptstyle J}L\ll\Phi_0$, where $I_{\rm\scriptscriptstyle J}$ 
denotes the (Josephson) critical current of the loop and $\Phi_0 = 
hc/2e$ is the quantum unit of flux. Such a loop cannot trap magnetic flux 
($\Phi = 0$) and the gauge invariant phase differences $\phi_1$
and $\phi_2$ across the two junctions are slaved to each other,
as the uniqueness of the wave function requires that $\phi_1 - 
\phi_2 = 2\pi \Phi/\Phi_0$. Combining a SDS' 
junction with a coupling energy $E_d$ and a conventional 
$s$-wave junction (coupling $E_s$) into a SDS' SQUID loop, 
we obtain a potential energy
\begin{equation}
V (\phi) =  E_d (1-\cos 2 \phi) + E_s (1-\cos \phi),
\label{Pot-Energy}
\end{equation}
exhibiting two minima at $\phi = 0, \pi$, see Fig.\ 2. The switch 
$s$ allows us to manipulate their energy separation, choosing 
between minima which are either degenerate or separated by $2E_s$.

In the quantum case, the phase fluctuates as a consequence of the 
particle--phase duality \cite{tinkham}. The phase fluctuations are
driven by the electrostatic energy required to move a Cooper pair
across the junction and are described by the kinetic energy
\be
T(\dot\phi) = (\hbar/2e)^2 C {\dot\phi}^2/2,
\ee
where $C$ denotes the 
loop capacitance. The dynamics of $\phi$ is manipulated by inserting 
a large switchable (switch $c$) capacitance $C_{\rm ext}$ into 
the loop acting in parallel with the capacitances $C_d$ and $C_s$ 
of the $d$- and $s$-wave junctions. Note that the Lagrangian 
$L = T - V$ of our loop is formally equivalent to that of a 
particle with `mass' $m \propto C$ moving in the potential $V(\phi)$. 

With the switch settings $c$ {\it on} and $s$ {\it off}, see Fig.\ 2(a),
the loop capacitance is large and the junction exhibits a doubly 
degenerate ground state which we characterize via the phase coordinate 
$\phi$, $|0\rangle$ and $|\pi\rangle$.
Closing the switch $s$, see Fig.\ 2(b), the degeneracy is lifted
and while $|0\rangle$ becomes the new ground state, the
$|\pi\rangle$-state is shifted upwards by the energy $2E_s$
of the $s$-wave junction, the latter being frustrated when
$\phi = \pi$. On the other hand, opening the switch $c$,
see Fig.\ 2(c), completely isolates the $d$-wave junction
and leads to the new ground and excited states $|\pm\rangle =
[|0\rangle \pm |\pi\rangle]/\sqrt{2}$ separated by the tunneling
gap $2\Delta_d$. The latter relates to the barrier $2E_d$ and
the capacitance $C_d$ of the $d$-wave junction via \cite{tinkham}
\be 
\Delta_d \propto E_d \exp(-2\sqrt{C_d E_d/e^2}).
\ee
Closing the switch $c$, the capacitance is increased by $C_{\rm ext}$ and 
the tunneling gap is exponentially suppressed. Using the above 
three settings, we can perform all the necessary single qubit 
operations:

{\it Idle-state:} The switch settings $c$-{\it on} and
$s$-{\it off} define the qubit's idle-state. While the large
capacitance $C_{\rm ext}$ inhibits tunneling, the degeneracy of
$|0\rangle$ and $|\pi\rangle$ guarantees a parallel time evolution
of the two states. This idle-state is superior to other
designs, where the two states of the qubit have {\it different}
energies and one has to keep track of the relative phase accumulated
between the basis states.

{\it Phase shifter:} Closing the switch $s$ separates the energies
of the basis states $|0\rangle$ and $|\pi\rangle$ by an amount
$2E_s$. Using a spinor notation for the two-level system, the relative
time evolution of the two states is described by the Hamiltonian
${\cal H}_s = -E_s \sigma_z$, with $\sigma_z$ a Pauli matrix.
Keeping the switch $s$ {\it on} during the time $t$, the time
evolution of the two states is given by the unitary rotation
$u_z(\varphi) = \exp (-i\sigma_z \varphi/2)$ with $\varphi =
-2 E_s t/\hbar$.
\begin{figure}[!tb]
\centerline{\psfig{file=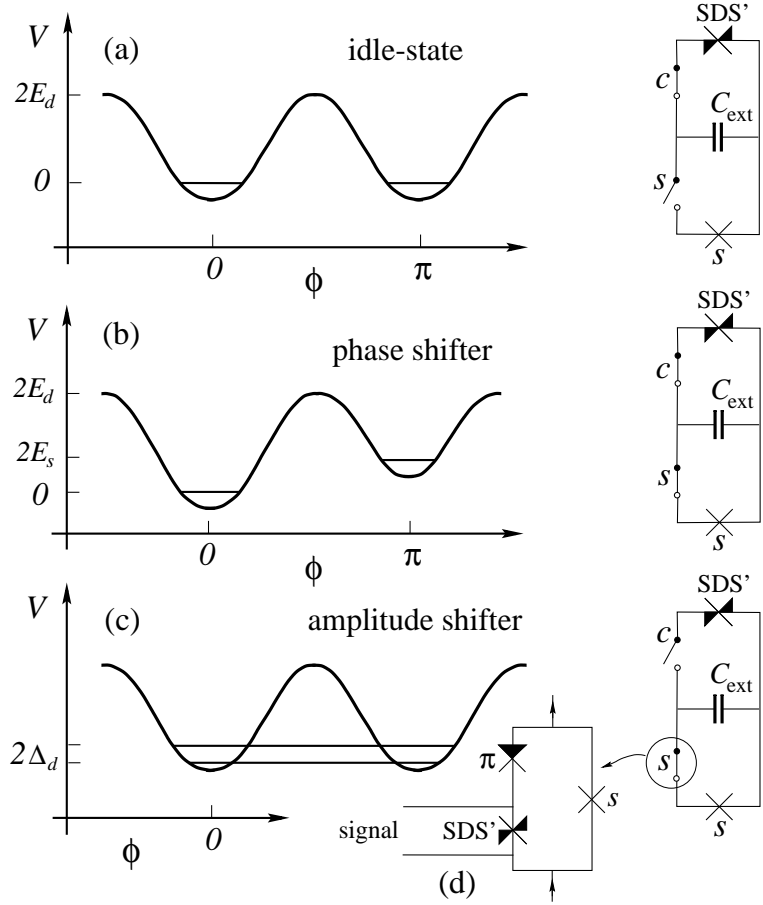,width=10cm,height=12cm}}
\vspace{4mm}
\caption{ Energy--phase diagrams for the SDS' SQUID loop.
(a) Idle-state:
The switches are set to $c$-{\it on} and $s$-{\it off} --- the
relative dynamics is quenched, leaving the state unchanged. 
(b) Phase-shifter:
With the switch settings $c$-{\it on} and $s$-{\it on} the 
relative phase between $|0\rangle$ and $|\pi\rangle$ increases
linearly with time.
(c) Amplitude-shifter:
The switch setting $c$-{\it off} isolates the $d$-wave junction. 
An initial state $|0\rangle$ oscillates back and forth between 
$|0\rangle$ and $|\pi\rangle$, allowing for a shift of amplitude.
(d) A SDS' junction, a $\pi$ junction, and a $s$-wave junction 
combined into a SQUID loop and serving as a switch.}
\end{figure}

{\it Amplitude shifter:} Assume we have prepared the loop in 
the ground state $|0\rangle$ and wish to produce a superposition 
by shifting some weight to the $|\pi\rangle$ state. Opening the 
switch $c$ in the loop, see Fig.\ 2(c), the time evolution
generated by the Hamiltonian ${\cal H}_d = \Delta_d \sigma_x$
of the open loop induces the rotation $u_x(\vartheta) = \exp 
(-i\sigma_x \vartheta/2)$ with $\vartheta = 2 \Delta_d t /\hbar$.
The system then oscillates back and forth between $|0\rangle$ 
and $|\pi\rangle$ with frequency $\omega = \Delta_d/\hbar$ 
and keeping the switch $c$  open for an appropriate time interval 
$t$ we obtain the desired shift in amplitude (note that the 
qubit remains isolated from the environment during these Rabi 
oscillations). 

Imposing the condition $E_d \gg E_s,~\Delta_d$ 
on the coupling energies, we make sure that the two states 
$|0\rangle$ and $|\pi\rangle$ are well defined while 
simultaneously involving only the low energy states 
$|0\rangle$ and $|\pi\rangle$ of the system. Furthermore,
all times involved should be smaller than the decoherence 
time $\tau_{\rm dec}$, requiring $E_s,~\Delta_d \gg \hbar/
\tau_{\rm dec}$.

The present setup differs significantly from the conventional 
(large inductance) SQUID loop design, where the low-lying states 
are distinguished via the different amount of trapped flux 
and their manipulation involves external magnetic fields $H$
or biasing currents $I$ cite{blatter:qtunneling}. 
SQUID loops of this type are being 
used in the design of classical Josephson junction computers 
\cite{Likharev} and have been proposed for the realization of
quantum computers, too, see \cite{Bosco}. However, this setup suffers
from the generic problem that the flux moving between the loops leads
to a magnetic field mediated long-ranged interaction between the
individual loops and further produces an unwanted coupling to the
environment. By contrast, our device remains decoupled from the 
environment, the operating states do not involve currents, 
and switching between states can be triggered with a minimal 
contact to the external world --- we therefore call our qubit 
implementation a `{\it quiet}' one.

Next, we discuss how to perform two-qubit operations within an array
of SDS' SQUID loops. A two-qubit state is a coherent superposition 
of single qubit states and can be expressed in the basis
$\{|xy\rangle\}$, where $x,y \in \{0,\pi\}$ denote the phases 
on the $d$-wave junctions of the first ($x$) and second ($y$) qubit, 
respectively. Unitary operations acting on these states are 
represented as $4\times 4$ unitary matrices. Single-qubit operations
$u$ acting on the second qubit take the block-matrix form
\begin{eqnarray}
{\cal U}_2 =
\left(\begin{array}{cc}\!
u & 0
\! \\ \!
0 & u
\! \end{array}\right),
\label{single_2}
\end{eqnarray}
and a similar block form selecting odd and even rows and columns 
defines the single-qubit operations on the first qubit. As all
logic operations on two qubits can be constructed from combinations
of single-qubit operations and the Controlled-NOT gate \cite{Ekert}
it is sufficient to define the operational realization of the latter.
The Controlled NOT gate performs the following action on two qubits: 
with the first (controller) qubit in state $|x\rangle$ and the
second (target qubit) in state $|y\rangle$ the operation shall
leave the target qubit unchanged if $x = 0$, while flipping it
between $0$ and $\pi$ when $x = \pi$, in matrix notation 
\begin{eqnarray}
{\cal U}_{\rm\scriptscriptstyle CNOT} =
\left(\begin{array}{cc}\!
1  & 0
\! \\ \!
0 & \sigma_x 
\! \end{array}\right).
\label{CNOT}
\end{eqnarray}
The above Controlled NOT operation can easily be constructed
from the two-qubit `phase shifter': Connecting two individual qubits in 
their idle-state over a $s$-wave junction into a SQUID loop, 
the states $|00\rangle$ and $|\pi\pi\rangle$ become separated 
from the states $|0\pi\rangle$ and $|\pi 0\rangle$ by the 
energy $2E_{s_b}$ of the $s$-wave junction. Keeping the two 
qubits connected during the time $t$ introduces a phase shift 
$\chi = -2E_{s_b}t/\hbar$ between the two pairs of states,
\begin{eqnarray}
{\cal U}_{\rm ps} (\chi) =
\left(\begin{array}{cc}\!
u_z(\chi) & 0 
\! \\ \!
0 & u_z(-\chi) 
\! \end{array}\right).
\label{phase_shifter}
\end{eqnarray}
The Controlled NOT gate (\ref{CNOT}) then can be constructed from the 
phase-shifter (\ref{phase_shifter}) via the following sequence of
single- and two-qubit operations (see \cite{Loss} for a similar 
realization of the CNOT gate),
\begin{equation}
{\cal U}_{\rm\scriptscriptstyle CNOT} = \exp(-i\pi/4)
{\cal U}_{2y}(\pi/2){\cal U}_{1z}(-\pi/2){\cal U}_{2z}(-\pi/2)
{\cal U}_{\rm ps}(\pi/2){\cal U}_{2y}(-\pi/2),
\label{reconstruct_CNOT}
\end{equation}
where the single qubit operations ${\cal U}_{i\mu} (\theta)$ rotate 
the qubit $i$ by an angle $\theta$ around the axis 
$\mu$ ($u_\mu(\theta) = \exp(-i\sigma_\mu\theta/2)$ acting on $i$) while 
leaving the other qubit unchanged. 

A key element in our design are the switches and a valid suggestion
is the single electron transistor discussed in the literature \cite{Joyez}. 
Here we propose a quiet switch design optimally adapted to our 
SDS' qubits. The underlying idea is to insert a SQUID loop which allows 
to either stiffen or relax the phase, thus producing a phase switch. 
Combining a SDS' junction with energy $E_d$, a $\pi$-junction 
with $E_\pi \ll E_d$, and a $s$-wave junction with $E_s = E_\pi$ 
into a (small inductance) SQUID loop, see Fig.\ 2(d), 
we obtain the following switching behavior: 
The phase $\phi = 0$ on the SDS' junction frustrates the 
remaining junctions, the loop's energy-phase relation 
is a constant, $E_{\rm sw} (\phi_\pi = \phi_s-\pi) \equiv 0$, and 
the switch is open. A voltage pulse coming down the signal lines and 
switching the SDS' junction into the $|\pi\rangle$ state changes 
the phase relation between the $\pi$- and the $s$-wave
junctions and closes the switch: the energy $E_{\rm sw} 
(\phi_\pi = \phi_s) = 2 E_\pi (1-\cos\phi_\pi)$ implies the
current-phase relation $I = (2e/\hbar) \partial_{\phi_\pi} 
E_{\rm sw}$ that stiffens the phase. The appropriate voltage pulses 
can be generated by driving an external SDS' SQUID loop unstable.

\section{Discussion}

The quiet device concept proposed above heavily relies on the 
double periodicity of the SD junction. As the second harmonic is
strongly suppressed in a SID tunnel junction, a more feasible 
suggestion for the realization of a $\cos 2\phi$ junction is 
the SND sandwich, where the superconductors are separated by 
a thin metallic layer N. For a clean metallic layer, the coupling 
energies for the $n$-th harmonic are large and of order 
$E_{\rm\scriptscriptstyle J} \sim k_{\rm\scriptscriptstyle F}^2 
{\cal A}\, \hbar v_{\rm\scriptscriptstyle F}/d$, producing the 
well known saw-tooth shape in the current-phase relation 
\cite{bib:Ishii} ($v_{\rm\scriptscriptstyle F}$ denotes the 
Fermi velocity in the N layer while $d$ and ${\cal A}$ are its 
width and area). In reality, it seems difficult to deposit 
a {\it clean} metallic film on top of a $d$-wave superconductor 
and we have to account for the reduction in the coupling 
$E_{\rm\scriptscriptstyle J}$ due the finite scattering length $l$ 
in the metal layer. Using quasi-classical techniques to describe 
a dirty SN$_{\rm\scriptscriptstyle D}$D junction, we obtain a second
harmonic coupling energy $E_d \sim k_{\rm\scriptscriptstyle F}^2 
{\cal A} \, (\hbar v_{\rm\scriptscriptstyle F}/d) (l/d)^3 \sim 
(R_{\rm\scriptscriptstyle Q}/R)(l/d)E_{\rm\scriptscriptstyle T}$, 
where $l$ denotes the scattering length in the normal metal, 
$R_{\rm\scriptscriptstyle Q} = \hbar/e^2$ is the quantum resistance,
and $E_{\rm\scriptscriptstyle T} \sim (\hbar 
v_{\rm \scriptscriptstyle F}/d) (l/d)$ is the Thouless energy.

The second important device parameter is the tunneling gap $\Delta_d$,
which depends quite sensitively on the coupling to the environment.
The usual reduction in the tunneling probability produced by the
environment \cite{CaldeiraLeggett} is modified if the system is 
effectively gapped at low energies \cite{AmbegaokarEckernSchon}.  
This is the case for our SNDN'S' junction where the low-energy
quasi-particle excitations in the metal are gapped over the 
Thouless energy $E_{\rm\scriptscriptstyle T}$ \cite{Golubov}.
The dynamics of the junction is only affected by the 
presence of virtual processes involving energies larger 
than $E_{\rm\scriptscriptstyle T}$, leading to a renormalized 
capacitance $C_{\rm ren}\sim\hbar/R E_{\rm\scriptscriptstyle T}$ 
(cf.\ \cite{schoenzaikin}) and resulting in the reduced
tunneling gap
\be
\Delta_d \propto E_d \exp[-\nu (R_{\rm\scriptscriptstyle Q}/R) \sqrt{l/d}], 
\ee
with $\nu$ of order unity. Consistency requires that the tunneling 
process is `massive' and hence slow, $\hbar/\tau < E_{\rm
\scriptscriptstyle T}$.  With a tunneling time $\tau \sim S/E_d$
($S\sim \hbar (R_{\rm\scriptscriptstyle Q}/R) \sqrt{l/d} =$ 
tunneling action) we find that the constraint $\hbar/\tau
E_{\rm\scriptscriptstyle T} \sim \sqrt{l/d} < 1$ is satisfied. 
The condition $\Delta_d \ll E_d$ requires the tunneling gap
$\Delta_d$ to be small, but large enough in order to allow for
reasonable switching times, requiring $(R_{\rm\scriptscriptstyle Q}/R)
\sqrt{l/d}$ to be of order 10. With typical device dimensions $d \sim
1000$ \AA, $l \sim 10$ \AA, and $R/R_{\rm\scriptscriptstyle Q} \sim
(d/l)(1/{\cal A}k_{\rm\scriptscriptstyle F}^2) \sim 1/100$, this
condition can be realized. Finally, the operating temperature 
$T$ is limited by the constraint $S/\hbar > E_d/T$, guaranteeing 
that our device operates in the quantum regime, and the requirement 
$T < E_{\rm\scriptscriptstyle T}$ that thermal quasi-particle 
excitations be absent. The first condition takes the form 
$T\ll \hbar/\tau \sim \sqrt{l/d} \, E_{\rm\scriptscriptstyle T}$ 
and is the more stringent one. Using the above parameters 
and a typical value $v_{\rm\scriptscriptstyle F} \sim 10^8$ cm/s,
we obtain a Thouless energy $E_{\rm \scriptscriptstyle T} \sim 1$ K 
and hence $T \ll 0.1$ K.

In conclusion, we have discussed a novel device concept for logic
gates in superconducting computers. The SDS' SQUID loop realizes a
number of attractive features which are potentially relevant both in
classical Josephson computers based on RSFQ logics as well as in
superconducting quantum computers. The most obvious advantage over
previous designs is the quietness of the device: The SDS' SQUID loop is
a naturally bistable device and does not involve external bias
currents or magnetic fields. Second, the basic states of the loop 
do not involve currents or trapped flux, hence long-range 
interactions between various elements of the computer are eliminated.  
Third, the qubits do not accumulate phase differences during idle 
time. And fourth, all operations can be carried out via simple 
switching processes.

\chapter{Conclusions and Outlook}
\label{conclusion}

In this thesis, we have discovered the variety of facets of mesoscopic
superconductivity induced by proximity, ranging from the non-equilibrium 
transport to the thermodynamic screening properties. A tremendous progress 
in nanofabrication technology and the perfection of low-temperature 
cryostats in the last two decades have made the small scales
of the Andreev or Thouless energies accessible to experiment. 
Proximity induced superconductivity combines the macroscopic phase of
the superconductor to the quantum coherence of the microscopic electron 
wavefunctions, allowing for the observation of single-particle properties 
on a macroscopic scale. The term 'mesoscopic' coins the importance of the 
phase coherence of the involved quasi-particles preserving quantum 
interference effects.

The study of mesoscopic superconductivity has repeatedly lead us to the 
microscopic process of the Andreev reflection (AR) governing the quasi-particle 
exchange across the normal-metal--superconductor contact. The AR pins the 
single-particle physics to the Fermi level, combining electrons above and 
holes below symmetrically. The non-trivial AR phase-shift $\pi/2$ translates 
into the suppression of the quasi-particle spectrum at the Fermi energy, 
the resonance structure in transport and the diamagnetic response in magnetism. 
These phenomena are understood in the idealized free electron gas description 
of the proximity metal, in good quantitative agreement with experiments. 

In this concluding chapter, we give a perspective on open questions and future 
work on Andreev physics. On one hand we discuss some specific problems 
that can be addressed based on the framework provided in this thesis and are 
of direct relevance to experiments. On the other hand, we focus on some questions 
which appear of long term interest, in connection to our work and the 
present developments in the field of mesoscopic superconductivity.

(i) The scattering matrix approach used in chapter \ref{nonlinearity} gives 
a direct theoretical access to the finite voltage shot noise in NS junctions. 
The numerical realization of disorder average based on the shot noise 
expression would allow for a quantitative comparison with shot noise experiments 
on diffusive NS junctions presently in progress \cite{jehl,kozhevnikov}.

(ii) In chapters \ref{magresponse} and \ref{nonlocalsns} we have found several 
signatures of non-locality in the proximity effect, which have just started to 
be traced experimentally \cite{bib:Heida}. We predicted a finite field cross-over 
from a $\Phi_0$ to a $2\Phi_0$ current--flux periodicity in ballistic SNS 
junctions, and the detachment of the current pattern as we reach the 
non-local regime. Future experiments on more extended ballistic junctions will 
require the self-consistent solution of the screening problem, 
where the over-screening and the sensitivity to impurity or boundary scattering 
crystallizes in the critical-current--flux relation.
Non-locality is also an issue in High-$T_c$ superconductors, where the quasi-
particle gap vanishes due to the d-wave symmetry of the order parameter, 
see \cite{kogan:96}.

(iii) The experimental feasibility of ballistic SNS weak links provides an 
access to non-equilibrium phenomena: The superconducting reservoirs act as a 
confining potential on the Andreev quasi-particle population which can be 
driven out of equilibrium in a controlled fashion \cite{elke}. This allows 
for the observation of supercurrents at large temperatures \cite{argaman} 
or the 
manipulation of the Josephson relation by a repopulation of the Andreev levels 
\cite{vanweesnature}. The study of phase relaxation processes in this system 
are within experimental reach.

(iv) The rapidly developing area of quantum computing relies on the 
existence of phase coherent devices as it exploits the time evolution of 
a quantum two-level system (qubit). The Josephson junctions offer themselves 
as a scalable, solid state implementation of a qubit, by making use of the 
quantization of the charge \cite{schoenqc1} or of the phase in mesoscopic weak 
links, 
see chapter \ref{quietsds}. Before speculating on the efficient use of 
quantum algorithms, the near future challenge lies in the experimental 
realization of single- and two-qubit operations and in the theoretical 
investigation of the phase coherence problem, as it arises from the coupling with 
the environment. 

(v) In chapters \ref{breakdown} and \ref{paramagnetic} we have addressed the 
nonlinear magnetic response in proximity structures, revealing a rich 
behavior in the low-temperature low-field corner of the $H-T$ phase diagram. 
The breakdown of the diamagnetic state studied in Chp.\ \ref{breakdown} 
was found to agree quantitatively with experimental data \cite{mota:89}. 
The polarization of the spontaneous moments discussed in Chp.\ \ref{paramagnetic} 
(see also below) imply a similar phase boundary at zero field. We predict a 
latent heat as a signature of both first order transitions. \\

Much of the underlying motivation for the detailed study of the orbital 
magnetism in this thesis was provided by the paramagnetic reentrance observed 
in the susceptibility of normal-metal coated superconducting cylinders 
\cite{mota:90}. While the studies within the free electron gas approximation 
of Chps.\ \ref{magresponse} and \ref{breakdown} 
could not account for the low-temperature anomaly, they allowed a good
characterization of the typical mean free path and the quality of the 
NS interface, see also \cite{mmb}. The effect of the cylindrical topology 
had already been previously considered in \cite{belzigdipl} using the 
quasi-classical technique, as well as more recently in \cite{bi}, failing to
produce either a non-monotonic temperature dependence or the order of 
magnitude observed in experiment. This prompted us to consider the possibility 
of a repulsive electron-electron interaction in the normal metal in 
Chp.\ \ref{paramagnetic}. We found that a repulsive coupling constant 
in the normal metal layer frustrates the NS contact by a phase difference 
$\pi$ across it. A density of states peak is accumulated at the Fermi energy 
due to the $\pi$-states trapped at the NS interface. This peak induces an 
paramagnetic instability towards spontaneous interface currents which 
naturally give rise to magnetic moments. The experimentally observed signatures 
being the paramagnetic reentrance in the susceptibility, a hysteretic behavior, 
dissipative response and creep qualitatively support our interpretation in 
terms of spontaneous magnetic moments and indicate the presence of repulsive 
electron-electron interactions in these material (Ag, Cu). In order to test 
our ideas, the following points merit consideration. \\

(vi) The local spectroscopy of the NS contact by means of a (scanning) tunneling 
microscope should reveal a (split) density of states peak at the chemical 
potential. 

(vii) The experimental measurement of the $dc$-magnetization curve would provide 
a direct evidence for our findings. For a quantitative comparison, the numerical 
solution of the Eilenberger equation is required, accounting for 
the self-consistency of the pair potential, the non-locality of the 
current--field relations that enter the Maxwell equation, and the presence 
of impurity scattering. 

(ix) Recent theoretical work also addresses the low-temperature corrections 
to the proximity effect due to interactions \cite{zhouspivak,nazarov}. First 
experiments on ferromagnet--superconductor junctions show non-trivial 
behavior \cite{courtoisferro} challenging our understanding within the free 
electron gas approximation. A density of states peak at the Fermi level may 
affect the transport properties beyond the perturbative level that has been 
considered so far. \\

As experiments progress to ultra-low temperatures we unravel new facets of 
Andreev physics. The correlations induced by the proximity to a bulk 
superconductor could allow for the observation of metallic electron-electron 
interaction effects on a macroscopic scale. The low-temperature anomalies 
provide the opportunity to determine both the size and the sign of the 
coupling in these materials. 

\addtocontents{toc}{\newpage}

\appendix

\chapter{Bogoliubov-de Gennes equations}
\label{bdgapp}

We diagonalize the BCS-Hamiltonian for superconductivity by a Bogoliubov 
transformation, considering a spatially inhomogeneous system.
The transformation is carried out by determining the quasiparticle 
wavefunctions that fulfill the Bogoliubov-de-Gennes (BdG) equations. 
We show that the symmetry in the solutions of the (BdG) 
equations with respect to reversing the energies $\epsilon_{\alpha}\to 
-\epsilon_{\alpha}$ is a consequence of the spin degeneracy. Making use of the 
spin-reversal symmetry, we express the Hamiltonian, the density, and the 
current operators in terms of the quasiparticle operators $\gamma_{\alpha}$, 
and arrive at the current expression used in chapter \ref{nonlinearity}.

The mean-field, spin singlet BCS Hamiltonian for superconductivity 
can be expressed in the form ($h_0=\left(-i{\bf\nabla}+e{\bf A}\right)^2/2m$)
\be
{\cal H} = \int d^3x : {\bf \hat{\Psi}}^{\dag}\left({\bf x}\right) 
\left( \begin{array}{cc}
   h_0  -\mu & \Delta \\
   \Delta^* & -h_0^* + \mu \end{array} \right)
{\bf \hat{\Psi}}\left({\bf x}\right) :\, 
\label{bcs} \ee
using the Nambu spin-up annihilation operator 
\be \bf{\hat{\Psi}}\left(\bf{x}\right)= \left( \begin{array}{c}
   \hat{\Psi}_{\uparrow}\left(\bf{x}\right) \\
    \hat{\Psi}_{\downarrow}^{\dag}\left(\bf{x}\right) \end{array}  \right).
\label{Nambupsi}\ee
The pair potential is given by $\Delta({\bf x})=\lambda \langle 
\hat{\Psi}_{\downarrow}\left({\bf x}\right)\hat{\Psi}_{\uparrow}\left({\bf x}
\right) \rangle$ (coupling constant $\lambda$), the colon ($:$) denoting  
normal ordering with respect to $\hat{\Psi}_{\uparrow}$ and 
$\hat{\Psi}_{\downarrow}$. 
The Hamiltonian (\ref{bcs}) can be diagonalized by a basis transformation,
\ba 
{\bf \hat{\Psi}}\left({\bf x}\right) &=& 
\sum_{\alpha} {\bf \Phi}_{\alpha}\left({\bf x}\right) \gamma_{\alpha},   
\label{transf} \\
\gamma_{\alpha} &=& \int d^3x\, {\bf \Phi}_{\alpha}^{\dag}
\left({\bf x}\right)  {\bf \hat{\Psi}}\left({\bf x}\right),    
\label{invtransf} 
\ea 
with the eigenfunctions ${\bf\Phi}_{\alpha}\left({\bf x}\right)$ of the BdG 
equations (which follow from the insertion of the Hamiltonian 
(\ref{bcs}) into $[{\cal H},\gamma_{\alpha}]
=- \epsilon_{\alpha} \gamma_{\alpha}$),
\ba \left( \begin{array}{cc}
   h_0 -\mu & \Delta \\
   \Delta^* & -h_0^* +\mu \end{array} 
\right)
  {\bf \Phi}_{\alpha}\left({\bf x}\right) = \epsilon_{\alpha} 
{\bf \Phi}_{\alpha}\left({\bf x}\right), \label{eq:bdg} \\
{\bf\Phi}_{\alpha}\left({\bf x}\right) = \left( \begin{array}{l}u_{\alpha}\\ 
                                        v_{\alpha} \end{array} \right)  . 
\nonumber 
\ea
Note that $\gamma_{\alpha}$ are annihilation operators of spin-up states 
and thus the excitations are all described in terms of spin-up quasi-particles. 
In order to preserve the (fermionic) commutation relations we need a complete 
orthonormal set of wavefunctions of the hermitian operator in 
(\ref{eq:bdg}),
\ba \sum_{\alpha}  {\bf \Phi}_{\alpha}\left({\bf x}\right) 
{\bf \Phi}_{\alpha}^{\dag}\left({\bf x}'\right) &=& 
{\bf 1} \delta\left({\bf x}-{\bf x}'\right),  
\label{completeness} \\
\int d^3x \,{\bf \Phi}_{\alpha}^{\dag}\left({\bf x}\right) 
{\bf \Phi}_{\alpha'}\left({\bf x}\right) &=& \delta_{\alpha,\alpha'}
\label{orthonormality} 
\ea
(involving both positive and negative energy eigenstates)\cite{kuemmel}.

Consider the spin-reversal transformation $\cal S$, 
\be {\cal S} \,{\bf\hat{\Psi}} \,{\cal S}^{-1} = 
\left( \scriptstyle{\begin{array}{cc}
0 & 1 \\ -1 & 0 \end{array}} \right) 
\left({\bf\hat{\Psi}}^{\dag}\right)^{\top},
\label{spinrev} \ee
$\cal S$ linear (${\cal S}\hat{\psi}_{\uparrow}{\cal S}^{-1}\!=\!
\hat{\psi}_{\downarrow}$, ${\cal S}\hat{\psi}_{\downarrow}{\cal S}^{-1}\!=\!
-\hat{\psi}_{\uparrow}$). Noting that the order parameter $\Delta({\bf x})$ is
invariant under the transformation (\ref{spinrev}), it is easily seen that the
Hamiltonian (\ref{bcs}) is spin-reversal symmetric, $[{\cal H},{\cal S}]=0$
(this symmetry extends also to finite magnetic field, if the Zeeman splitting 
is neglected).
By means of spin-reversal we may attribute to each quasi-particle operator 
$\gamma_{\alpha}$ a linearly independent operator 
$\gamma_{\bar{\alpha}}$ through 
\be \gamma_{\bar{\alpha}}^{\dag}= {\cal S} \gamma_{\alpha} 
{\cal S}^{-1} = \int d^3x \,{\bf \hat{\Psi}}^{\dag}\left({\bf x}\right) 
\left( \scriptstyle{\begin{array}{rr} 0 & -1 \\ 1 & 0 \end{array}} \right)
{\bf \Phi}_{\alpha}^{*}\left({\bf x}\right). 
\label{relatedgamma} 
\ee
$\gamma_{\bar{\alpha}}$ describes an excitation with opposite 
energy $\epsilon_{\bar{\alpha}}=-\epsilon_{\alpha}$ 
(according to $[{\cal H},\gamma_{\bar{\alpha}}]= 
\epsilon_{\alpha} \gamma_{\bar{\alpha}}$).
From Eqs.\ (\ref{relatedgamma}) and (\ref{invtransf}) we infer the effect of 
spin-reversal on the electron-hole wavefunction,
\be 
{\bf \Phi}_{\bar{\alpha}}\left({\bf x}\right) = 
\left( \scriptstyle{\begin{array}{cc} 0 & -1 \\ 1 & 0 \end{array}} \right) 
{\bf \Phi}_{\alpha}^{*}\left({\bf x}\right) 
= \left( \begin{array}{c} -v_{\alpha}^* \\ u_{\alpha}^* \end{array} \right) .
\label{spinrevwave}
\ee
We arrive at a complete set of quasi-particle 
excitations in spin-up space, which are grouped into pairs 
${\gamma_{\alpha},\gamma_{\bar{\alpha}}}$ with energies $\pm \epsilon_{\alpha}$, as a direct consequence of spin degeneracy. Note that within the 
Nambu picture, all quasi-particles carry spin up 
[spin $\uparrow e$ and spin $\uparrow h$ (instead of spin $\downarrow e$)], 
explaining the opposite energy of the related wavefunction (\ref{spinrevwave}). 
The Hamiltonian (\ref{bcs}) takes the form,
\be {\cal H} = \sum_{\epsilon_{\alpha}>0} \epsilon_{\alpha} \left( \gamma_{\alpha}^{\dag} 
\gamma_{\alpha} + \gamma_{\bar{\alpha}} \gamma_{\bar{\alpha}}^{\dag}  -2 \int d^3x\, |v_{\alpha}|^2 \right). 
\label{Halpha}\ee
The ground state is realized by filling all the negative energy 
quasi-particle states.
 
The spin-reversal symmetry  allows us to express all equations using only half 
of the eigenstates, i.e., 
one representative ${\bf \Phi}_n$ from each pair of states 
$\left\{{\bf \Phi}_{\alpha},{\bf \Phi}_{\bar{\alpha}}\right\}$. 
In the following we choose the positive energy eigenstates, expressing the
negative energy eigenstates through (\ref{spinrevwave}). We keep 
the positive energy states for the spin-up quasi-particles ($\gamma_n=\gamma_{n\uparrow}$), 
and reinterpret the related quasi-particle states of opposite (negative) energy 
as spin-down excitations ($\gamma_{\bar{n}}=\gamma_{\bar{n}\downarrow}^{\dag}$). 
The Bogoliubov transformation (\ref{transf}) then takes the well-known form
\be {\bf\hat{\Psi}}\left({\bf x}\right) = 
\sum_n \left( \begin{array}{l}u_{n}\\  
v_{n} \end{array} \right) \gamma_{n\uparrow} 
+ \left( \begin{array}{l}-v_{n}^*\\  
u_{n}^* \end{array} \right) \gamma_{\bar{n}\downarrow}^{\dag},
\label{btstandard} 
\ee
and the Hamilton operator is expressed by
${\cal H} = \sum_{\epsilon_{n}>0} \epsilon_{n} 
( \gamma_{n\uparrow}^{\dag} 
\gamma_{n\uparrow} + \gamma_{\bar{n}\downarrow}^{\dag} 
\gamma_{\bar{n}\downarrow}  
-2 \int d^3x\, |v_{n}|^2 )$,
displaying the spin degeneracy in the usual fashion. In the same way, we give 
the density and current operators in both representations,
using all (indexed by $\alpha$) or only the positive energy eigenstates 
(indexed by $n$), respectively ($ u \stackrel{\leftrightarrow}
{\nabla} v = u \nabla v - (\nabla u) v$),
\ba \rho\left({\bf x}\right) &=& 
-e : {\bf \hat{\Psi}}^{\dag}\left({\bf x}\right) 
\left( \scriptstyle{\begin{array}{cc} 1 & 0 \\ 0 & -1 \end{array}} \right) 
{\bf \hat{\Psi}}\left({\bf x}\right) : 
\nonumber \\
&=& -e \sum_{\alpha,\alpha'} \left( 
u_{\alpha}^*({\bf x}) u_{\alpha'}({\bf x}) \gamma^{\dag}_{\alpha}
\gamma_{\alpha'} + v_{\alpha}^*({\bf x}) v_{\alpha'}({\bf x}) \gamma_{\alpha'}
\gamma^{\dag}_{\alpha} \right)
\label{rhoalpha} \\
&=& -e \sum_{\epsilon_m,\epsilon_n > 0} 
\left[u_m^*({\bf x}) u_n({\bf x}) - v_m^*({\bf x}) v_n({\bf x}) \right] 
\left( \gamma_{m\uparrow}^{\dag} \gamma_{n\uparrow} + 
\gamma_{\bar{m}\downarrow}^{\dag} \gamma_{\bar{n}\downarrow} \right) 
\nonumber \\ 
&& -2e \sum_{\epsilon_n>0} |v_n({\bf x})|^2 \label{rhon} \\
&& -e \sum_{\epsilon_m,\epsilon_n > 0} 
\left\{ \left[ u_m({\bf x}) v_n({\bf x}) 
+ v_m({\bf x}) u_n({\bf x}) \right] \gamma_{m\uparrow} 
\gamma_{\bar{n}\downarrow} \, + \mbox{ h.c. } \right\}, \nonumber
\ea
\ba 
{\bf j}\left({\bf x}\right) &=& -\frac{e}{2mi} : {\bf \hat{\Psi}}^{\dag}
\left({\bf x}\right)  {\bf \stackrel{\leftrightarrow}{\nabla} } 
{\bf \hat{\Psi}}\left({\bf x}\right) : 
\nonumber \\
&=& -\frac{e}{2mi} \sum_{\alpha,\alpha'} 
u_{\alpha}^*({\bf x}) {\bf \stackrel{\leftrightarrow}{\nabla} } u_{\alpha'}({\bf x}) 
\gamma^{\dag}_{\alpha} \gamma_{\alpha'} 
- v_{\alpha}^*({\bf x}) {\bf \stackrel{\leftrightarrow}{\nabla} } v_{\alpha'}({\bf x}) 
\gamma_{\alpha'} \gamma^{\dag}_{\alpha} 
\label{jalpha} \\
&=& \displaystyle  -\frac{ e}{2mi} \left\{ \sum_{\epsilon_m,\epsilon_n > 0} 
\left(u_m^*({\bf x}) {\bf \stackrel{\leftrightarrow}{\nabla} } u_n({\bf x}) 
+ v_m^*({\bf x}) {\bf \stackrel{\leftrightarrow}{\nabla} } v_n({\bf x}) 
\right) \left( \gamma_{m\uparrow}^{\dag} \gamma_{n\uparrow} 
+ \gamma_{\bar{m}\downarrow}^{\dag} \gamma_{\bar{n}\downarrow} \right) 
\right. \nonumber \\
&& \;\;\;\;\;\;\;\;\;\;\;\;\;\; -2 \sum_{\epsilon_n>0} 
v_n^*({\bf x}){\bf \stackrel{\leftrightarrow}{\nabla} } v_n({\bf x})  
 \label{jn}  \\
&& \left. + \sum_{\epsilon_m,\epsilon_n > 0} \left[
\left(- u_m({\bf x}) {\bf \stackrel{\leftrightarrow}{\nabla} } v_n({\bf x}) 
+ v_m({\bf x}) {\bf \stackrel{\leftrightarrow}{\nabla} } u_n({\bf x}) \right) 
\gamma_{m\uparrow} \gamma_{\bar{n}\downarrow} \, - \mbox{ h.c. } \right] 
\right\} .
\nonumber
\ea
The current operator is easily generalized to finite magnetic field. We note
that current and density operators obey the continuity equation
\be \frac{\partial\rho({\bf x})}{\partial t} + {\bf \nabla \cdot j}({\bf x}) = 
2ie {\bf \hat{\Psi}}^{\dag}\left({\bf x}\right) 
\left( \scriptstyle{\begin{array}{cc} 0 & \Delta({\bf x}) \\ -\Delta^*({\bf x}) & 0 
\end{array}} \right)  {\bf \hat{\Psi}}\left({\bf x}\right).
\label{continuity}
\ee
Clearly, in the presence of the pair potential $\Delta$ the single 
quasi-particle currents are not conserved. However, as the right hand side 
of (\ref{continuity}) vanishes when taking the self-consistent expectation 
value, the total current is conserved, the quasi-particle currents being 
balanced by the current of the condensate. The Andreev reflection represents
a good example, where the quasi-particle current of $2e$ entering the 
superconductor decays and is converted to supercurrent.

\chapter{Quasi-classical Green's function technique}
\label{quasiclassicaltheory}

\newcommand{\vhp}{\ensuremath{\hat{\mathbf{p}}}}
\newcommand{\vr}{\ensuremath{\mathbf{r}}}
\newcommand{\vR}{\ensuremath{\mathbf{R}}}

\newcommand{\PsiDag}{\ensuremath{\Psi^{\dag}}}
\newcommand{\PsiTil}{\ensuremath{\tilde{\Psi}}}
\newcommand{\PsiTilA}{\ensuremath{\PsiTil_{\alpha}}}
\newcommand{\PsiTilB}{\ensuremath{\PsiTil_{\beta}}}
\newcommand{\PsiTilDag}{\ensuremath{\PsiTil^{\dag}}}
\newcommand{\PsiTilDagA}{\ensuremath{\PsiTilDag_{\alpha}}}
\newcommand{\PsiTilDagB}{\ensuremath{\PsiTilDag_{\beta}}}
\newcommand{\Gab}{G_{\alpha\beta}}
\newcommand{\Gqc}{\breve{g}}
\newcommand{\fDag}{f^{\dag}}
\newcommand{\DStar}{\Delta^{\ast}}
\newcommand{\DNull}{\Delta_{0}}
\newcommand{\komTil}{\tilde{\omega}_{n}}
\newcommand{\mA}{\mathcal{A}}
\newcommand{\GOm}{\Omega}
\newcommand{\GOmNull}{\GOm_{0}}
\newcommand{\FluxQ}{\Phi_{0}}
\newcommand{\vdd}{d^{\prime}}
\newcommand{\vdv}{d_{V}}
\newcommand{\Rez}{\mathrm{Re}}
\newcommand{\Imz}{\mathrm{Im}}
%
%

Here we give give a short, but self-contained derivation of the Eilenberger 
equation (\ref{eq:eilenberger}) and the corresponding quasi-classical 
current expression (\ref{eq:current}). We show how the renormalization 
procedure takes care of the cancellation of bulk diamagnetic and Fermi 
surface paramagnetic currents in the normal metal. 
Eq.\ (\ref{eq:current}) depends on the quasi-classical Green's function at 
the Fermi surface and its deviation from the normal state. The treatment 
given here follows the spirit of \cite{kopnin}.

\section{Eilenberger equations}

We define the $2\times 2$ matrix Green's function in imaginary time 
[$x=({\bf x},\tau)$]
\be
\hat{G} (x_{1},x_{2})=
    -\langle T_{\tau} {\bf \hat{\Psi}} (x_{1}) 
   {\bf \hat{\Psi}}^{\dag} (x_{2}) \rangle
=\left( \begin{array}{cc}
             G(x_{1},x_{2})        &   F(x_{1},x_{2}) \\
             F^{\dag}(x_{1},x_{2}) &  -G(x_{2},x_{1} )   
           \end{array}
    \right)
,
\label{bdg:defg}
\ee
by the Nambu-space operator (\ref{Nambupsi}) in the Heisenberg
picture with respect to the spin-symmetric Hamiltonian (\ref{bcs}).
$\hat{G}$ fulfills the equation of motion
\begin{equation}
  \hat{G}^{-1} \ast \hat{G} (x_{1},x_{2}) =\delta(x_{1}-x_{2}).
  \label{eq:GorkovMR}
\end{equation}
where $\hat{G}^{-1}$ is the matrix operator
\begin{equation}
  \hat{G}^{-1}(x_1,x_2)= - \left( 
      \begin{array}{cc}
        \frac{\partial}{\partial\tau_1} +h_0 -\mu & \Delta(x_1) \\
        \Delta^{\ast}(x_1) & \frac{\partial}{\partial\tau_1} -h_0^* +\mu
      \end{array}
    \right) \delta(x_1-x_2),
\label{bdg:g-1}
\end{equation}
$h_0 = -\frac{1}{2m} ( {\bf \nabla}_1+ie {\bf A}(x_1))^2$, and 
the contraction is defined by 
$A \ast B (x_1,x_2)=\int dx' A (x_1,x') B(x',x_2)$. Similarly, the conjugate
equation of motion takes the form
\be
\hat{G}\ast \hat{G}^{-1} (x_{1},x_{2}) = \delta(x_{1}-x_{2}).
  \label{eq:GorkovML}
\ee
In the following we assume homogeneity in time 
$\partial_{\tau_1}=-\partial_{\tau_2}$. We also want to neglect the 
${\bf A}^2$ term in (\ref{bdg:g-1}).
While the Green's function $\hat{G}$ varies on the Fermi wavelength $1/k_F$, 
the vector potential varies on a much larger scale, which we denote by 
$\zeta$. With 
\begin{equation}
  \left({\bf \nabla}+ ie {\bf A}(x)\right)^2=
    \nabla^2 + ie \left({\bf \nabla} \cdot {\bf A}(x) \right)
    + 2ie {\bf A}(x) \cdot {\bf \nabla}- e^2 {\bf A}(x)^2,
\end{equation}
and typically ${\bf \nabla} \sim k_F$, we find that the term
$e^2 {\bf A}^2 \sim e^2 H^2 \zeta^2$ is negligible in comparison to
$2e {\bf A}\cdot {\bf \nabla} \sim e H k_F\zeta$ under the condition 
\begin{equation}
  \zeta \ll \frac{k_F c}{eH} \equiv r_{L},
  \label{eq:VPNeglCond}
\end{equation}
where $r_{L}$ denotes the Larmor radius, which is the cyclotron radius of an 
electron traveling at Fermi velocity $v_F$.

The quasi-classical approximation is best done in the Fourier representation
\begin{equation}
  \tilde{G}({\bf p}_1,{\bf p}_2;\tau)=\int d^3 x_{1} d^3 x_{2}\,
    G ({\bf x}_1,{\bf x}_2;\tau) e^{-i{\bf p}_1\cdot{\bf x}_1+
    i{\bf p}_2\cdot{\bf x}_2},
\end{equation}
where the equation of motion for the first Gorkov Green's function 
in (\ref{bdg:defg}) takes the form
\begin{eqnarray}
  \left(-\frac{\partial}{\partial\tau}-\frac{{\bf p}_1^2}{2m}+\mu\right)
    \tilde{G}({\bf p}_1,{\bf p}_2;\tau)
  -\int d^3 k^{\prime}\,\Delta({\bf k}^{\prime}) \,
    \tilde{F}^{\dag}({\bf p}_1-{\bf k}^{\prime},{\bf p}_2;\tau) && 
  \nonumber\\
  -\frac{e}{2m}\int d^3 k^{\prime}\;
    {\bf k}^{\prime}\cdot{\bf A}({\bf k}^{\prime}) \,
\tilde{G}({\bf p}_1-{\bf k}^{\prime},{\bf p}_2;\tau) \quad\quad && 
  \label{eq:GorkovFT} \\
\  -\frac{e}{m}\int d^3 k^{\prime}\,{\bf A}({\bf k}^{\prime}) 
  \cdot({\bf p}_1-{\bf k}^{\prime}) \,
    \tilde{G}({\bf p}_1-{\bf k}^{\prime},{\bf p}_2;\tau) \,
 = \, (2\pi)^3\delta({\bf p}_1-{\bf p}_2)\delta(\tau). && \nonumber
\end{eqnarray}
Introducing the center-of-mass and relative coordinates
\begin{equation}
  {\bf x}=\frac{1}{2}({\bf x}_1+{\bf x}_2),\;\; 
  {\bf r}={\bf x}_1-{\bf x}_2,
\end{equation}
and their conjugate momenta
\begin{equation}
  {\bf k}={\bf p}_1-{\bf p}_2,\;\; 
  {\bf p}=\frac{1}{2}({\bf p}_1+{\bf p}_2),
  \label{eq:DefCmRelKoor}
\end{equation}
we find that ${\bf r}$ varies on scale $1/k_F$, while $\bf x$ varies 
on the much larger scale $\zeta$. Eq.\ (\ref{eq:GorkovFT}) translates 
to [$G({\bf k},{\bf p}; \tau) = \tilde{G}({\bf p}_1,{\bf p}_2; \tau)$]
\begin{eqnarray} 
  \left(-\frac{\partial}{\partial\tau}-\frac{({\bf p}+\frac{{\bf k}}{2})^2}{2m}
    +\mu\right) G({\bf k},{\bf p};\tau)
  -\int d^3 k^{\prime}\, \Delta({\bf k}^{\prime})\, F^{\dag} 
({\bf k}-{\bf k}^{\prime},{\bf p}-\frac{{\bf k}^{\prime}}{2};\tau) \quad &&
 \label{eq:GorkovCmRelKoor}  \\
  -\frac{e}{2m}\int d^3 k^{\prime}(2{\bf p}+{\bf k}-{\bf k}^{\prime})\cdot
    {\bf A}({\bf k}^{\prime}) \, G({\bf k}-{\bf k}^{\prime},
 {\bf p}-\frac{{\bf k}^{\prime}}{2};\tau) \, = \,  
  (2\pi)^3\delta({\bf k})\delta(\tau). &&
\nonumber
\end{eqnarray}
In the quasi-classical approximation, we retain only contributions for 
$k^{\prime}\ll k_F$, assuming a smooth variation of $\Delta$ and $\bf A$ on 
Fermi wavelength ($k_F\zeta \ll 1$), and obtain
\begin{eqnarray}
  \left(-\frac{\partial}{\partial\tau}-\frac{{\bf p}^2+{\bf p}\cdot{\bf k}}
    {2m}+\mu\right)\, G({\bf k},{\bf p};\tau)
  -\int d^3 k^{\prime}\,\Delta({\bf k}^{\prime})\, 
    F^{\dag}({\bf k}-{\bf k}^{\prime},{\bf p};\tau) \quad &&
  \label{eq:GorkovApprox} \\
  -\frac{e}{m}\int d^3 k^{\prime}\;{\bf p}\cdot
    {\bf A}({\bf k}^{\prime})\, G({\bf k}-{\bf k}^{\prime},{\bf p};\tau)
    \, = \, (2\pi)^3\delta({\bf k})\delta(\tau), && \nonumber
\end{eqnarray}
Transforming back to the real space center-of-mass coordinate 
and carrying out a Fourier expansion in time, we arrive at
\begin{eqnarray}
  \left(i\omega_{n}-\frac{{\bf p}\,^2-
    i{\bf p}\cdot {\bf \nabla}}{2m}+\mu-
 \frac{e}{m}{\bf p}\cdot{\bf A}({\bf x})\right)G({\bf x},{\bf p};\omega_{n})
\nonumber \\   -\Delta({\bf x})\, F^{\dag}({\bf x},{\bf p};\omega_{n}) = 1.
\end{eqnarray}
The fast oscillations of $G$ on Fermi wavelength are described 
by the ${\bf p}^2$ term. The above procedure is carried
out similarly for the Green's functions $F$ and $F^{\dag}$, obtaining 
the quasi-classical equation of motion for the matrix Green's function 
$\hat{G}$ and its conjugate
\begin{eqnarray}
\hat{G}^{-1}({\bf x},{\bf p};\omega_{n}) \, 
 \hat{G}({\bf x},{\bf p};\omega_{n})  &=&1,  \label{eq:GorkovQR} \\
\hat{G}({\bf x},{\bf p};\omega_{n}) \, 
 \hat{G}^{-1} ({\bf x},{\bf p};\omega_{n})  &=&1, \label{eq:GorkovQL}
\end{eqnarray}
where ($\xi_{p}=p^2/2m-\mu$)
\begin{equation} \!\!\!\!\!\!\!\!\!\!
\hat{G}^{-1} ({\bf x},{\bf p};\omega_{n})= 
  i\omega_{n}-\frac{e}{mc}{\bf p}\cdot{\bf A}
 + \left(-\xi_{p}+i\frac{{\bf p}\cdot\nabla}{2m}\right) \hat{\tau_{3}}
    -\Rez \Delta\hat{\tau_{1}} +\Imz\Delta\hat{\tau_{2}},
\label{bdg:gop}
\end{equation}
and the gradient $\bf \nabla$ is to be taken with a negative sign
when acting to the left in (\ref{eq:GorkovQL}). We have thus arrived 
at first-order differential equations with respect to the center-of-mass
coordinate $\bf x$, the dependence of which accounts for the inhomogeneity of 
the slowly varying potential and field. The momentum dependence of $\hat{G}$ 
lies at the Fermi surface $|{\bf p}|=k_F$.

\begin{figure}[t]
  \begin{center}\begin{picture}(200,120)(0,0)
    \put(0,60){\vector(1,0){200}}
    \put(100,0){\vector(0,1){120}}
    \put(65,58){\line(0,1){3}}
    \put(135,58){\line(0,1){3}}
    \put(140,45){$E_{C}$}
    \put(117,40){$\frac{1}{2}$}
    \put(117,75){$\frac{1}{2}$}
    \put(80,85){$\gamma_{0}$}
    \put(95,87){\line(3,-1){20}}
    \put(95,85){\line(1,-2){22}}
    \ArrowArc(100,65)(35,0,180)
    \ArrowArcn(100,55)(35,0,180)
    \ArrowLine(65,65)(135,65)
    \ArrowLine(65,55)(135,55)
    \put(50,42){$\frac{1}{2}$}
    \put(50,75){$\frac{1}{2}$}
    \put(10,90){$\gamma_{\infty}$}
    \put(25,95){\line(3,-2){20}}
    \put(25,90){\line(1,-2){22}}
    \ArrowArcn(100,65)(39,180,360)
    \ArrowArc(100,55)(39,180,360)
    \ArrowLine(139,65)(200,65)
    \ArrowLine(139,55)(200,55)
    \ArrowLine(0,65)(61,65)
    \ArrowLine(0,55)(61,55)
  \end{picture}\end{center}
\caption{Integration paths: $\gamma_0$ denotes the two inner half circles, 
$\gamma_{\infty}$ the two outer high energy contributions. Together $\gamma_0$ 
and $\gamma_{\infty}$ reduce to the integration over the real line.}
  \label{f:IntPathCurrent}
\end{figure}
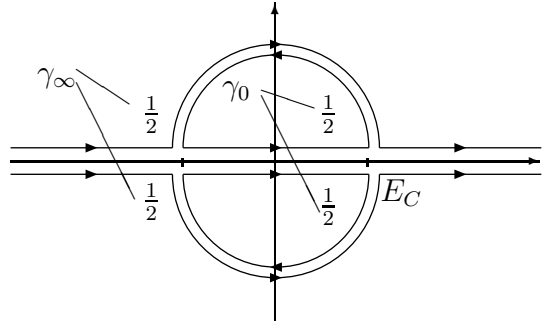

We are now ready to define the quasi-classical Green's function 
by the integral over the momentum energy $=\xi_p$,
\be
\hat{g}_{\omega_n}({\bf x}, {\bf v}_F) = 
  \frac{i}{\pi} \hat{\tau_3} \int_{\gamma_{0}} d\xi_{p}\,
      \hat{G}({\bf x},{\bf p};\omega_{n}) ,
  \label{eq:DefQcGf}
\ee
keeping the dependence on the momentum direction as denoted by the 
Fermi velocity ${\bf v}_F$. To provide a cutoff for the high energy 
contributions at $E_{C}$, we use the integration path 
$\gamma_0$ shown in Fig.\  \ref{f:IntPathCurrent} 
($T, \Delta \ll E_{C} \ll E_{F}$). 
From the subtraction of the equation of motion (\ref{eq:GorkovQR})
and its conjugate (\ref{eq:GorkovQL}), appropriately multiplied with 
$\hat{\tau_3}$, we eliminate the $\xi_p$ term in
(\ref{bdg:gop}) and after integration over the momentum energy we find
the Eilenberger equation
\ba  \lefteqn {
-({\bf v}_F\cdot \mathbf{\nabla}) \hat{g}_{\omega_n} 
({\bf x},{\bf v}_F) \, = } \\
&& \Big[ \{\omega_{n}+ie {\bf v}_F\cdot{\bf A}({\bf x})\} \hat{\tau_3}
  +\Rez\Delta({\bf x})\hat{\tau_2} +\Imz\Delta({\bf x})\hat{\tau_1}\, ,\,
  \hat{g}_{\omega_n} ({\bf x},{\bf v}_F)  \Big]. \nonumber
  \label{eq:Eilenb}
\ea
The quasi-classical equations need to be complemented by a normalization 
condition, as the source term on the right hand side of (\ref{eq:GorkovQR})
and (\ref{eq:GorkovQL}) has been canceled by the subtraction. The 
normalization is provided by 
\be
\hat{g}^2=1,
\label{bdg:norm}
\ee
as follows from the homogeneous solution and the fact that 
$\hat{g}^2$ is a constant of motion of (\ref{eq:Eilenb}). 
The following symmetry relations for the quasi-classical Green's functions
\be g(\omega_n)^{\ast} = -g(-\omega_n), \;\; 
    f(\omega_n)^{\ast} = f^{\dag} (-\omega_n),
\label{bdg:symm1}
\ee
and
\begin{equation}
  g({\bf v}_F)^{\ast}=g(-{\bf v}_F), 
  \;\;f({\bf v}_F)^{\ast}=\fDag(-{\bf v}_F),
\label{bdg:symm2}
\end{equation}
simplify the solution of (\ref{eq:Eilenb}). Eq.\ (\ref{bdg:symm1}) 
follows from the definition (\ref{eq:DefQcGf}) and Eq.\  (\ref{bdg:symm2}) 
directly from the equation of motion (\ref{eq:Eilenb}). The quasi-classical
Green's function takes the form
\be
\hat{g} = \left( \begin{array}{cc} g & f \\ f^{\dag} & -g \end{array}
          \right).
\ee
For convenience, we redefine the 
pair potential $\Delta \to i\Delta$ and obtain the form of the equations 
used in this thesis, see (\ref{eq:eilenberger}),
\be  
-({\bf v}_F\cdot \mathbf{\nabla}) \hat{g}  = 
 \Big[ \{\omega_{n}+ie {\bf v}_F\cdot{\bf A}({\bf x})\} \hat{\tau_3}
  +\Rez\Delta({\bf x})\hat{\tau_1} - \Imz\Delta({\bf x})\hat{\tau_2}\, ,\,
  \hat{g}  \Big].
  \label{eq:Eilenbnew}
\ee

\section{Quasi-classical current expression}

The current density in terms of the usual Green's function 
(with respect to center-of-mass and relative momentum) is given by
\be
{\bf j}({\bf k}) = -\frac{2e}{m}\int\frac{d^3 p}{(2\pi)^3} \, 
 {\bf p}\, G ({\bf k},{\bf p};0-)- \frac{ne^2}{m} {\bf A}({\bf k}).
\label{bdg:usualj}
\ee
The first term in (\ref{bdg:usualj}) gives the paramagnetic current  
which is induced by the perturbation of the Green's function $G$ in 
the presence of an applied field and lives at the Fermi surface. The 
second term in (\ref{bdg:usualj}) gives the diamagnetic current giving
the rigid response of the bulk density $n$ to the vector potential.
In the normal state $G=G^{(n)}$, 
the paramagnetic current is known to cancel the
diamagnetic current up to small correction 
of the order of Landau diamagnetism, see e.g. \cite{tinkham}.
Using this fact, we easily find that the current
\be
{\bf j}({\bf k}) = -\frac{e}{m} \int_{-\infty}^{+\infty} d\xi_p\, N(\xi_p)
\int \frac{d\Omega_p}{4\pi} \,  {\bf p}\, 
\left( G({\bf k},{\bf p};0-) - G^{(n)}({\bf k},{\bf p};0-) \right),
\ee
can be expressed by the deviation of the Green's function from the
normal state value ($N(\xi_p)$ is the density of states). Splitting the 
integration of the momentum energy $\xi_p$ along the paths $\gamma_0$ and
$\gamma_{\infty}$, we obtain
\ba
 {\bf j}({\bf k})&=&
  -e N_0 \int\frac{d\Omega_{p}}{4\pi }\, \frac{{\bf p}_F}{m} 
   \int_{\gamma_{0}} d\xi_{p}\, 
  \left(G({\bf k},{\bf p};0-) - G^{(n)}({\bf k},{\bf p};0-) \right) \\
&& - \frac{e}{m} \int_{\gamma_{\infty}} d\xi_{p}\, N(\xi_p)
  \int\frac{d\Omega_{p}}{4\pi }\,  {\bf p}\, 
 \left( G -  G^{(n)} \right).
\label{bdg:decomposition}
\ea
The high energy contribution in (\ref{bdg:decomposition}) vanishes as 
the Green's function $G$ only deviates from $G^{(n)}$ at energies 
of the order $\xi_p \sim T, \Delta \ll E_C$. Inserting the Fourier expansion 
in time into the first we find
\be
{\bf j}({\bf k}) = 
  -e N_0 \int\frac{d\Omega_{p}}{4\pi }\, \frac{{\bf p}_F}{m} 
   \int_{\gamma_{0}} d\xi_{p}\, T \sum_{\omega_n} 
  \left(G({\bf k},{\bf p};\omega_n) - G^{(n)}({\bf k},{\bf p};\omega_n) 
\right).
\ee
Having introduced the high-energy cutoff by $\gamma_0$ we can exchange 
the Matsubara sum and the energy integration, which otherwise would have
lead to a divergence, see \cite{agd}. Using the definition of 
the quasi-classical Green's functions (\ref{eq:DefQcGf}), the current takes the 
form
\begin{equation}
{\bf j}({\bf x})=ie N_0 \pi T \sum_{\omega_n} \int\frac{d\Omega_{p}}{4\pi}
    {\bf v}_F \left( g_{\omega_n}({\bf x},{\bf v}_F) - 
    g^{(n)}_{\omega_n}({\bf x},{\bf v}_F) \right).
\label{bdg:currentexp}
\end{equation}
The quasi-classical Green's function for the normal state 
$g={\rm sgn} (\omega_n)$ that is found from $G({\bf p})=1/(i\omega_n-\xi_p)$
conveniently drops out of Eq.\ (\ref{bdg:currentexp}) and we obtain the
quasi-classical current expression (\ref{eq:current}).

\chapter{Response kernel}
\label{kernel} 

Starting from Eqs.\ (\ref{eilenfirst1}) and (\ref{eilenfirst2}) 
we derive the linear response kernel $K(x,x')$.
The transformation of $g$, $f$, and $f^{\dag}$ introduced 
in \cite{schopohl,schopohl:98},
\begin{equation}
 \label{schopohltrafo}
 a(x,v_{x}) =
 \frac{f(x,v_{x})}{1+g(x,v_{x})}, \quad \quad
 a^{\dag}(x,v_{x}) =
 \frac{f^{\dag}(x,v_{x})}{1+g(x,v_{x})},
\end{equation}
decouples the full equations of motions (\ref{eq:eilenberger})
to two Riccati differential equations. For the zeroth order equations 
we obtain
\begin{eqnarray}
  \nonumber
-{v}_{x} \partial_x a_{0}(x,v_x) &=& 
   2\tilde{\omega}(x) a_{0}(x,v_{x}) + \tilde{\Delta}(x) 
   \Big( a^2_0(x,v_x)-1 \Big) \\  
v_x \partial_x a^{\dag}_{0}(x,v_x) &=&
   2\tilde{\omega}(x) a^{\dag}_{0}(x,v_x)+ \tilde{\Delta}(x)
   \Big( a^{\dag 2}_0(x,v_x) - 1 \Big).
\end{eqnarray} 
After linearization of (\ref{schopohltrafo}) we obtain
\begin{eqnarray}
 f_{1}(x,v_{x},v_{y})&=&
 2\frac{a_{1}(x,v_{x},v_{y})-
 a_{0}^{2}(x,v_{x})
 a^{\dag}_{1}(x,v_{x},v_{y})}{
 (1+a_{0}(x,v_{x})a^{\dag}_{0}(x,v_{x}))^{2}}\\
 f^{\dag}_{1}(x,v_{x},v_{y})&=&
 2\frac{a^{\dag}_{1}(x,v_{x},v_{y})
 -a^{\dag 2}_{0}(x,v_{x})
 a_{1}(x,v_{x},v_{y})}{
 (1+a_{0}(x,v_{x}) a^{\dag}_{0}(x,v_{x}))^{2} } ,
\end{eqnarray}
and Eqs.\ (\ref{eilenfirst1}) and (\ref{eilenfirst2}) are decoupled to
\begin{eqnarray}
\label{riccatifirst}
 -\frac{v_{x}}{2} \partial_x
    a_{1}(x,v_{x},v_{y}) &=&
   \left[ \tilde{\omega}(x)+\tilde{\Delta}(x)a_{0}(x,v_{x})\right]
   a_{1}(x,v_{x},v_{y}) \\ 
&& +  ev_y A_y(x) a_{0}(x,v_{x}) \nonumber \\
\frac{v_{x}}{2}\partial_x a^{\dag}_{1}
  (x,v_{x},v_{y}) &=& 
  \left[ \tilde{\omega}(x)+ \tilde{\Delta}(x)a^{\dag}_{0}(x,v_{x})\right]
 a^{\dag}_{1}(x,v_{x},v_{y}) \\
&& + ev_y A_y(x) a^{\dag}_{0}(x,v_{x}). \nonumber
\end{eqnarray}
As a consequence of Eq.~(\ref{riccatifirst}) we find
$a_{1}(-v_{y})=-a_{1}(v_{y})$ and the same for
$a^{\dag}_{1}$, implying $\langle f_{1}\rangle =\langle
f^{\dag}_{1}\rangle =0$. Furthermore,
since $a^{\dag}_{1}(v_{x})=a_{1}(-v_{x})$, we only
have to consider one of the two equations (e.g. the first one).
Equation (\ref{riccatifirst}) is an inhomogeneous first-order
differential equation, which can be integrated analytically. Assuming
that $f$ and $f^{\dag}$ do not change sign as a function of $x$, 
with the help of (\ref{eilenreal}) the solution can be written as
\begin{eqnarray}
\label{solution}
a_{1}(x,v_{x},v_{y}) &=& c(v_{x},v_{y})
   \frac{m(v_{x},x,x_{0})}{f^{\dag}_{0}(x,v_{x})} \\ &&
 -\frac{2ev_{y}}{v_{x} f^{\dag}_{0}(x,v_{x})}
   \int_{x_{0}}^{x} [1-g_{0}(x',v_{x})] m(v_{x},x,x') A_y(x') dx'
\nonumber
\end{eqnarray}
where 
\begin{equation}
 \label{defm}
 m(v_{x},x,x^{\prime}) = 
 \exp\left(\frac{2}{v_{x}}\int_{x}^{x'}
 \frac{\tilde{\Delta}(x'')}{
 f^{\dag}_{0}(x'',v_{x})} dx''\right)
\end{equation}
In this equation $x_{0}$ is an arbitrary reference point and the
constant $c$ has to be determined by the appropriate boundary
conditions. $m$ satisfies the relations of a propagator, 
$m(v_x,x,x^{\prime}) = m(u,x^{\prime},x)^{-1}$ and $
m(v_x,x,x^{\prime\prime}) m(u,x^{\prime\prime},x^{\prime})=
m(v_x,x,x^{\prime})$.  Now we determine the constant $c$ for a system of
size $[-d_s,d]$. We assume specular reflection at two boundaries at
$x=-d_s,d$ and ideal interfaces between different materials inside the
system. The appropriate boundary conditions are the continuity of the
Green's function along the classical trajectories, i.e., 
$\hat{g}(x,v_{x},v_{y}) = \hat{g}(x,-v_{x},v_{y})$ at $x=-d_s,d$ 
for the specular reflection and 
$\hat{g}(x\!=\! 0-,v_{x},v_{y}) = \hat{g}(x\!=\! 0+,-v_{x},v_{y})$
at the NS interface. The same conditions are valid for $a_{1}$ and
$a^{\dag}_{1}$. We thus obtain
\begin{eqnarray}
 \label{constant}
 c(v_{x},v_{\rm{y}}) &= &
 2e\frac{v_{\rm{y}}}{v_{x}}\int_{-d_{s}}^{d}
 \frac{m(v_{x},d,x^{\prime})+m(-v_{x},d,x^{\prime})}{ 
 m(v_{x},d,-d_{s})-m(-v_{x},d,-d_{s})}\nonumber\\&&
 \times [1-g_{0}(x',v_{x})] 
 A(x^{\prime})dx^{\prime}\;\hfill .
\end{eqnarray}
The current is determined by Eq.~(\ref{eq:current}), 
expressed by the solution (\ref{solution}). We
obtain the following general result for the linear current functional,
\begin{equation}
 \label{jyapp}
 j_{y}(x) = -\int\limits_{-d_{s}}^{d} K(x,x^\prime)
 A(x^{\prime})dx^{\prime}\; ,
\end{equation}
where the kernel $K(x,x^{\prime})$ is given by
\begin{eqnarray} 
 K(x,x^\prime) &=& 
 e^{2} v_F N_0 \pi T \sum_{\omega_n>0}
 \int\limits_{0}^{v_{F}} du \, \frac{v_{F}^{2}-u^{2}}{{v_{F}^{2}}u}
 [1+g_{0}(x,u)][1-g_{0}(x',u)] \nonumber \\
&& \bigg[\Theta(x-x^{\prime})m(u,x,x^{\prime})+
 \Theta(x^{\prime}-x)m(-u,x,x^{\prime})\nonumber \\
 &&
 +\frac{ m(-u,x,d)m(u,d,x^{\prime})}{ 1- m(u,d,-d_s) m(-u,-d_s,d) } 
\label{fullkernel} \\ &&
 + \frac{m(u,x,-d_s)m(-u,-d_s,x^{\prime})}{ 1- m(u,d,-d_s) m(-u,-d_s,d) }
\nonumber\\ && 
 +\frac{m(-u,x,d)m(u,d,-d_s)m(-u,-d_s,x^{\prime})}{1-m(u,d,-d_s) m(-u,-d_s,d)}
\nonumber \\ && 
+\frac{m(u,x,-d_s)m(-u,-d_s,d)m(u,d,x^{\prime})}{1-m(u,d,-d_s) m(-u,-d_s,d) } 
\bigg]. \nonumber 
\end{eqnarray}
\noindent
Equation (\ref{fullkernel}) gives the exact linear-response kernel of
any quasi-one-dimensional system, consisting of a combination of
normal and superconducting layers extending from $x=-d_s$ to $x=d$.
The kernel is expressed in terms of the quasi-classical Green's
functions in absence of the fields, which may be specified for the
particular problem of interest. We note two characteristic features
of Eq.~(\ref{fullkernel}): The factor $1-g_{0}$ measures the deviation 
from the normal state Green's function $g_0\equiv 1$, which produces
no screening current. The propagator
$m(u,x,x')$ shows up in six summands which represent all the
ballistic paths from $x$ to $x^{\prime}$, accounting for multiple
reflection at the walls at $-d_s$ and $d$. Thus the first two
summands connecting $x$ and $x'$ directly constitute the
bulk contribution, while the additional four summands are specific
to a finite system (assuming specular reflection at the boundary).
We note that a form similar to (\ref{fullkernel}) may be derived for
non-ideal interfaces between the normal and superconducting layers,
if the appropriate boundary conditions following \cite{zaitsev} 
are taken into account.

For the NS system considered in Chapter \ref{magresponse}, 
the linear response kernel (\ref{fullkernel}) may be simplified
using $m(u,x,-\infty) \to 0$ and $m(-u,-\infty,x) \to 0$ 
as $d_s\to \infty$ ($u>0$), obtaining
\begin{eqnarray} 
K(x,x^\prime)& =& e^{2} v_F N_0 \pi T \sum_{\omega_n>0}
 \int\limits_{0}^{v_{F}}du\frac{v_{F}^{2}-u^{2}}{v_{F}^{2}u}
 \left[ 1+g_{0}(x,u)\right] \left[1-g_{0}(x',u)\right] \nonumber \\
 \nonumber & &
 \bigg[\Theta(x-x^{\prime})m(u,x,x^{\prime})+
 \Theta(x^{\prime}-x)m(-u,x,x^{\prime}) \label{jyinf} \\ &&
 + m(-u,x,d)m(u,d,x^{\prime}) \bigg]\;. \nonumber
\end{eqnarray}

\backmatter

\chapter{List of publications}
\vspace{-.5cm}
\begin{itemize}
\item
A.~L.\ Fauch\`ere, W.\ Belzig, and G.\ Blatter,\\
{\it Paramagnetic Instability at Normal-Metal - Superconductor Interfaces},\\
Physical Review Letters 82, 3336 (1999).

\item
L.~B.~Ioffe, V.~B.~Geshkenbein, M.~V.~Feigel'man, A.~L.\ Fauch\`ere,\\ 
and G.~Blatter, 
{\it Quiet SDS Josephson Junctions for Quantum Computing},\\
Nature 398, 679 (1999).

\item
U.\ Ledermann, A.~L.\ Fauch\`ere, and G.\ Blatter,\\
{\it Nonlocality in mesoscopic Josephson junctions with strip geometry},\\
Physical Review B 59, R9027 (1999).

\item
A.~L.\ Fauch\`ere, V.~B.\ Geshkenbein, and G.\ Blatter,\\
{\it Comment on ``Orbital Paramagnetism of Electrons in Proximity to a \\
Superconductor''}, Physical Review Letters 82, 1796 (1999).

\item
W.\ Belzig, C.\ Bruder, A.~L.\ Fauch\`ere,\\
{\it Diamagnetic Response at Arbitrary Impurity Concentration},\\
Physical Review B 58, 14531 (1998).

\item
A.~L.~Fauch\`ere, G.~Lesovik, and G.\ Blatter,\\
{\it Finite Voltage Shot Noise in Normal-Metal - Superconductor Junctions},\\
Physical Review B 58, 11177 (1998).

\item
A.~L.\ Fauch\`ere and G.\ Blatter\\
{\it Magnetic Breakdown in a Normal-Metal - Superconductor Proximity 
Sandwich},\\
Physical Review B 56, 14102 (1997).

\item
G.\ Lesovik, A.~L.\ Fauch\`ere, and G.\ Blatter,\\
{\it Nonlinearity in NS transport: Scattering Matrix Approach},\\
Physical Review B 55, 3146 (1997).

\item
J.~Boutin and A.~L.\ Fauch\`ere,\\
{\it Peptide Libraries: Statistical Evaluation of the Peptide Distribution},\\
Trends in Pharmacological Sciences (TiPS) 17, 8 (1996).\\
\end{itemize}
\newpage
\thispagestyle{empty}

\chapter{Curriculum Vitae}

\begin{tabular}{rl}

1/1996 -- present & Ph.~D.\ research work and teaching assistance at the 
                   Institute for\\ & Theoretical Physics, Eidgen\"ossische  
                  Technische Hochschule (ETH),\\ & Zurich, Switzerland \\
3/1999  & Dissertation on {\it Transport and Magnetism in Mesoscopic}\\ & 
     {\it Superconductors}, under the supervision of Prof.\ Dr.\ G.\ Blatter \\
\\
10/1990 -- 10/1995 &  Physics studies at the Eidgen\"ossische Technische 
                     Hochschule   (ETH), \\
                     & Zurich, Switzerland \\
10/1995      &   Masters in theoretical physics, with distinction, thesis on \\
        &  {\it Transport in Mesoscopic Metallic Cylinders}, \\ &
            with Prof.\ Dr.\ G.\ Blatter\\
\\
12/1993     & Nomination to the Swiss Study Foundation --- Fonds for  
             talented\\ &  young people \\
\\
4/1984 -- 9/1990 &  High school, Kantonsschule Limmattal, 
                    Urdorf, Switzerland \\
9/1990 & Diploma (Matura type B), with distinction\\
\\
8/1988 -- 7/1989 & Student exchange program (AFS), Walnut Hills High School, \\
                 &  Cincinnati, Ohio, USA\\
5/1989       & Advanced Placement Exams of College Board\\
\\
4/1978 -- 3/1984 & Elementary school, Geroldswil and Urdorf, Switzerland\\
\\
21/5/1971 & Born in Sion, Valais, Switzerland
\\
\end{tabular}

\end{document}